\documentclass[12pt, a4paper]{article}
\usepackage[utf8]{inputenc}
\pdfoutput=1

\usepackage[top=1in, bottom=1in, left=0.75in, right=0.75in]{geometry}
\usepackage[T1]{fontenc}
\usepackage{lmodern}
	\DeclareFontFamily{OMX}{lmex}{}
	\DeclareFontShape{OMX}{lmex}{m}{n}{<-> lmex10}{}	
\usepackage{abstract}
\usepackage{amsmath, amssymb, amsfonts, mathtools}	
	\numberwithin{equation}{section}	
\usepackage{graphicx}       
	\graphicspath{ {./figures/} }
\usepackage[caption=false]{subfig} 
\usepackage{caption}   
\usepackage[dvipsnames]{xcolor}         
\usepackage{physics}
\usepackage{tensor}
\usepackage{mathrsfs}
\usepackage{bm}
\usepackage{enumitem}
\usepackage{cite}
\usepackage{bigints}    
\usepackage[bottom]{footmisc}	

\usepackage[subfigure]{tocloft}

\usepackage[normalem]{ulem}

\usepackage{hyperref}		
\definecolor{myRed}{rgb}{0.545,0,0}
\definecolor{myDarkBlue}{rgb}{0,0,0.5}
\hypersetup{colorlinks=true, linktocpage, linkcolor=blue, citecolor={myRed}, urlcolor=blue}  

\def \be {\begin{equation}}
\def \ee {\end{equation}}
\def \nn {\nonumber}
\def \del {\partial}

\DeclarePairedDelimiterX\inp[2]{\langle}{\rangle}{#1\,\delimsize\vert\,\mathopen{}#2}	
\newcommand{\overbar}[1]{\mkern 1.5mu\overline{\mkern-1.5mu#1\mkern-1.5mu}\mkern 1.5mu}
\DeclareMathOperator{\sign}{sgn}

\begin{document}

\begin{titlepage}
\vspace*{-10mm}   
\baselineskip 10pt      
\baselineskip 20pt   

\begin{center}
\noindent
{\LARGE\bf
Towards Hamiltonian Formalism for \\ String Field Theory and Nonlocality
\par}
\vskip 10mm
\baselineskip 20pt

\renewcommand{\thefootnote}{\fnsymbol{footnote}}

{\large
Chih-Hao~Chang$^a$\,\footnote[1]{\texttt{\url{r13222016@ntu.edu.tw}}},
Pei-Ming~Ho$^{a, b}$\,\footnote[2]{\texttt{\url{pmho@phys.ntu.edu.tw}}},
I-Kwan~Lee$^a$\,\footnote[3]{\texttt{\url{b10202045@ntu.edu.tw}}},
Wei-Hsiang~Shao$^a$\,\footnote[4]{\texttt{\url{whsshao@gmail.com}}}
}

\renewcommand{\thefootnote}{\arabic{footnote}}

\vskip5mm

{\normalsize \it  
$^a$Department of Physics and Center for Theoretical Physics, National Taiwan University, \\
No. 1, Sec. 4, Roosevelt Road, Taipei 106319, Taiwan
\\
\vspace{.2cm}
$^b$Physics Division, National Center for Theoretical Sciences, \\
No. 1, Sec. 4, Roosevelt Road, Taipei 106319, Taiwan
}  

\vskip 15mm
\begin{abstract}
\vspace{-3mm}
\normalsize

String field theories exhibit exponential suppression of interactions 
among the component fields at high energies 
due to infinite-derivative factors such as $e^{\ell^2 \Box / 2}$ in the vertices. 
This nonlocality has hindered the development of a consistent Hamiltonian formalism, leading some to question whether such a formalism is even viable.
To address this challenge, we introduce a toy model inspired by string field theory
and construct its Hamiltonian formalism by demanding that it reproduce 
all correlation functions derived from the path-integral formalism.
Within this framework, 
we demonstrate for this toy model that 
physical-state constraints can be imposed to eliminate negative-norm states, 
while zero-norm states decouple from the physical state space.
This approach provides a novel perspective on 
the nonlocality inherent in string field theories.

\end{abstract}
\end{center}

\end{titlepage}

\newcommand\afterTocSpace{\bigskip\medskip}
\newcommand\afterTocRuleSpace{\bigskip\bigskip}

\hrule
\tableofcontents
\afterTocSpace
\hrule
\afterTocRuleSpace

\section{Introduction}

String field theories~\cite{Kaku:1974zz, Witten:1985cc, Zwiebach:1992ie, deLacroix:2017lif} are off-shell formulations of string theories.
In principle, an action for a string field theory can be constructed in terms of the component spacetime fields.
A salient feature of such an action is the presence of infinite-order time derivatives.
Schematically, the action takes the form~\cite{Cremmer:1974ej, Neveu:1985gx, Hata:1985zu, Hata:1986jd, Cremmer:1986if, Samuel:1986wp, Gross:1986ia, Eliezer:1989cr, Kostelecky:1989nt, Taylor:2003gn, Pius:2016jsl, Sen:2016qap}
\be
\label{SFT_action}
S = 
\bigintssss d^{D + 1} X
\Biggl[
\frac{1}{2}
\sum_j \phi_j \left( \Box - m_j^2 \right) \phi_j
- \sum_{j_1, \, \cdots, \, j_n} 
\frac{1}{n!} \, \lambda_{j_1 \cdots j_n} \, \tilde{\phi}_{j_1} \cdots \tilde{\phi}_{j_n}
\Biggr] 
\, ,
\ee
where each field $\phi_j$ appears in the interaction terms only through the nonlocal form
\be
\label{phi_tilde}
\tilde{\phi}_j 
\equiv
e^{\ell^2 \Box / 2} \, \phi_j \, .
\ee
Here,
$\ell^2 \sim \ell_s^2 \equiv \alpha'$ 
is of the same order of magnitude 
as the Regge slope parameter $\alpha'$.\footnote{
The value of the proportionality constant 
depends on the specific theory under consideration. 
For instance, in Witten's bosonic open string field theory~\cite{Witten:1985cc}, 
we have $\ell^2 = 2 \alpha' \ln\left( 3 \sqrt{3} / 4 \right)$. 
}
The action~\eqref{SFT_action} has been employed~\cite{Tomboulis:2015gfa, Pius:2016jsl, Carone:2016eyp, Briscese:2018oyx, Chin:2018puw, Pius:2018crk, Buoninfante:2018mre, DeLacroix:2018arq, Buoninfante:2018gce, Briscese:2021mob, Koshelev:2021orf, Erbin:2021hkf} 
to study various aspects of string field theories, 
such as causality, UV finiteness, and unitarity of the perturbative $S$-matrix. 

The infinite derivatives appearing in the interaction terms 
through $\tilde{\phi}_j$~\eqref{phi_tilde}
are crucial for the removal of ultraviolet (UV) divergences,
and they indicate nonlocality as a fundamental feature of string theory~\cite{Calcagni:2013eua}.\footnote{
Derivative couplings are omitted from the action~\eqref{SFT_action}.
While derivative couplings of finite order do not significantly affect the results discussed in this work, we cannot draw definitive conclusions regarding infinite-order derivative couplings which are also present in string field theories. We simply assume that such terms are absent in the models under consideration.}
They are also believed to be at the root of stringy properties such as the space-time uncertainty relation~\cite{Yoneya:1987gb, Yoneya:1989ai, Yoneya:1997gs, Yoneya:2000bt} 
as well as the generalized uncertainty principle~\cite{Amati:1988tn, Konishi:1989wk, Guida:1990st}.\footnote{Moreover, it has been suggested~\cite{Lowe:1994ns, Lowe:1995ac, Ho:2023tdq, Ho:2024tby} that such stringy nonlocality could play a significant role 
in addressing the black hole information paradox~\cite{Hawking:1976ra}.}
On the other hand, the presence of infinite derivatives in string theories has led to skepticisms about the feasibility of developing a well-defined Hamiltonian formalism~\cite{Eliezer:1989cr}.
Indeed, in the presence of nonlocality in the time direction, how can one define a Hamiltonian as an operator that generates infinitesimal time evolution? 
This leads to a more fundamental question: 
Can we assert that string theory qualifies as a well-defined quantum theory 
without a Hamiltonian formalism?
Motivated by the significance of this issue,
the primary purpose of this paper is to resolve the long-standing tension 
between the nonlocality in the action~\eqref{SFT_action} 
and a consistent Hamiltonian formalism.

Numerous efforts have been made~\cite{Llosa94, Barci:1995ad, Woodard:2000bt, Calcagni:2007ef, Ferialdi:2012, Talaganis:2017vsp, Heredia:2021wja, Heredia:2022mls} to establish Hamiltonian formalisms for nonlocal theories, even at the level of free field theories.
Approaches based on expanding the theory in powers of derivatives following the ideas of Ostrogradski~\cite{Ostrogradsky:1850fid, Woodard:2015zca} typically result in ill-defined Hamiltonian formalisms, where either the energy is unbounded from below or negative-norm states are present (unless the phase space is projected to the low-energy subspace~\cite{Cheng:2001du, Cheng:2002rz}).
Here, we propose a different approach to constructing Hamiltonian formalisms for models with the same type of nonlocality as the string field theory action~\eqref{SFT_action}. 
Since the path-integral formalism for such theories 
has been extensively studied through analyses of their $S$-matrix properties~\cite{Pius:2016jsl, Carone:2016eyp, Briscese:2018oyx, Chin:2018puw, Pius:2018crk, Buoninfante:2018mre, DeLacroix:2018arq, Buoninfante:2018gce, Briscese:2021mob, Koshelev:2021orf, Buoninfante:2022krn}, 
our strategy is to define a Hamiltonian formalism 
aimed at reproducing the path-integral results. 

Note that the nonlocal interactions in string field theories 
lead to subtleties even in the path-integral formalism.
As is apparent from the action~\eqref{SFT_action}, 
the exponential operator $e^{\ell^2 \Box / 2}$ 
responsible for the suppression of vertices in the limit of large space-like momenta 
is exponentially large for large time-like momenta. 
A common way to regulate such behavior in $S$-matrix calculations 
is to carry out integrations over momenta in the Euclidean momentum space. 
The resulting correlation functions can be analytically continued to Lorentzian signature 
by suitably deforming the integration contours, 
as explained in ref.~\cite{Pius:2016jsl} 
(see also refs.~\cite{Pius:2018crk, Buoninfante:2022krn}). 
Nevertheless, it remains difficult to 
introduce a time-dependent background in this framework, 
as the Wick rotation scheme may fail 
depending on the temporal profile of the background field~\cite{Visser:2017atf}.\footnote{
An example is the interaction with a 
Gaussian background field $b(t) \propto e^{-t^2 / 2 L^2}$ 
whose Fourier transform $\mathcal{F}[b](\omega) \propto e^{- L^2 \omega^2 / 2}$ 
blows up exponentially as $\omega$ approaches infinity away from the real axis.
}
The need to Wick-rotate the time direction 
in order to make sense of the exponential operator $e^{\ell^2 \Box / 2}$ 
presents an additional challenge in 
formulating a Hamiltonian formalism, 
which relies on a \textit{real} time coordinate.

As an alternative to Euclideanizing the momentum space, 
we adopt a recently proposed prescription 
in which the string length parameter $\ell$ 
undergoes an analytic continuation~\cite{Ho:2023tdq}:
\be 
\label{ellE}
\ell^2 \to i \, \ell_E^2
\qquad 
\text{with}
\quad 
\ell_E^2 > 0
\, .
\ee 
Under this continuation, the exponential operator in eq.~\eqref{SFT_action} 
becomes $e^{i \, \ell_E^2 \, \Box / 2}$, which also induces suppression at high energies. 
This prescription allows for the incorporation of interactions 
with non-stationary background fields while preserving the Lorentzian signature.
In the context of string amplitudes, 
it corresponds to an analytic continuation of 
the modular parameters of the string worldsheet~\cite{Berera:1992tm, Witten:2013pra}.
The validity of this analytic continuation for extracting physical results 
will be justified in section~\ref{sec:toy} and appendix~\ref{app:AC}.
Building on this framework, 
we shall extend the formalism developed in ref.~\cite{Ho:2023tdq} 
into a more complete Hamiltonian formalism.

Another important aspect of the Hamiltonian formalism for the action~\eqref{SFT_action} 
is that a perturbation theory in the coupling constants $\lambda_{j_1\cdots j_n}$ 
is only possible in terms of the field variables $\tilde{\phi}_j$
(see section~\ref{sec:toy_Ham} and appendix~\ref{app:pert} for details).
Since a Hamiltonian formalism is sensitive to the order of time derivatives ---
and higher derivatives appear only in the interaction terms of the action~\eqref{SFT_action} ---
the difference between the zeroth-order $\mathcal{O}(\lambda^0_{j_1\cdots j_n})$ 
and first-order $\mathcal{O}(\lambda^1_{j_1\cdots j_n})$ expansions
in the Hamiltonian formalism is not small 
when expressed in terms of the field variables $\phi_j$,
even for arbitrarily small couplings $\lambda_{j_1\cdots j_n}$. 
In constrast,
when formulated in terms of $\tilde{\phi}_j$,
the nonlocality is entirely encoded in 
the free-field part of the action~\eqref{SFT_action} 
(see eq.~\eqref{SFT_action2} below),
and a perturbative expansion of the interaction terms 
does not introduce additional higher derivatives.
Therefore, it is natural to conjecture that 
the details of the interaction terms are irrelevant 
when we focus on the nonlocal features of the theory~\eqref{SFT_action} 
in terms of the variables $\tilde{\phi}_j$.
This conjecture remains to be verified in future works.

As a first step toward understanding 
the Hamiltonian formalism for string field theory, 
this paper focuses on a simplified toy model, 
defined below in eq.~\eqref{S_phi}. 
This model exhibits the same type of nonlocality as string field theories 
but is restricted to quadratic interactions with background fields.
Although the toy model~\eqref{S_phi} considered here 
is quadratic in the field and can thus be viewed as a free field theory, 
the problem of a consistent Hamiltonian formalism 
remains highly nontrivial due to nonlocality. 
Indeed, similar models have been studied~\cite{Llosa94, Barci:1995ad, Woodard:2000bt, Calcagni:2007ef} in past attempts to construct Hamiltonian descriptions of nonlocal theories, 
yet a fully consistent Hamiltonian formalism 
has not been achieved at the quantum level, 
largely due to Ostrogradskian instabilities~\cite{Eliezer:1989cr, Woodard:2000bt, Woodard:2015zca}.
In this work, we explicitly demonstrate how 
a well-defined Hamiltonian formalism can be constructed 
for this toy model~\eqref{S_phi} despite the nonlocality.
While we leave the full action~\eqref{SFT_action} with generic interactions 
for future investigations, 
our results suggest a new paradigm for the quantization of nonlocal theories.

Specifically, 
the toy model under consideration in this work has the action
\be 
\label{S_phi}
S_{\phi}
= 
\int d^2 X 
\left[ 
\frac{1}{2} \, \phi \, \Box \, \phi
+
2 \lambda \, 
\widetilde{B}(V) \, 
\tilde{\phi}^2
\right]
\ee 
in the light-cone frame $(U, V)$
of $(1 + 1)$-dimensional Minkowski space $X^{\mu} = (t, x)$, 
where
\be 
\label{UandV}
U \equiv t - x \, ,
\qquad 
V \equiv t + x 
\ee 
are the light-cone coordinates.
It is equivalent to
\be
\label{S-phi-2}
S_{\phi}[\tilde{\phi}]
= 
\int d^2 X 
\left[ 
\frac{1}{2} \, \tilde{\phi} \, \Box \, e^{-\ell^2 \Box} \, \tilde{\phi}
+
2 \lambda \, 
\widetilde{B}(V) \, 
\tilde{\phi}^2
\right]
\ee
in terms of the field variable $\tilde{\phi} \equiv e^{\ell^2 \Box / 2} \phi$.
Here, $\widetilde{B}(V)$ represents a fixed background profile, 
and the infinite-derivative modification appears through 
$\tilde{\phi}$ in the interaction term of~\eqref{S_phi}. 
Since we will focus on the high-energy limit in the $U$-direction, 
where nonlocal effects are most significant,
it is reasonable to neglect the $U$-dependence 
of the background field $\widetilde{B}$ 
and consider only its variation along $V$.
The action~\eqref{S_phi} serves as a prototype for nonlocalities 
introduced by the operator $e^{\ell^2 \Box / 2}$
and can be used as a simplified model for studying the nonlocal features 
of the string field theory action~\eqref{SFT_action}.

A key reason why the light-cone frame~\eqref{UandV} is favored is that 
the d'Alembertian operator $\Box = -4 \partial_U \partial_V$ 
is first order in light-cone derivatives, 
making the degree of nonlocality milder 
compared to the usual Minkowski coordinates $(t, x)$~\cite{Erler:2004hv, Erbin:2021hkf}.
It turns out that 
for each Fourier mode with light-cone frequency $\Omega$ 
defined with respect to the retarded coordinate $U$, 
the effect of the infinite-derivative operator $e^{\ell^2 \Box / 2}$ 
reduces to a finite shift $\propto \ell^2 \Omega$ 
in the advanced time direction $V$~\cite{Ho:2023tdq}.  

Our method provides a \textit{nonperturbative} treatment of the nonlocality in eq.~\eqref{S_phi} (i.e., as opposed to expanding $e^{\ell^2 \Box / 2}$ in powers of $\ell^2 \Box$).
Crucially, the extra degrees of freedom introduced by the infinite-derivative modification will be shown to decouple from the physical Hilbert space in the Hamiltonian formalism.
Moreover, an important feature of the quantized theory is the emergence of the uncertainty bound $\Delta U \Delta V \gtrsim \ell^2$ on the physical states~\cite{Ho:2023tdq}, which can be interpreted as the light-cone analog of the stringy \textit{space-time uncertainty relation} 
$\Delta t \, \Delta x \gtrsim \ell^2$~\cite{Yoneya:1987gb, Yoneya:1989ai, Yoneya:1997gs, Yoneya:2000bt}.

The nonlocal model~\eqref{S_phi} studied here 
incorporates interactions with a time-dependent background field. 
Advancing our understanding of such nonlocal theories in time-dependent backgrounds represents a significant step toward formulating string theory in weakly curved spacetimes, with potential implications for black hole physics and cosmology, 
where quantum field theories are also typically considered at the free field level.

This paper is organized as follows.
In section~\ref{sec:toy}, we show that the toy model~\eqref{S_phi} in the light-cone frame can be interpreted as a collection of infinitely many independent copies of a one-dimensional model described by the action (see eq.~\eqref{S_a_tilde} below)
\be
S[\tilde{a}, \tilde{a}^{\dagger}]
=
\int dt 
\left[
i \, \tilde{a}^{\dagger}(t - \sigma / 2) \, 
\del_t \, \tilde{a}(t + \sigma / 2)
+
\lambda \, b(t) \, 
\tilde{a}(t) \, \tilde{a}^{\dagger}(t)
\right] 
,
\ee
where $\sigma$ is a constant parameter characterizing the scale of nonlocality,
and $b(t)$ is a fixed background field.
A Hamiltonian formalism for this nonlocal theory is established in section~\ref{sec:toy_Ham} based on its path-integral correlation functions, where we show that despite the nonlocality in time, which could potentially compromise unitarity, physical-state constraints can be imposed to remove negative-norm states and render the zero-norm states spurious.
In section~\ref{sec:stringy}, 
we apply the developed Hamiltonian formalism 
to the two-dimensional toy model~\eqref{S_phi}, 
and demonstrate as an explicit example how 
the calculation of Hawking radiation is modified, 
as suggested in refs.~\cite{Ho:2023tdq, Ho:2024tby}.
Section~\ref{sec:summary} concludes with a summary of findings and their implications.

Throughout this work, 
we adopt natural units $\hbar = c = 1$ 
and use the mostly plus convention $(-, +, \cdots, +)$ for the metric signature. 
In the $(D + 1)$-dimensional Minkowski space, the d'Alembertian operator is given by 
$\Box = - \del_t^2 + \vec{\grad}^2$, 
with $\vec{\grad}^2$ denoting the $D$-dimensional Laplacian. 

\section{Toy Model of String Field Theories}
\label{sec:toy}

In the string field theory action schematically presented as eq.~\eqref{SFT_action},
all interaction vertices are dressed with exponential operators of the form $e^{\ell^2 \Box / 2}$. 
Consequently, the detection of any field $\phi_j$ by an Unruh-DeWitt detector~\cite{Unruh:1976db, DeWitt:1980hx} is determined by the Wightman functions $\ev{\hat{\tilde{\phi}}_j (X) \hat{\tilde{\phi}}_j (X')}{0}$ associated with $\tilde{\phi}_j$, rather than $\phi_j$ itself. 
More generally, measurements of physical observables are essentially probing the correlation functions $\ev{\tilde{\phi}_i (X) \cdots \tilde{\phi}_j (X')}$ of the fields $\left\{ \tilde{\phi}_i \right\}$ as all measurements rely on interactions~\cite{Ho:2023tdq}. 
As we will see in section~\ref{sec:toy_Ham}, 
a perturbative Hamiltonian formalism for such nonlocal theories
can only be formulated in terms of $\tilde{\phi}_j$ instead of $\phi_j$.
Therefore, we rewrite the action via the change of variables~\eqref{phi_tilde} as~\cite{Siegel:2003vt}
\be
\label{SFT_action2}
S = 
\bigintssss d^{D + 1} X
\Biggl[
\frac{1}{2}
\sum_j \tilde{\phi}_j 
\left( \Box - m_j^2 \right) 
e^{- \ell^2 \Box} \, 
\tilde{\phi}_j
- \sum_{j_1, \, \cdots, \, j_n} 
\frac{1}{n!} \, 
\lambda_{j_1 \cdots j_n} \, 
\tilde{\phi}_{j_1} \cdots \tilde{\phi}_{j_n}
\Biggr]
\, ,
\ee
so that the infinite-derivative factor $e^{\ell^2 \Box / 2}$ appears only in kinetic terms.

In the path-integral formalism, there is no difference between expressing the action in the form of eq.~\eqref{SFT_action} or eq.~\eqref{SFT_action2}.
The correlation functions of $\{ \tilde{\phi}_i \}$ and those of $\{ \phi_i \}$ 
are simply related by
\be
\ev{\tilde{\phi}_i (X) \cdots \tilde{\phi}_j(X')}
=
e^{\ell^2 \Box/2} \cdots e^{\ell^2 \Box'/2}
\ev{\phi_i (X) \cdots \phi_j (X')}
\ee
according to the field redefinition~\eqref{phi_tilde}.
It is also often falsely asserted that applying the Hamiltonian formalism to either $\tilde{\phi}_i$ or $\phi_i$ makes no difference.
The common argument~\cite{Barci:1995ad, Kajuri:2017jmy} is that, in both cases, canonical quantization proceeds in the standard way as in local quantum field theories, since the infinite-derivative operator $e^{-\ell^2 \Box}$ in the kinetic terms of~\eqref{SFT_action2} does not alter the solution space of the wave equation $\Box \, \phi_i = 0$, nor does it affect the Cauchy problem for the field $\phi_i$ at the perturbative level~\cite{Barnaby:2007ve}.
However, as will be discussed in section~\ref{sec:toy_Ham}, 
the choice of variables significantly impacts the viability of a Hamiltonian formalism.


For simplicity, we shall focus our discussions on a toy model with a single massless field in $(1 + 1)$-dimensional Minkowski spacetime.
Inspired by eq.~\eqref{SFT_action2}, the kinetic term of the toy model is defined by
\be 
\label{S0}
S_0 =
\frac{1}{2}
\int d^2 X \, 
\phi \, \Box \, \phi 
= 
\frac{1}{2}
\int d^2 X \, 
\tilde{\phi} \, \Box \, e^{-\ell^2\Box} \, \tilde{\phi}
\, ,
\ee 
where
\be
\tilde{\phi} 
\equiv 
e^{\ell^2 \Box/2} \phi
= 
e^{-2 \ell^2 \del_U \del_V} \phi
\ee
in terms of the light-cone coordinates defined in eq.~\eqref{UandV}.
When acting on eigenfunctions $e^{i\Omega_U U}$ 
of the light-cone derivative $\del_U$, 
the exponential factor 
$\mathrm{exp} (- 2 \ell^2 \del_U \del_V) = \mathrm{exp}(- 2 i \ell^2 \Omega_U \del_V)$ 
behaves as a shift operator in the $V$-direction.

In momentum space $k^{\mu} = (k^0, k^1)$,
the propagator of $\tilde{\phi}$ can be inferred from the action~\eqref{S0} as 
\be 
\label{prop_p}
-i \, 
\frac{e^{-\ell^2 k^2}}{k^2 - i \epsilon}
=  
\frac{i \exp\bigl[ \ell^2 (k^0)^2 - \ell^2 (k^1)^2 \bigr]}
{(k^0)^2 - (k^1)^2 + i \epsilon}
\, .
\ee 
Given the factor $\exp\bigl[ \ell^2 (k^0)^2 \bigr]$ 
that is exponentially large for large time-like momenta,
a sensible path-integral formalism
requires some form of analytic continuation. 
For example, in perturbative $S$-matrix calculations,
loop integrals are usually defined over Euclidean momenta 
with Wick-rotated energies $k^0 = i k_E^0$ ($k_E^0 \in \mathbb{R}$),
where the originally diverging term 
$\exp\bigl[ \ell^2 (k^0)^2 \bigr] = \exp\bigl[ - \ell^2 (k_E^0)^2 \bigr]$
becomes exponentially suppressed in the UV regime.

Alternatively, one may retain the Lorentzian signature of momentum space
while complexifying the string length $\ell$ according to eq.~\eqref{ellE} as~\cite{Ho:2023tdq}
\be 
\label{ell_to_ellE (sec2)}
\ell^2 = i \, \ell_E^2
\qquad 
\text{with}
\quad 
\ell_E^2 > 0
\, .
\ee 
Under this prescription, 
the factor $\exp\bigl[ \ell^2 (k^0)^2 \bigr] = \exp\bigl[ \pm i \, \ell_E^2 \, (k^0)^2 \bigr]$ 
becomes a rapidly oscillating phase in the UV regime,
which also leads to exponential suppression. 
A detailed discussion on the validity of this analytic continuation can be found in appendix~\ref{app:AC}.
It is further demonstrated in appendix~\ref{app:continuation_schemes} that 
both continuation schemes --- 
whether through a complexified string length~\eqref{ell_to_ellE (sec2)}
or via Euclideanized momentum space ---
yield the same propagator in position space.
Additionally, 
as illustrated recently in ref.~\cite{Ho:2023tdq}, 
implementing the analytic continuation~\eqref{ell_to_ellE (sec2)} 
in the light-cone coordinates $(U, V)$ 
with corresponding outgoing and ingoing frequencies
\be 
\label{lightcone_freq}
\Omega_U 
\equiv 
\frac{k^0 + k^1}{2} 
\, ,
\qquad 
\Omega_V 
\equiv 
\frac{k^0 - k^1}{2}
\ee 
offers a potentially viable route to 
establishing a Hamiltonian formalism for string field theories.

The two methods of analytic continuation described above 
(Wick-rotating the time-like momentum $k^0$ or the string tension parameter $\ell^{-2}$) 
are merely different mathematical prescriptions for regularizing UV divergences. 
When both approaches are applicable, 
we shall demand that the outcomes of the two prescriptions be equivalent. 

In addition to the kinetic term~\eqref{S0}, 
we introduce an interaction term between the massless field $\tilde{\phi}$ 
and an external background field $\widetilde{B}(X) \equiv e^{\ell^2 \Box / 2} B(X)$, 
and work with the action
\begin{align}
S_{\phi}[\tilde{\phi}]
&= 
\int d^2 X 
\left[ 
\frac{1}{2} \, \tilde{\phi} \, \Box \, e^{-\ell^2 \Box} \, \tilde{\phi}
+
2 \lambda \, 
\widetilde{B}(X) \, 
\tilde{\phi}^2
\right]
\label{S-phi-3}
\\
&=
\int dU \int dV 
\left[ 
( \del_U \tilde{\phi} ) \, 
e^{4 \ell^2 \del_U \del_V}
\del_V
\tilde{\phi}
+
\lambda \, \widetilde{B}(U, V) \, \tilde{\phi}^2
\right]
,
\label{SFT_action_lightcone}
\end{align}
where $\lambda$ is a small, dimensionless coupling constant. 
This can be viewed as a simplified version of the general string field theory action~\eqref{SFT_action2} in the light-cone frame $(U, V)$.

We shall treat the light-cone coordinate $V$ 
as the time coordinate in the light-cone frame.
The outgoing sector of the massless field,
in terms of $\phi$ and $\tilde{\phi}$~\eqref{phi_tilde},
can be expanded in Fourier modes as
\begin{align}
\label{phi_Fourier}
\phi(U, V)
&= 
\int_0^{\infty} 
\frac{d\Omega}{\sqrt{4\pi\Omega}} 
\left[ 
a_{\Omega}(V) \, e^{- i \Omega U} 
+ 
a^{\dagger}_{\Omega}(V) \, e^{i \Omega U} 
\right]
,
\\
\label{phi_tilde_Fourier}
\tilde{\phi}(U, V)
&= 
\int_0^{\infty} 
\frac{d\Omega}{\sqrt{4\pi\Omega}} 
\left[
\tilde{a}_{\Omega}(V) \, e^{- i \Omega U} 
+ 
\tilde{a}^{\dagger}_{\Omega}(V) \, e^{i \Omega U} 
\right]
,
\end{align}
respectively.\footnote{
Since the ingoing modes $e^{- i \Omega_V V}$
are omitted in the discussion below,
we will now refer to $\Omega_U$ simply as $\Omega$ for convenience.
}
Due to eq.~\eqref{phi_tilde}, 
the Fourier components are related by 
\begin{align}
\label{a_tilde_vs_a}
\tilde{a}_{\Omega}(V) 
&= 
a_{\Omega}(V - \sigma_{\Omega}/2)
\, ,
\\
\tilde{a}^{\dagger}_{\Omega}(V) 
&= 
a^{\dagger}_{\Omega}(V + \sigma_{\Omega}/2)
\, ,
\end{align}
where
\be 
\label{sigma_Om}
\sigma_{\Omega} 
\equiv 
- 4 i \ell^2 \Omega 
\, .
\ee 
Upon analytic continuation~\eqref{ell_to_ellE (sec2)} 
of the string tension~\cite{Ho:2023tdq},
the parameter 
\be 
\label{sigma_Om_ellE}
\sigma_{\Omega} \rightarrow 
4 \ell_E^2 \, \Omega
\ee 
is positive-definite,
and it quantifies the length scale of nonlocality 
associated to a given mode with outgoing light-cone frequency $\Omega$.
There is a larger nonlocality for a higher-frequency mode.

For simplicity, we shall focus on the case when the background field $B(U, V)$ is $U$-independent so that $\widetilde{B}(V) = e^{-2 \ell^2 \del_U \del_V} B(V) = B(V)$
(hence reducing eq.~\eqref{S-phi-3} to eq.~\eqref{S-phi-2}).
This is also a good approximation in the high-energy limit 
where $\Omega$ is much larger than the characteristic frequency 
of the background field $B(U, V)$ in the $U$-direction.
Plugging in the mode expansions~\eqref{phi_Fourier} and~\eqref{phi_tilde_Fourier},
the action~\eqref{SFT_action_lightcone} can be expressed as
\be 
\label{S_2d_phi}
S_{\phi}[\phi]
= 
\int dV \int_0^{\infty} d\Omega 
\left[
i \, a^{\dagger}_{\Omega}(V) \, 
\del_V \, a_{\Omega}(V)
+ 
\frac{\lambda}{\Omega} \, 
B(V) \, 
a_{\Omega}(V - \sigma_{\Omega}/2) \, 
a^{\dagger}_{\Omega}(V + \sigma_{\Omega}/2)
\right]
,
\ee 
or equivalently,
\be 
\label{S_2d_phi_tilde}
S_{\phi}[\tilde{\phi}]
= 
\int dV \int_0^{\infty} d\Omega 
\left[
i \, \tilde{a}^{\dagger}_{\Omega}(V - \sigma_{\Omega} / 2) \, 
\del_V \, \tilde{a}_{\Omega}(V + \sigma_{\Omega} / 2)
+ 
\frac{\lambda}{\Omega} \, 
B(V) \, 
\tilde{a}_{\Omega}(V) \, 
\tilde{a}^{\dagger}_{\Omega}(V)
\right]
.
\ee 
Although the action of the toy model considered here is quadratic in the fields, 
it falls within a class of nonlocal theories 
that were previously shown to exhibit instabilities 
in the Ostrogradskian framework~\cite{Woodard:2000bt} 
(taking the infinite-derivative limit).
Nevertheless, we will explicitly demonstrate that the Hamiltonian description we construct for this model remains free of such instabilities --- a crucial and nontrivial feature of the formalism proposed in this work. 

The system described by the action~\eqref{S_2d_phi_tilde}
is merely infinite copies of decoupled Fourier modes 
$\tilde{a}_{\Omega}(V)$ and $\tilde{a}^{\dagger}_{\Omega}(V)$.
To understand this system,
it is sufficient to focus on a single Fourier mode.
Hence,
we shall omit the label $\Omega$
and make the following replacements:
\begin{align}
\bigl( a_{\Omega}(V) \, , a_{\Omega}^{\dagger}(V) \bigr)
&\to 
\bigl( a(t) \, , a^{\dagger}(t) \bigr)
\, ,
\label{corres_a}
\\
\bigl( \tilde{a}_{\Omega}(V) \, , \tilde{a}_{\Omega}^{\dagger}(V) \bigr)
&\to 
\bigl( \tilde{a}(t) \, , \tilde{a}^{\dagger}(t) \bigr)
\, ,
\label{corres_a_tilde}
\\
B(V) / \Omega 
&\to 
b(t)
\, ,
\\
\sigma_{\Omega} 
&\to
\sigma \, ,
\label{corres_B_b}
\end{align}
where we have also renamed $V$ as $t$.
This reduces the toy model~\eqref{S_2d_phi} to a one-dimensional model governed by the action
\be
\label{S_a}
S[a, a^{\dagger}]
=
\int dt 
\left[
i \, a^{\dagger}(t) \, \del_t \, a(t)
+
\lambda \, b(t) \, 
a(t - \sigma / 2) \, a^{\dagger}(t + \sigma / 2)
\right]
,
\ee
or in terms of $\tilde{a}$ and $\tilde{a}^{\dagger}$:
\be
\label{S_a_tilde}
S[\tilde{a}, \tilde{a}^{\dagger}]
=
\int dt 
\left[
i \, \tilde{a}^{\dagger}(t - \sigma / 2) \, 
\del_t \, \tilde{a}(t + \sigma / 2)
+
\lambda \, b(t) \, 
\tilde{a}(t) \, \tilde{a}^{\dagger}(t)
\right]
,
\ee
where $\sigma > 0$ represents the length scale of nonlocality.
Based on eq.~\eqref{a_tilde_vs_a}, the variable sets $(a, a^{\dagger})$ and $(\tilde{a}, \tilde{a}^{\dagger})$ in this 1D model are related via
\be
\label{a_tilde_vs_a (toy)}
\tilde{a}(t) 
= 
a(t - \sigma/2)
\qquad \text{and} \qquad
\tilde{a}^{\dagger}(t) 
= 
a^{\dagger}(t + \sigma/2)
\, .
\ee
In the section below, we shall study this 1D model in the perturbation theory with respect to the coupling constant $\lambda$.

Let us comment that since $\ell^2$ enters the theory~\eqref{SFT_action_lightcone} 
exclusively through the factor $e^{4 \ell^2 \del_U \del_V}$, 
analytically continuing $\ell$ as in~\eqref{ell_to_ellE (sec2)} is equivalent to 
Wick-rotating the light-cone time coordinate as $V \rightarrow -i V_E$ ($V_E \in \mathbb{R}$) 
while keeping the string length $\ell$ real. 
Consequently, 
all results derived under the analytic continuation of the string length 
with a real time coordinate have a one-to-one correspondence with 
those obtained using a Wick-rotated time coordinate $V_E$ and a real string length $\ell$. 
This equivalence is explicitly demonstrated in appendix~\ref{app:continuation_schemes}
for the free propagator of $\tilde{\phi}$. 
Similarly, the physical observables in the nonlocal 1D model~\eqref{S_a_tilde} have a one-to-one correspondence with those derived from the action 
\be
\label{S_a_tilde_tE}
S_E [\tilde{a}, \tilde{a}^{\dagger}]
=
\int dt_E
\left[
i \, \tilde{a}^{\dagger}(t_E - \sigma_E / 2) \, 
\del_{t_E} \, \tilde{a}(t_E + \sigma_E / 2)
+
\lambda \, b(t_E) \, 
\tilde{a}(t_E) \, \tilde{a}^{\dagger}(t_E)
\right]
,
\ee
where 
\be
\label{Wick-V}
t = - i t_E \qquad (t_E \in \mathbb{R}) 
\, ,
\ee
and $\sigma_E \equiv i \sigma > 0$.

\section{Hamiltonian Formalism for Nonlocal 1D Model}
\label{sec:toy_Ham}

The goal of this section is to construct a Hamiltonian formalism that reproduces the path-integral correlation functions of the nonlocal 1D model~\eqref{S_a_tilde}.
Typically, the path integral and Hamiltonian formalisms are required to yield 
consistent correlation functions, i.e.,
\be
\langle 
\cdots O_i (X_i) 
\cdots O_j (X_j) \cdots 
\rangle
= 
\ev{\mathcal{T} 
\bigl\{ 
\cdots \hat{O}_i (X_i) 
\cdots \hat{O}_j (X_j) \cdots
\bigr\}
}{0}
,
\ee 
where $\mathcal{T}$ denotes time ordering,
and $\hat{\cal O}_i(X_i)$ are time-dependent operators in the Heisenberg picture.\footnote{
Here we assume that the operators $\hat{O}_i$ are defined 
in terms of the fundamental fields without time derivatives.
Otherwise, 
the time derivatives should be carried out 
after the time-ordering operation on the right-hand side.
}

Naively, one may attempt to quantize the 1D model~\eqref{S_a} in terms of $a(t)$ and $a^{\dagger}(t)$ as a perturbation theory in the coupling constant $\lambda$.
However, as shown in appendix~\ref{app:pert}, this straightforward treatment is unable to reproduce the path-integral correlation functions in the Heisenberg picture.
In particular, it is demonstrated that 
\be 
\label{mismatch}
\ev{\mathcal{T} 
\bigl\{ 
\hat{a}(t_1) \, \hat{a}^{\dagger}(t_2)
\bigr\}
}{0}
\neq 
\ev*{a(t_1) \, a^{\dagger}(t_2)}
\ee 
in the perturbation theory.
This mismatch arises for the following reasons.
In terms of $a(t)$ and $a^{\dagger}(t)$, 
the unperturbed part of the action~\eqref{S_a} is local, 
with the nonlocality confined to the interaction term. 
Consequently, in the canonical quantization of $a(t)$ and $a^{\dagger}(t)$, 
a perturbative treatment of the interaction implies 
treating the nonlocality perturbatively as well.
Since perturbation theory introduces only small corrections 
to the zeroth-order operators $\hat{a}_0$ and $\hat{a}_0^{\dagger}$ 
in the unperturbed local theory, 
it systematically ignores the extra structures 
associated with the nonlocality appearing in the interaction. 
Indeed, we show in appendix~\ref{app:pert} that 
the mismatch~\eqref{mismatch} occurs starting at first order in $\lambda$ 
precisely because the operators $\hat{a}_0$ and $\hat{a}_0^{\dagger}$ 
are time-independent and fail to capture the effects of 
the infinite time derivatives present in the full theory.

In Ostrogradski's formulation~\cite{Ostrogradsky:1850fid, Woodard:2015zca} 
of higher-derivative theories,
the dimension of the phase space depends on the order of time derivatives.
When all higher-order derivatives are introduced solely through interaction terms, 
as in~\eqref{S_a},
turning on even a slight interaction 
leads to an abrupt change in the phase space dimension.\footnote{
Strictly speaking, 
Ostrogradski's framework does not apply to 
theories with infinite time derivatives.
In fact, as we will see in section~\ref{sec:spurious}, 
the nonlocality in the 1D model~\eqref{S_a_tilde} 
does not change the dimension of the physical state space 
compared to ordinary local theories. 
} 
This discontinuity is not a small correction to the canonical structure, 
and thus the perturbative approximation is expected to fail.

On the other hand, by working with $\tilde{a}(t)$ and $\tilde{a}^{\dagger}(t)$,
the nonlocality is addressed directly 
at the level of the free theory (see eq.~\eqref{S_a_tilde}), 
while the interaction term can be handled using perturbation theory.

In this work, 
we do not seek to present a systematic Hamiltonian formalism 
applicable to all nonlocal theories.
Instead, our focus is on developing a Hamiltonian formalism 
tailored to~\eqref{S_a_tilde} in terms of $\tilde{a}(t)$ and $\tilde{a}^{\dagger}(t)$,
such that it aligns with the path-integral formalism. 
This is accomplished by requiring the correspondence 
\be
\label{path_op_corres}
\ev{\mathcal{T} 
\bigl\{
\cdots \hat{\tilde{a}}(t_i) \cdots \hat{\tilde{a}}^{\dagger}(t_j) \cdots
\bigr\}
}{0}
=
\bigl\langle 
\cdots \tilde{a}(t_i) \cdots \tilde{a}^{\dagger}(t_j) \cdots 
\bigr\rangle
\, ,
\ee 
which, as we will see later, 
can only be satisfied through analytic continuation.

\subsection{Operator algebra from two-point function}
\label{sec:2pt_function}

As the nonlocal 1D model~\eqref{S_a_tilde} has a quadratic action, 
all higher-point correlation functions are determined by 
the two-point correlation function via Wick's theorem.
Therefore, the complete information of the theory is encoded in the two-point correlation function.
In this subsection, we evaluate the two-point correlation function in the path-integral formalism and use it to determine the commutation relations for the Hamiltonian formalism.

Using the path-integral formalism, 
one can derive the Schwinger-Dyson equations
\begin{align}
\del_t \, 
\bigl\langle 
\hat{\tilde{a}}(t) \, \hat{\tilde{a}}^{\dagger} (t')
\bigr\rangle
-
i \lambda \, 
b(t - \sigma) \, 
\bigl\langle 
\hat{\tilde{a}}(t - \sigma) \, \hat{\tilde{a}}^{\dagger} (t')
\bigr\rangle 
&=
\delta(t - t' - \sigma)
\, , 
\label{SD1}
\\ 
\del_{t'} \, 
\bigl\langle 
\hat{\tilde{a}}(t) \, \hat{\tilde{a}}^{\dagger} (t')
\bigr\rangle 
+
i \lambda \, 
b(t' + \sigma) \, 
\bigl\langle  
\hat{\tilde{a}}(t) \, \hat{\tilde{a}}^{\dagger} (t' + \sigma)
\bigr\rangle 
&=
-
\delta(t - t' - \sigma)
\label{SD2}
\end{align}
for the two-point correlation function from the action~\eqref{S_a_tilde}.
In particular, the zeroth-order $\mathcal{O}(\lambda^0)$ two-point function 
can be inferred from these equations as
\be 
\label{D0}
\bigl\langle 
\tilde{a}(t) \, \tilde{a}^{\dagger}(t')
\bigr\rangle_0
=
\Theta(t - t' - \sigma) 
\equiv
i \int_{-\infty}^{\infty} \frac{d \omega}{2 \pi} \, 
\frac{e^{-i \omega \left( t - t' - \sigma \right)}}{\omega + i \epsilon}
\, ,
\ee 
where $\Theta(z)$ is the unit step function.\footnote{
Recall that the momentum-space propagator~\eqref{prop_p} has a denominator 
\be
(k^0)^2 - (k^1)^2 + i \epsilon 
= 
4 \Omega_U \Omega_V + i \epsilon 
= 
4 \Omega_U 
\left( 
\Omega_V + i \epsilon / 4 \Omega_U
\right)
\ee
containing the Feynman $i \epsilon$ term.
By identifying $\Omega_U$ with $\Omega$ and $\Omega_V$ with $\omega$, the $i \epsilon$ term carries over to the propagator~\eqref{D0} in the 1D model since $\Omega$ is positive by definition (see eq.~\eqref{phi_tilde_Fourier}).
}
Throughout this work, unless otherwise stated, the subscript number on physical observables will be used to denote the order of the coupling constant $\lambda$ involved.

Strictly speaking, 
since a real string length parameter $\ell \in \mathbb{R}$
in the 2D model of interest~\eqref{S_2d_phi_tilde}
corresponds to analytically continuing $\sigma > 0$ in this 1D model
to an imaginary value (see eqs.~\eqref{sigma_Om} and~\eqref{sigma_Om_ellE}),
the step function in eq.~\eqref{D0} 
should ultimately be interpreted as a shorthand notation for 
the integral on the right-hand side of the equation.

In the presence of the background interaction term in the action~\eqref{S_a_tilde}, 
the perturbative expansion of the correlation function to all orders in $\lambda$ is given by  
\be 
\label{D}
\bigl\langle 
\tilde{a}(t) \, \tilde{a}^{\dagger}(t')
\bigr\rangle
=
\bigl\langle 
\tilde{a}(t) \, 
\tilde{a}^{\dagger}(t') \, 
e^{i S_{\mathrm{int}}}
\bigr\rangle_0
\equiv 
\sum_{n = 0}^{\infty} \, 
\bigl\langle 
\tilde{a}(t) \, \tilde{a}^{\dagger}(t')
\bigr\rangle_n
\, ,
\ee 
where
\be 
\label{Dn}
\begin{aligned}
\bigl\langle 
\tilde{a}(t) \, \tilde{a}^{\dagger}(t')
\bigr\rangle_n
=
(i \lambda)^n 
\bigintssss_{-\infty}^{\infty} 
\Biggl[ \, 
\prod_{j = 1}^n 
& \, 
d t_j \, b(t_j) 
\Biggr] \, 
\Theta(t - t_n - \sigma) \, 
\Theta(t_1 - t' - \sigma) 
\\ 
&
\times 
\Theta(t_n - t_{n - 1} - \sigma) 
\cdots 
\Theta(t_2 - t_1 - \sigma) 
\end{aligned}
\ee 
represents the $n$th-order correction term.

Intuitively, for $\sigma > 0$, 
the right-hand side of eq.~\eqref{Dn} leads to an overall step function 
$\Theta \bigl( t - t' - (n + 1) \, \sigma \bigr)$. 
This implies that, for a given time separation $t - t'$, 
higher-order contributions to the two-point correlation function 
vanish for $n > (t - t') / \sigma - 1$. 
Furthermore, $\ev{\tilde{a}(t) \, \tilde{a}^{\dagger}(t')}$ 
identically vanishes if $t - t' < \sigma$, 
regardless of the background profile $b(t)$. 
However, in applications to the 2D model~\eqref{S_2d_phi_tilde}, 
where an analytic continuation $\sigma = -i \sigma_E$ 
with $\sigma_E \in \mathbb{R}$ is required at the end of the calculation, 
this suppression is no longer straightforwardly apparent 
since it hinges on the analytic properties of the background field $b(t)$.
This point remains to be clarified in future works.

We will now construct the Hamiltonian formalism perturbatively 
to reproduce the correlation functions~\eqref{Dn} 
in the Heisenberg picture of quantum mechanics.
This is achieved by expanding the operators in terms of $\lambda$:
\be
\label{a_expand}
\hat{\tilde{a}}(t) 
= 
\hat{\tilde{a}}_0(t)
+ 
\hat{\tilde{a}}_1(t) 
+ 
\mathcal{O}(\lambda^2)
\, ,
\qquad
\hat{\tilde{a}}^{\dagger}(t) 
= 
\hat{\tilde{a}}_0^{\dagger}(t) 
+ 
\hat{\tilde{a}}_1^{\dagger}(t) 
+ 
\mathcal{O}(\lambda^2)
\, ,
\ee
and then defining them order by order such that 
the correspondence~\eqref{path_op_corres} with the path-integral results is obeyed.
For the two-point function, 
the correspondence \eqref{path_op_corres} states that 
\be
\label{path_op_corres_2pt}
\Theta(t - t') 
\ev{\hat{\tilde{a}}(t) \, 
\hat{\tilde{a}}^{\dagger}(t')}{0}
+ 
\Theta(t' - t) 
\ev{
\hat{\tilde{a}}^{\dagger}(t') \, 
\hat{\tilde{a}}(t) 
}{0}
=
\bigl\langle 
\tilde{a}(t) \, 
\tilde{a}^{\dagger}(t') 
\bigr\rangle
\, .
\ee

At the zeroth order, 
with the vacuum state $\ket{0}$ defined as
\be
\label{vacuum}
\hat{\tilde{a}}_0(t) 
\ket{0} 
= 
0
\qquad 
\forall \ 
t
\, ,
\ee
eqs.~\eqref{D0} and~\eqref{path_op_corres_2pt} imply that the lowering and raising operators $\hat{\tilde{a}}_0(t)$ and $\hat{\tilde{a}}_0^{\dagger}(t)$ satisfy
\be
\label{aa_t>t'}
\langle 0 | \, 
\comm*{\hat{\tilde{a}}_0(t)}{\hat{\tilde{a}}_0^{\dagger}(t')}
| 0 \rangle
= 
\Theta(t - t' - \sigma)
\qquad 
\text{for}
\quad 
t > t' 
\, .
\ee
As usual, 
assuming that the commutator $[\hat{\tilde{a}}_0(t), \hat{\tilde{a}}_0^{\dagger}(t')]$ 
is a $c$-number, we conclude that 
\be
\label{aa_comm_t>t'}
\comm*{\hat{\tilde{a}}_0(t)}{\hat{\tilde{a}}_0^{\dagger}(t')}
= 
\Theta(t - t' - \sigma)
\qquad 
\text{for}
\quad 
t > t' 
\, .
\ee

In the Hamiltonian formalism, 
we also need $\comm*{\hat{\tilde{a}}_0(t)}{\hat{\tilde{a}}_0^{\dagger}(t')}$
for $t < t'$, although this is not required 
for the time-ordered product to fulfill the equality~\eqref{path_op_corres_2pt}.
To resolve this ambiguity, we choose to impose the involution under which 
\be 
\label{involution}
\hat{\tilde{a}}_0(t) 
\mapsto 
\hat{\tilde{a}}_0^{\dagger}(t)
\qquad 
\text{and}
\qquad
\sigma \mapsto \sigma \, ,
\ee 
together with complex conjugation and reversing the ordering of operators, 
as a symmetry involution on the operator algebra.
From eq.~\eqref{aa_comm_t>t'},
this then implies that
\begin{align}
\label{aa_comm_t<t'}
\comm*{\hat{\tilde{a}}_0(t)}{\hat{\tilde{a}}_0^{\dagger}(t')}
= 
\Theta(t' - t - \sigma)
\qquad 
\text{for}
\quad 
t < t' 
\, .
\end{align}
At this point, the symmetry under~\eqref{involution} 
appears to be an arbitrary choice, 
and the commutator~\eqref{aa_comm_t<t'} for $t < t'$ 
could have been chosen differently without affecting the agreement~\eqref{path_op_corres} 
with the path-integral formalism.\footnote{
Note that prior to the analytic continuation $\ell^2 \to i \ell_E^2$ in the 2D model~\eqref{S_2d_phi_tilde}, the symmetry under time reversal involves a complex conjugation under which $\sigma_{\Omega} \mapsto - \sigma_{\Omega}$ (cf. eq.~\eqref{sigma_Om}).
However, it turns out that imposing this symmetry featuring the transformation $\sigma \mapsto - \sigma$ in our Hamiltonian formalism for the 1D model does not work.
}
However, we will see below that this choice is 
algebraically convenient for the Hamiltonian formalism 
and leads to a consistent Hilbert space representation without ghosts.

Combining eqs.~\eqref{aa_comm_t>t'} and~\eqref{aa_comm_t<t'}, 
we arrive at 
\be
\label{comm0}
\comm*{\hat{\tilde{a}}_0(t)}{\hat{\tilde{a}}_0^{\dagger}(t')}
= 
\Theta (|t - t'| - \sigma)
\, .
\ee 
This expression is incompatible with the assumption of 
time-independent operators $\hat{\tilde{a}}_0$ and $\hat{\tilde{a}}_0^{\dagger}$.
In other words, 
the form~\eqref{D0} of the path-integral correlation function
excludes the possibility of a perturbative formulation 
in which $\hat{\tilde{a}}_0(t)$ and $\hat{\tilde{a}}_0^{\dagger}(t)$ 
satisfy the equations of motion $\del_t \, \hat{\tilde{a}}_0(t) 
= \del_t \, \hat{\tilde{a}}_0^{\dagger}(t)
= 0$ at the zeroth order.
We will elaborate on this point further in section~\ref{sec:phys}. 

We note that, in the canonical quantization of the action~\eqref{S_a_tilde},
the conjugate momentum of $\tilde{a}(t)$
is defined as $\tilde{\pi}(t) \equiv \delta S/\delta [\partial_t \, \tilde{a}(t)]
= i \tilde{a}^{\dagger}(t - \sigma)$.
Therefore, eq.~\eqref{comm0} is consistent with 
the ``equal-time'' commutation relation
\be 
\label{comm_eq}
\comm*{\hat{\tilde{a}}(t)}{\hat{\tilde{\pi}}(t)}
=
\comm*{\hat{\tilde{a}}(t)}{i \hat{\tilde{a}}^{\dagger}(t - \sigma)}
= 
i
\, .
\ee 
To address the ambiguity of the step function $\Theta(z)$ at $z = 0$, we should in principle modify eq.~\eqref{comm0} as
\be
\label{comm0_epsilon}
\comm*{\hat{\tilde{a}}_0(t)}{\hat{\tilde{a}}_0^{\dagger}(t')}
= 
\Theta (|t - t' | \, - \, \sigma + \epsilon)
\, ,
\ee
where $\epsilon > 0$ is an infinitesimal parameter
analogous to Feynman's $i\epsilon$-prescription.

With the creation and annihilation operators 
$\hat{\tilde{a}}_0(t)$ and $\hat{\tilde{a}}_0^{\dagger}(t)$
constructed above,
we can define the Fock space of the theory as
\be
\label{fock}
\mathrm{span}
\left\{
\Pi_{i = 1}^n \, 
\hat{\tilde{a}}_0^{\dagger}(t_i)
|0\rangle
\right\}
.
\ee
Compared to the Fock space of the local theory with $\sigma = 0$, 
this appears to be a much larger space 
due to the time dependence of the creation operator. 
As is typical in a nonlocal theory, 
this enlarged Fock space is expected to contain negative-norm states
that would render the quantum theory pathological.
However, remarkably, we will demonstrate later that
after imposing the equations of motion as physical-state constraints, 
negative-norm states are removed (see section~\ref{sec:phys}) 
and zero-norm states decouple (see section~\ref{sec:spurious}).
Consequently, the resulting physical state space
turns out to be identical to the Fock space of the local theory,
and the quantum theory of this nonlocal 1D model is well-defined.

The operator algebra at higher orders in $\lambda$ 
can be derived in a similar fashion 
by demanding the correspondence~\eqref{path_op_corres_2pt}.
At $\mathcal{O}(\lambda)$, it reads
\begin{align}
\Theta(t - t') 
\left[
\langle 0|
\hat{\tilde{a}}_0(t) \, 
\hat{\tilde{a}}_1^{\dagger}(t')
|0 \rangle
+ 
\langle 0|
\hat{\tilde{a}}_1(t) \, 
\hat{\tilde{a}}_0^{\dagger}(t')
|0 \rangle
\right]
&=
\bigl\langle 
\tilde{a}(t) \, \tilde{a}^{\dagger}(t')
\bigr\rangle_1
\nn \\
&= 
i \lambda \, 
\Theta(t - t' - 2\sigma) 
\int_{t' \, + \, \sigma}^{t \, - \, \sigma} 
b(t_1) \, d t_1 
\label{D1}
\end{align}
according to eq.~\eqref{Dn}.
To derive this equation, 
we have employed the following algebraic identity:
\be
\label{step_combine}
\int_{-\infty}^{\infty} dt'' \,
\Theta(t - t'' - \sigma) \, 
b(t'') \, 
\Theta(t'' - t' - \sigma)
= 
\Theta(t - t' - 2\sigma) 
\int_{t' \, + \, \sigma}^{t \, - \, \sigma} 
dt'' \, b(t'') 
\, ,
\ee 
which allows for the combination of step functions.
This manipulation is straightforward for $\sigma \in \mathbb{R}$.
In the case relevant to the 2D toy model~\eqref{S_2d_phi_tilde} where $\sigma = - i \sigma_E$ with $\sigma_E \in \mathbb{R}$, the step functions in the above equation should be replaced with their integral representations given on the right-hand side of eq.~\eqref{D0}.
Nevertheless, it can be straightforwardly verified that eq.~\eqref{step_combine} 
agrees with the result obtained by carrying out the $\omega$-integral in eq.~\eqref{D0}
in the Euclideanized ``momentum'' space with $\omega = i \omega_E$ ($\omega_E \in \mathbb{R}$)
and real $\sigma_E$.

To proceed with defining the first-order operators, 
let us start with the perturbative ansatz
\begin{align}
\hat{\tilde{a}}_1(t)
&=
i \lambda 
\int_{-\infty}^{\infty} dt'' \, 
g_1(t, t'') \, 
b(t'') \, 
\hat{\tilde{a}}_0 (t'')
\, , 
\label{a1}
\\
\hat{\tilde{a}}_1^{\dagger} (t)
&=
- i \lambda 
\int_{-\infty}^{\infty} dt'' \, 
g_1^{c} (t, t'') \, 
b(t'') \, 
\hat{\tilde{a}}_0^{\dagger} (t'')
\, .
\label{a1_dagger}
\end{align}
Eq.~\eqref{D1} determines the two functions $g_1(t ,t')$ and $g_1^{c}(t, t')$ to be 
\be 
\label{g1}
g_1 (t, t')
=
g_1^{c} (t, t')
=
\xi \, 
\Theta(t - t' - \sigma)
+
\left( \xi - 1 \right)
\Theta(t' - t - \sigma)
\, , 
\ee 
where $\xi \in \mathbb{R}$ is a free parameter that cannot be fixed by either the correspondence~\eqref{D1} or the conjugate symmetry~\eqref{involution}.\footnote{
This ambiguity is already present in the local theory with $\sigma = 0$.
However, in this case, 
it corresponds solely to a time-independent term 
$\propto \int_{-\infty}^{\infty} b(t) \, dt$ in the first-order operators, 
which can be absorbed by a constant phase factor.
}
A detailed derivation of the expressions for $g_1(t ,t')$ and $g_1^{c}(t, t')$ 
can be found in appendix~\ref{app:derive_g1}.

Nonetheless,
the operator algebra is uniquely determined.
Specifically,
the $\mathcal{O}(\lambda)$ correction to the commutator $\comm*{\hat{\tilde{a}}_0(t)}{\hat{\tilde{a}}_0^{\dagger}(t')}$~\eqref{comm0} can be obtained as
\begin{align}
\comm*{\hat{\tilde{a}} (t)}{\hat{\tilde{a}}^{\dagger} (t')}_1
&=
\comm*{\hat{\tilde{a}}_1(t)}{\hat{\tilde{a}}_0^{\dagger}(t')}
+
\comm*{\hat{\tilde{a}}_0(t)}{\hat{\tilde{a}}_1^{\dagger}(t')}
\nn \\
&= 
i \lambda \, \Theta(\abs{t - t'} - 2 \sigma)
\int_{t' \, + \, \sign(t - t') \sigma}^{t \, - \, \sign(t - t') \sigma}  
b(t'') \, d t''
\, ,
\label{comm1}
\end{align}
which is independent of the parameter $\xi$.

Causality demands setting $\xi = 1$ to ensure that both 
$\hat{\tilde{a}}(t)$ and $\hat{\tilde{a}}^{\dagger}(t)$ 
depend only on the background profile $b(t'')$ in the past ($t'' < t$). 
Otherwise, physical observables associated with the physical states of this theory 
(discussed further in section~\ref{sec:phys}) 
would depend acausally on the background field.

The procedure outlined above allows us to construct the operators $\hat{\tilde{a}}(t)$ and $\hat{\tilde{a}}^{\dagger}(t)$, along with their commutators, order by order in a manner that is consistent with the path-integral correlation function 
$\ev{\tilde{a}(t) \, \tilde{a}^{\dagger}(t')}$.
More explicitly, from eqs.~\eqref{SD1} and~\eqref{SD2}, 
we find that the $\mathcal{O}(\lambda^n)$ correction to the correlation function obeys
\begin{align}
\del_t \, 
\bigl\langle 
\tilde{a}(t) \, \tilde{a}^{\dagger}(t')
\bigr\rangle_n
&=
i \lambda \, b(t - \sigma) \, 
\bigl\langle 
\tilde{a}(t - \sigma) \, \tilde{a}^{\dagger}(t')
\bigr\rangle_{n - 1}
\, ,
\\
\del_{t'} \, 
\bigl\langle 
\tilde{a}(t) \, \tilde{a}^{\dagger}(t')
\bigr\rangle_n
&=
- i \lambda \, b(t' + \sigma) \, 
\bigl\langle 
\tilde{a}(t) \, \tilde{a}^{\dagger}(t' + \sigma)
\bigr\rangle_{n - 1}
\, .
\end{align}
Given the quadratic nature of the action~\eqref{S_a_tilde}, 
$\hat{\tilde{a}}(t)$ is expected to be a linear functional of $\hat{\tilde{a}}_0$, 
while $\hat{\tilde{a}}^{\dagger} (t)$ is anticipated to be 
a linear functional of $\hat{\tilde{a}}_0^{\dagger}$, 
to all orders in $\lambda$. 
Thus, in order for the correspondence~\eqref{path_op_corres_2pt} 
to be fulfilled at $\mathcal{O}(\lambda^n)$, the commutator 
\be 
W_n (t, t')
\equiv 
\bigl[ 
\hat{\tilde{a}} (t) 
\, , 
\hat{\tilde{a}}^{\dagger} (t')
\bigr]_n
=
\sum_{j = 0}^n
\ev{\hat{\tilde{a}}_j (t) \, \hat{\tilde{a}}_{n - j}^{\dagger} (t')}{0}
\ee 
in the operator formalism must satisfy
\begin{align}
\del_t
\left[
\Theta(t - t') \, 
W_n (t, t')
\right] 
&=
i \lambda \, b(t - \sigma) \, 
\bigl\langle 
\tilde{a}(t - \sigma) \, \tilde{a}^{\dagger}(t')
\bigr\rangle_{n - 1}
\, ,
\\
\del_{t'}
\bigl[
\Theta(t - t') \, 
W_n (t, t')
\bigr] 
&=
- i \lambda \, b(t' + \sigma) \, 
\bigl\langle 
\tilde{a}(t) \, \tilde{a}^{\dagger}(t' + \sigma)
\bigr\rangle_{n - 1}
\, ,
\end{align}
which are equivalent to the following equations:
\begin{align}
W_n (t, t)
&=
0
\, , 
\label{SD_n_1}
\\ 
\Theta(t - t') \, 
\del_t \, 
W_n (t, t')
&=
i \lambda \, b(t - \sigma) \, 
\bigl\langle 
\tilde{a}(t - \sigma) \, \tilde{a}^{\dagger}(t')
\bigr\rangle_{n - 1}
\, ,
\label{SD_n_2}
\\
\Theta(t - t') \, 
\del_{t'} 
W_n (t, t')
&=
- i \lambda \, b(t' + \sigma) \, 
\bigl\langle 
\tilde{a}(t) \, \tilde{a}^{\dagger}(t' + \sigma)
\bigr\rangle_{n - 1}
\, .
\label{SD_n_3}
\end{align}
Again, these equations are only capable of determining the commutator for $t > t'$:
\be 
\label{comm_n_t>t'}
W_n (t, t')
=
i \lambda 
\int_{-\infty}^{t \, - \, \sigma}
d t'' \, b(t'') \, 
\bigl\langle 
\tilde{a}(t'') \, \tilde{a}^{\dagger}(t')
\bigr\rangle_{n - 1}
=
\bigl\langle 
\tilde{a}(t) \, \tilde{a}^{\dagger}(t')
\bigr\rangle_n
\qquad 
\text{for}
\quad 
t > t' 
\, ,
\ee 
similar to the situation in eq.~\eqref{aa_comm_t>t'} at the zeroth order.

Nevertheless, it follows from eq.~\eqref{comm_n_t>t'} that 
\begin{align}
W_n^{\ast} (t, t') 
\big\vert_{t \, > \, t'}
&=
- i \lambda \int_{-\infty}^{\infty} 
dt'' \, b(t'') \, 
\bigl\langle 
\tilde{a}(t) \, \tilde{a}^{\dagger}(t'')
\bigr\rangle_0^{\ast} 
\, 
\bigl\langle 
\tilde{a}(t'') \, \tilde{a}^{\dagger}(t')
\bigr\rangle_{n - 1}^{\ast}
\\
&=
- i \lambda \int_{-\infty}^{\infty} 
dt'' \, b(t'') 
\left[ \Theta(t - t'') \, W_0^{\ast} (t, t'') \right]
\left[ \Theta(t'' - t') \, W_{n - 1}^{\ast} (t'', t') \right]
,
\end{align}
where we have utilized the correspondence~\eqref{path_op_corres_2pt} in the second line. 
If both $W_0(t, t')$ and $W_{n - 1}(t, t')$ respect conjugate symmetry, i.e., $W_0^{\ast} (t, t') = W_0 (t', t)$ and $W_{n - 1}^{\ast} (t, t') = W_{n - 1} (t', t)$, we can further express
\be 
W_n^{\ast} (t, t') 
\big\vert_{t \, > \, t'}
=
- i \lambda \int_{t'}^{t} 
dt'' \, b(t'') \, 
W_0 (t'', t) \, W_{n - 1} (t', t'')
=
W_n (t', t) 
\big\vert_{t' \, < \, t}
\, .
\ee 
Hence, the counterpart of~\eqref{comm_n_t>t'} in the domain $t < t'$ is uniquely determined for all $n \geq 1$ once the involution~\eqref{involution} is imposed on the algebra $W_0 (t, t') = \bigl[ 
\hat{\tilde{a}}_0 (t) 
\, , 
\hat{\tilde{a}}_0^{\dagger} (t')
\bigr]$ at zeroth order in $\lambda$. 
The $\mathcal{O}(\lambda^n)$ correction to the commutator can then be written as 
\begin{align}
W_n (t, t')
&=
\Theta(t - t') \, 
\bigl\langle 
\tilde{a}(t) \, \tilde{a}^{\dagger}(t')
\bigr\rangle_n
+
\Theta(t' - t) \, 
\bigl\langle 
\tilde{a}(t') \, \tilde{a}^{\dagger}(t)
\bigr\rangle_n^{\ast}
\, .
\label{comm_n}
\end{align}
Subsequently, the $\mathcal{O}(\lambda^n)$ operators $\hat{\tilde{a}}_n (t)$ and $\hat{\tilde{a}}_n^{\dagger} (t)$, as functions of $\hat{\tilde{a}}_0$ and $\hat{\tilde{a}}_0^{\dagger}$, can be determined.

With the ambiguity~\eqref{g1} in the first-order operators 
resolved by selecting $\xi = 1$ to respect causality, 
the operators satisfying eq.~\eqref{comm_n} are completely fixed 
to all orders in $\lambda$ as 
(see details in appendix~\ref{app:a_allorder})
\begin{align}
\hat{\tilde{a}}_n(t)
&=
i \lambda 
\int_{-\infty}^{\infty} dt'' \, 
\Theta(t - t'' - \sigma) \, 
b(t'') \, 
\hat{\tilde{a}}_{n - 1} (t'')
\nn \\
&= 
(i \lambda)^n 
\bigintssss_{-\infty}^{\infty} 
\Biggl[ \, 
\prod_{j = 1}^n 
d t_j \, b(t_j)
\Biggr] \, 
\Theta(t - t_n - \sigma) \, 
\Theta(t_n - t_{n - 1} - \sigma) 
\cdots 
\Theta(t_2 - t_1 - \sigma) \, 
\hat{\tilde{a}}_0 (t_1)
\, , 
\label{an}
\\
\hat{\tilde{a}}_n^{\dagger} (t)
&=
- i \lambda 
\int_{-\infty}^{\infty} dt'' \, 
\Theta(t - t'' - \sigma) \, 
b(t'') \, 
\hat{\tilde{a}}_{n - 1}^{\dagger} (t'')
\nn \\
&= 
(- i \lambda)^n 
\bigintssss_{-\infty}^{\infty} 
\Biggl[ \, 
\prod_{j = 1}^n 
d t_j \, b(t_j)
\Biggr] \, 
\Theta(t - t_n - \sigma) \, 
\Theta(t_n - t_{n - 1} - \sigma) 
\cdots 
\Theta(t_2 - t_1 - \sigma) \, 
\hat{\tilde{a}}_0^{\dagger} (t_1)
\label{an_dag}
\end{align}
for $n \geq 1$. 
We have thus established the correspondence~\eqref{path_op_corres_2pt} 
with the path-integral correlation function to all orders in the interacting theory.

\subsection{Physical-state constraints from equations of motion}
\label{sec:phys}

The equations of motion obtained from variations of the nonlocal 1D model action~\eqref{S_a_tilde} are given by 
\begin{align}
i \del_t \, 
\tilde{a}(t + \sigma) 
+ 
\lambda \, b(t) \, \tilde{a}(t)
&=
0
\, ,
\label{eom_a}
\\
- i \del_t \, 
\tilde{a}^{\dagger}(t - \sigma) 
+ 
\lambda \, b(t) \, \tilde{a}^{\dagger}(t)
&=
0
\, .
\label{eom_a_dagger}
\end{align}
As mentioned in the previous section, 
the Heisenberg operators $\hat{\tilde{a}}(t)$ and $\hat{\tilde{a}}^{\dagger}(t)$ 
that we have constructed based on the path-integral correlation function
do not satisfy the equations of motion.
The equations of motion should instead be implemented as constraints on the physical states.
It is a common feature of theories with infinite time derivatives that 
the equations of motion appear as constraints in the Hamiltonian formalism~\cite{Llosa94, Woodard:2000bt}.

To recover the equations of motion in the classical limit, 
they are realized as requirements on the expectation values in the physical Hilbert space:
\be
\label{H_phys}
\mathcal{H}_{\mathrm{phys}}
\equiv 
\Bigl\{
\ket{\Psi} \in 
\mathrm{span} 
\bigl\{
\Pi_{i = 1}^n \, 
\hat{\tilde{a}}_0^{\dagger}(t_i)
|0\rangle
\bigr\}
\ \Big\vert \ 
\ev{(\text{equations of motion})}{\Psi}
=
0
\Bigr\}
\, .
\ee
We define physical states $\ket{\Psi}$ to be those satisfying the constraint
\be 
\label{eom_a_full}
\left[
i \del_t \, 
\hat{\tilde{a}}(t + \sigma) 
+ 
\lambda \, b(t) \, \hat{\tilde{a}}(t)
\right]
\ket{\Psi} 
=
0
\, ,
\ee 
with their conjugates $\bra{\Psi}$ defined by 
\be 
\label{eom_a_dag_full}
\bra{\Psi} 
\bigl[
- i \del_t \, 
\hat{\tilde{a}}^{\dagger}(t - \sigma) 
+ 
\lambda \, b(t) \, \hat{\tilde{a}}^{\dagger}(t)
\bigr] 
=
0
\, .
\ee 

At zeroth order in $\lambda$, the physical states satisfy
\be 
\label{eom_constraint0}
i \del_t \, 
\hat{\tilde{a}}_0(t + \sigma)
\ket{\Psi}
= 
0
\, .
\ee 
According to this definition, 
the vacuum $\ket{0}$~\eqref{vacuum} is a physical state.

Now consider a generic $n$-particle excited state
\be
\label{n_particle_state}
\big|
\{ \psi^{(j)} \}_{j = 1}^{n} 
\big\rangle 
=
\prod_{j = 1}^n
\hat{\tilde{\mathcal{A}}}^{\dagger} [\psi^{(j)}]
\ket{0}
,
\ee
where
\be
\label{A_dagger}
\hat{\tilde{\mathcal{A}}}^{\dagger} [\psi^{(j)}]
\equiv
\lim_{N \to \infty} \frac{1}{2N \sigma}
\int_{-N \sigma}^{N \sigma} 
d\tau \, \psi^{(j)}(\tau) \, 
\hat{\tilde{a}}_0^{\dagger}(\tau)
\ee
represents the creation operator which generates a generic one-particle state.
It involves a superposition of $\hat{a}_0^{\dagger}$ at different times 
weighted by the wavefunction $\psi^{(j)}(\tau) \in \mathbb{C}$.
In particular, 
the integral is regularized by an integer $N$ such that 
the norm of physical states would turn out finite (see section~\ref{sec:norm}).

At the zeroth order, 
the physical-state constraint~\eqref{eom_constraint0} is satisfied 
when the zeroth-order wavefunctions $\psi^{(j)}_0(\tau)$ obey
\be 
\lim_{N \to \infty} \frac{1}{2N \sigma}
\int_{-N \sigma}^{N \sigma} 
d \tau \, 
\psi_0(\tau) \, 
i \del_t \, 
\comm*{\hat{\tilde{a}}_0(t + \sigma)}{\hat{\tilde{a}}_0^{\dagger}(\tau)}
=
0
\, .
\ee 
Substituting the zeroth-order commutator~\eqref{comm0} into the equation above yields 
the periodicity condition on the unperturbed wavefunctions:
\be 
\label{periodic}
\psi_0(t + 2 \sigma) 
- 
\psi_0(t) 
= 
0
\qquad 
\forall 
\ t \, .
\ee 
Owing to this periodicity, 
integrals of $\psi_0(t)$ over (semi-)infinite time intervals 
need to be properly regularized, 
as we did in eq.~\eqref{A_dagger} 
by including the factor $(2 N \sigma)^{-1}$ 
and taking the limit $N \rightarrow \infty$.

In addition to eq.~\eqref{eom_constraint0}, 
the other equation-of-motion constraint 
\be 
\label{eom_constraint0_conj}
\bra{\Psi}
i \del_t \, 
\hat{\tilde{a}}^{\dagger}_0(t - \sigma)
=
0
\ee 
must be imposed on the dual space of physical states. 
For a generic dual state 
\be
\big\langle
\{ \psi^{(j) \ast} \}_{j = 1}^{n} 
\big|
=
\prod_{j = 1}^n
\bra{0}
\hat{\tilde{\mathcal{A}}} [\psi^{(j) \ast}]
\, ,
\ee
with $\hat{\tilde{\mathcal{A}}}$ given by
\be
\hat{\tilde{\mathcal{A}}} [\psi^{(j) \ast}]
\equiv
\lim_{N \to \infty} \frac{1}{2N \sigma}
\int_{-N \sigma}^{N \sigma} 
d\tau \, \psi^{(j) \ast}(\tau) \, 
\hat{\tilde{a}}_0 (\tau)
\, ,
\ee
it can be verified that implementing the constraint~\eqref{eom_constraint0_conj} again 
leads to the periodicity condition~\eqref{periodic} on the wave functions.
Hence, we can identify the conjugate $\langle \Psi |$ of a given physical state 
$| \Psi \rangle = \prod_{j = 1}^n
\hat{\tilde{\mathcal{A}}}^{\dagger} [\psi_0^{(j)}]
\ket{0}$~\eqref{n_particle_state} simply as 
$\langle \Psi | =
\bra{0}
\prod_{j = 1}^n
\hat{\tilde{\mathcal{A}}} [\psi_0^{(j) \ast}]$ at the zeroth order.

When the background interaction is turned on, 
the physical-state constraints take the forms 
\eqref{eom_a_full} and~\eqref{eom_a_dag_full}.
As it should have been obvious from the calculation above, 
it suffices to impose these constraints on the subspace of one-particle states 
\be 
\label{1_particle_state}
\ket{1_{\Psi}}
=
\hat{\tilde{\mathcal{A}}}^{\dagger}[\Psi] 
\ket{0}
\equiv 
\lim_{N \to \infty} \frac{1}{2N \sigma}
\int_{-N \sigma}^{N \sigma} 
d \tau \, 
\Psi(\tau) \, 
\hat{\tilde{a}}_0^{\dagger} (\tau)
\ket{0}
\ee 
and their corresponding dual states
\be 
\label{1_particle_dual}
\bra{1_{\Psi}}
\equiv
\lim_{N \to \infty} \frac{1}{2N \sigma}
\int_{-N \sigma}^{N \sigma} 
d \tau \, 
\Psi^{c}(\tau) 
\bra{0}
\hat{\tilde{a}}_0 (\tau)
\, ,
\ee 
respectively.
Having obtained the perturbative expansions~\eqref{a_expand} 
of the operators $\hat{\tilde{a}}$ and $\hat{\tilde{a}}^{\dagger}$, 
along with their commutator algebra, in section~\ref{sec:2pt_function}, 
the physical-state wavefunctions $\Psi(t)$ and $\Psi^c (t)$ 
in the interacting theory can be determined perturbatively from the constraints
\begin{align}
\left[
i \del_t \, 
\hat{\tilde{a}}(t + \sigma) 
+ 
\lambda \, b(t) \, \hat{\tilde{a}}(t)
\right]
\ket{1_{\Psi}} 
&=
0
\, , 
\label{eom_a_1particle}
\\ 
\bra{1_{\Psi}} 
\bigl[
- i \del_t \, 
\hat{\tilde{a}}^{\dagger}(t - \sigma) 
+ 
\lambda \, b(t) \, \hat{\tilde{a}}^{\dagger}(t)
\bigr] 
&=
0
\, .
\label{eom_a_dag_1particle}
\end{align}

Since the constraints~\eqref{eom_a_1particle} and~\eqref{eom_a_dag_1particle} 
are not Hermitian conjugates of each other for real $\sigma$, 
the dual physical states $\bra{1_{\Psi}}$ in the interacting theory 
are \textit{not} the Hermitian conjugates of the physical states $\ket{1_{\Psi}}$.
This distinction implies that the dual wavefunction $\Psi^{c}(t)$ 
is not equivalent to the usual complex conjugate $\Psi^{\ast}(t)$.
Instead, for a state $\ket{1_{\Psi}}$ with a given zeroth-order wavefunction $\psi_0$, 
we identify~\eqref{1_particle_dual} whose associated wavefunction $\Psi^c$ 
has the zeroth-order contribution $\psi^{\ast}_0$ as its dual state.

To proceed, 
we carry out the perturbative expansions
\begin{align}
\label{psi_expand}
\Psi(\tau)
&=
\psi_0(\tau)
+
\lim_{N \to \infty}
\left( 2 N \sigma \right)
\sum_{n = 1}^{\infty}
\psi_n (\tau)
\, ,
\\
\label{psi_expand_star}
\Psi^{c}(\tau)
&=
\psi_0^{\ast}(\tau)
+
\lim_{N \to \infty}
\left( 2 N \sigma \right)
\sum_{n = 1}^{\infty}
\psi_n^c (\tau)
\, ,
\end{align}
where $\psi_n (t)$ and $\psi_n^c (t)$ represent
corrections to the wavefunctions that satisfy 
the equation-of-motion constraints~\eqref{eom_a_1particle} and~\eqref{eom_a_dag_1particle}, respectively, at order $\lambda^n$.
The physical state $\ket{1_{\Psi}}$ and its conjugate $\bra{1_{\Psi}}$
can then be solved order by order in terms of 
any given zeroth-order wavefunction $\psi_0(\tau)$
that satisfies the periodic boundary condition~\eqref{periodic}.
The wavefunction expansions above are written in a way 
such that the state $\ket{1_{\Psi}}$~\eqref{1_particle_state} 
is essentially a functional of 
the \textit{normalized} zeroth-order wave function $\psi_0 / 2 N \sigma$
in perturbation theory.
Likewise, $\bra{1_{\Psi}}$~\eqref{1_particle_dual} 
is a functional of $\psi_0^{\ast} / 2 N \sigma$.

At first order in $\lambda$,
the constraint~\eqref{eom_a_1particle} requires that 
\be 
\begin{aligned}
0 
= 
\lim_{N \to \infty}
\int_{-N \sigma}^{N \sigma} 
d \tau \, 
\biggl\{
&
\frac{\psi_0 (\tau)}{2 N \sigma} 
\left[ 
i \del_t \, 
\comm*{\hat{\tilde{a}}_1 (t + \sigma)}{\hat{\tilde{a}}_0^{\dagger}(\tau)}
+
\lambda \, b(t) \, 
\comm*{\hat{\tilde{a}}_0 (t)}{\hat{\tilde{a}}_0^{\dagger}(\tau)}
\right]
\\
&
+ 
\psi_1(\tau) \, 
i \del_t \, 
\comm*{\hat{\tilde{a}}_0 (t + \sigma)}{\hat{\tilde{a}}_0^{\dagger}(\tau)}
\biggr\}
\, ,
\end{aligned}
\ee 
which leads to the condition
\be 
\label{psi_1_difference}
\psi_1 (t) - \psi_1 (t - 2 \sigma)
=
\lim_{N \to \infty}
\int_{-N \sigma}^{N \sigma} d \tau \, 
\frac{\psi_0 (\tau)}{2 N \sigma} \, 
\Bigl[
\del_t \, \hat{\tilde{a}}_1 (t - \sigma) 
- 
i \lambda \, b(t - 2 \sigma) \, \hat{\tilde{a}}_0 (t - 2 \sigma)
\, , \, 
\hat{\tilde{a}}_0^{\dagger} (\tau)
\Bigr]
\, .
\ee 
Given that the time dependence of $\psi_1(t)$ 
encodes the $\mathcal{O}(\lambda)$ response of
the wavefunction to the background field $b(t)$ sourcing the interaction,
by assuming that $b(t)$ vanishes as $t \to \pm \infty$,
the natural boundary conditions would then be 
$\psi_1( \pm \infty ) = \text{constant}$.
This in turn eliminates the contributions from 
the periodic homogeneous solutions to eq.~\eqref{psi_1_difference},
up to a constant term.
The first-order correction $\psi_1 (t)$ 
to the wavefunction can then be iteratively solved as
\begin{align}
\psi_1 (t)
=
&
\lim_{N \to \infty}
\int_{-N \sigma}^{N \sigma} d \tau \, 
\frac{\psi_0 (\tau)}{2 N \sigma} \,  
\sum_{j = 0}^{\infty} \, 
\Bigl[
\del_t \, 
\hat{\tilde{a}}_1 (t - 2 j \sigma - \sigma) 
- 
i \lambda \, b(t - 2 j \sigma - 2 \sigma) \, 
\hat{\tilde{a}}_0 (t - 2 j \sigma - 2 \sigma)
\, , \, 
\hat{\tilde{a}}_0^{\dagger} (\tau)
\Bigr] 
\nn \\
&
+ 
\lambda \, c_1
\, ,
\label{psi_1_iterative}
\end{align}
where $c_1$ is a constant.
Combining the form~\eqref{a1} of $\hat{\tilde{a}}_1 (t)$ 
with eq.~\eqref{g1},
we obtain 
\be 
\sum_{j = 0}^{\infty}
\left[
\del_t \, 
\hat{\tilde{a}}_1 (t - 2 j \sigma - \sigma) 
- 
i \lambda \, b(t - 2 j \sigma - 2 \sigma) \, 
\hat{\tilde{a}}_0 (t - 2 j \sigma - 2 \sigma)
\right]
=
- i \lambda 
\left( \xi - 1 \right)
b(t) \, \hat{\tilde{a}}_0 (t)
\, .
\ee 
Substituting this into the commutator in eq.~\eqref{psi_1_iterative}
results in 
\begin{align}
\psi_1 (t)
&=
- i \lambda 
\left( \xi - 1 \right)
b(t)
\lim_{N \to \infty}
\left[ 
\int_{-N \sigma}^{t \, - \, \sigma} d \tau \, 
\frac{\psi_0 (\tau)}{2 N \sigma} 
+
\int_{t \, + \, \sigma}^{N \sigma} d \tau \, 
\frac{\psi_0 (\tau)}{2 N \sigma} 
\right] 
+
\frac{\lambda \, c_1}{2 N \sigma} 
\, .
\end{align}
Since $\psi_0(t)$ is $2 \sigma$-periodic,
the expression above can be simplified as 
\be 
\psi_1 (t)
=
i \lambda 
\left( 1 - \xi \right)
b(t) \, 
\overbar{\psi}_0
+ 
\frac{\lambda \, c_1}{2 N \sigma} 
\, ,
\label{psi_1}
\ee 
where
\be
\label{psi0_avg}
\overbar{\psi}_0 
\equiv
\lim_{N \to \infty}
\int_{-N \sigma}^{N \sigma} d \tau \, 
\frac{\psi_0 (\tau)}{2 N \sigma}
\ee
is the average value of $\psi_0(t)$.\footnote{
We shall use the bar symbol to denote the time average throughout this paper.
}
As previously anticipated, 
$\psi_1 (t)$ is obtained as a functional of the zeroth-order term $\psi_0 / 2 N \sigma$.
The additive constant $\lambda \, c_1$ can be absorbed away 
by redefining the zeroth-order term $\psi_0(t)$,
as a constant is also periodic.
Therefore, without loss of generality, we set
\be
\label{c1}
c_1 = 0 \, .
\ee

By the same token,
the conjugate constraint~\eqref{eom_a_dag_1particle}
imposes a condition on the dual states~\eqref{1_particle_dual}.
At $\mathcal{O}(\lambda)$,
the constraint~\eqref{eom_a_dag_1particle} can be expressed as
\be 
\begin{aligned}
0 
=
\lim_{N \to \infty}
\int_{-N \sigma}^{N \sigma} 
d \tau \, 
\biggl\{
&
\frac{\psi_0^{\ast} (\tau)}{2 N \sigma}
\left[ 
i \del_t \, 
\comm*{\hat{\tilde{a}}_0 (\tau)}{\hat{\tilde{a}}_1^{\dagger}(t - \sigma)}
-
\lambda \, b(t) \, 
\comm*{\hat{\tilde{a}}_0 (\tau)}{\hat{\tilde{a}}_0^{\dagger}(t)}
\right]
\\
&
+ 
\psi_1^{c}(\tau) \, 
i \del_t \, 
\comm*{\hat{\tilde{a}}_0 (\tau)}{\hat{\tilde{a}}_0^{\dagger}(t - \sigma)}
\biggr\}
\, ,
\end{aligned}
\ee 
which leads to 
\be 
\psi_1^{c} (t) - \psi_1^{c} (t - 2 \sigma)
=
\lim_{N \to \infty}
\int_{-N \sigma}^{N \sigma} d \tau \, 
\frac{\psi_0^{\ast} (\tau)}{2 N \sigma} \, 
\Big[ 
\hat{\tilde{a}}_0 (\tau)
\, , \, 
\del_t \, \hat{\tilde{a}}_1^{\dagger} (t - \sigma) 
+
i \lambda \, b(t) \, \hat{\tilde{a}}_0^{\dagger} (t)
\Bigr] 
\, .
\ee 
Similar to how $\psi_1(t)$ was determined,
the resulting first-order correction to the dual wavefunction 
is found to be 
\begin{align}
\psi_1^{c}(t)
=
&
\lim_{N \to \infty}
\int_{-N \sigma}^{N \sigma} d \tau \, 
\frac{\psi_0^{\ast} (\tau)}{2 N \sigma} 
\sum_{j = 0}^{\infty} \, 
\Bigl[ 
\hat{\tilde{a}}_0 (\tau)
\, , \, 
\del_t \, \hat{\tilde{a}}_1^{\dagger} (t - 2 j \sigma - \sigma) 
+
i \lambda \, b(t - 2 j \sigma) \, 
\hat{\tilde{a}}_0^{\dagger} (t - 2 j \sigma)
\Bigr] 
\nn \\
&
+
\frac{\lambda \, c_1^{\prime}}{2 N \sigma} 
\label{psi_2_iterative}
\end{align}
for an arbitrary constant $c_1^{\prime}$.
Moreover, 
since 
\be 
\sum_{j = 0}^{\infty}
\left[
\del_t \, 
\hat{\tilde{a}}_1^{\dagger} (t - 2 j \sigma - \sigma) 
+ 
i \lambda \, b(t - 2 j \sigma) \, 
\hat{\tilde{a}}_0 (t - 2 j \sigma)
\right]
=
i \lambda \, 
\xi \, 
b(t) \, \hat{\tilde{a}}_0^{\dagger} (t)
\ee 
according to eqs.~\eqref{a1_dagger} and~\eqref{g1}, 
the expression~\eqref{psi_2_iterative} reduces to 
\begin{align}
\psi_1^{c}(t)
&=
i \lambda \, \xi \, b(t)
\lim_{N \to \infty}
\left[ 
\int_{- N \sigma}^{t \, - \, \sigma} d \tau \, 
\frac{\psi_0^{\ast} (\tau)}{2 N \sigma} 
+
\int_{t \, + \, \sigma}^{N \sigma} d \tau \, 
\frac{\psi_0^{\ast} (\tau)}{2 N \sigma} 
\right] 
+
\frac{\lambda \, c_1^{\prime}}{2 N \sigma} 
\\
&=
i \lambda \, \xi \, b(t) \, \overbar{\psi}_0^{\ast}
+
\frac{\lambda \, c_1^{\prime}}{2 N \sigma} 
\, ,
\label{psi_1_star}
\end{align}
where
\be
\label{psi0_star_avg}
\overbar{\psi}_0^{\ast} 
=
\lim_{N \to \infty}
\int_{- N \sigma}^{N \sigma} d \tau \, 
\frac{\psi_0^{\ast} (\tau)}{2 N \sigma} 
\, .
\ee
If $\langle 1_{\Psi} |$ defined above is interpreted as 
the conjugate of a given physical state $\ket{1_{\Psi}}$, 
the zeroth-order term $\psi_0^{\ast} (t)$ of the dual wavefunction $\Psi^c (t)$ 
is already determined once $\psi_0 (t)$ is fixed 
by resolving the constant $c_1$ in eq.~\eqref{c1}. 
Consequently, 
the constant term $\lambda \, c_1^{\prime}$ in eq.~\eqref{psi_1_star} 
cannot be absorbed by redefining $\psi^{\ast}_0(t)$ again. 
Instead, we will fix the free parameter $c_1^{\prime}$ in section~\ref{sec:norm} 
by requiring that the norm of a physical state is real.

Higher-order corrections to the wavefunctions $\Psi(t)$ and $\Psi^{c}(t)$ can be derived
from the physical-state constraints~\eqref{eom_a_1particle} and~\eqref{eom_a_dag_1particle}, respectively, by following similar steps as outlined above.
Let us begin by pointing out that the $\mathcal{O}(\lambda)$ correction $\psi_1 (t)$~\eqref{psi_1} to the wavefunction would be zero under the causal prescription $\xi = 1$ 
for the operators $\hat{\tilde{a}}_1 (t)$ and $\hat{\tilde{a}}_1^{\dagger} (t)$ 
constructed in section~\ref{sec:2pt_function}. 
As a matter of fact, 
with the choice $\xi = 1$, the lowering and raising operators 
$\hat{\tilde{a}}_n (t)$ and $\hat{\tilde{a}}_n^{\dagger} (t)$ at $\mathcal{O}(\lambda^n)$ 
derived in eqs.~\eqref{an} and~\eqref{an_dag} satisfy 
\begin{align}
\del_t \, \hat{\tilde{a}}_n (t)
&=
i \lambda \, 
b(t - \sigma) \, 
\hat{\tilde{a}}_{n - 1}(t - \sigma)
\, , 
\\
\del_t \, \hat{\tilde{a}}_n^{\dagger} (t)
&=
- i \lambda \, 
b(t - \sigma) \, 
\hat{\tilde{a}}_{n - 1}^{\dagger} (t - \sigma)
\end{align}
for all $n \geq 1$,
which imply the equalities
\begin{align}
i \del_t \, \hat{\tilde{a}} (t + \sigma) 
+
\lambda \, b(t) \, \hat{\tilde{a}} (t)
&=
i \del_t \, \hat{\tilde{a}}_0 (t + \sigma)
\, , 
\label{eom_a_reduce}
\\
- i \del_t \, \hat{\tilde{a}}^{\dagger} (t + \sigma) 
+
\lambda \, b(t) \, \hat{\tilde{a}}^{\dagger} (t)
&=
- i \del_t \, \hat{\tilde{a}}_0^{\dagger}(t + \sigma)
\label{eom_a_dag_reduce}
\end{align}
to all orders in $\lambda$.
In particular, due to the relation~\eqref{eom_a_reduce}, 
the physical-state constraint~\eqref{eom_a_1particle} 
reduces to just the zeroth-order constraint:
\be 
\label{eom_a_xi=1}
0
=
\left[
i \del_t \, 
\hat{\tilde{a}}(t + \sigma) 
+ 
\lambda \, b(t) \, \hat{\tilde{a}}(t)
\right]
\ket{1_{\Psi}} 
=
i \del_t \, 
\hat{\tilde{a}}_0 (t + \sigma) 
\ket{1_{\Psi}} 
\, .
\ee 
Thus, the all-order physical wavefunction $\Psi(t)$
is precisely given by the periodic wavefunction~\eqref{periodic},
i.e., 
\be 
\label{Psi_xi=1}
\Psi (t) = \psi_0 (t) 
\, , 
\qquad 
\text{where}
\quad 
\psi_0 (t + 2 \sigma)
=
\psi_0 (t) 
\quad 
\forall \ 
t
\, .
\ee 

On the other hand, 
making use of eq.~\eqref{eom_a_dag_reduce}, 
the constraint~\eqref{eom_a_dag_1particle} on the dual physical state becomes 
\be 
\label{eom_a_dag_xi=1}
0
=
\bra{1_{\Psi}} 
\bigl[
- i \del_t \, 
\hat{\tilde{a}}_0^{\dagger} (t - \sigma) 
+ 
\lambda \, b(t) \, \hat{\tilde{a}}^{\dagger}(t)
-
\lambda \, b(t - 2 \sigma) \, \hat{\tilde{a}}^{\dagger}(t - 2 \sigma)
\bigr] 
\, .
\ee 
It is shown in appendix~\ref{app:psi_n} that 
the $\mathcal{O}(\lambda^n)$ term $\psi_n^c (t)$ 
in the dual wave function $\Psi^c (t)$ 
determined from this condition takes the form
\begin{align}
\psi_n^c(t)
=
& \ 
(i \lambda)^n \, \overbar{\psi}_0^{\ast} \, b(t)
\bigintssss_{-\infty}^{\infty} 
\Biggl[ \, 
\prod_{j = 1}^{n - 1}
d t_j \, 
b(t_j) 
\Biggr] \, 
\Theta(t_{n - 1} - t_{n - 2} - \sigma) 
\cdots 
\Theta(t_2 - t_1 - \sigma) \, 
\Theta(t_1 - t - \sigma) 
\nn \\
& +
\sum_{j = 0}^{n - 2} 
i \, \lambda^{n - j} \, c_{n - j - 1}^{\prime} \, b(t)
\int_{-\infty}^{\infty} \frac{d \tau}{2 N \sigma} \, 
\bigl[ 
\hat{\tilde{a}}_0 (\tau) \, , 
\hat{\tilde{a}}_j^{\dagger} (t) 
\bigr]
+
\frac{\lambda^n \, c_n^{\prime}}{2 N \sigma}
\label{psi_n^c}
\end{align}
for $n \geq 1$.
Besides the arbitrary constant $c_n^{\prime}$ in the expression, $\psi_n^c(t)$ receives contributions from all the additive constants $\bigl\{ c_j^{\prime} \bigr\}_{j = 1}^{n - 1}$
that are present in the lower-order corrections $\bigl\{ \psi_j^c \bigr\}_{j = 1}^{n - 1}$
to the dual wavefunction.
Just like $c_1^{\prime}$,
these parameters $c_j^{\prime}$ will be fixed shortly in section~\ref{sec:norm}.

The fact that $\psi_n^c (t)$ depends only on the \textit{average} $\overbar{\psi}_0^{\ast}$~\eqref{psi0_star_avg} of the zeroth-order wavefunction signals the decoupling of the infinite number of extra degrees of freedom present in the naive Fock space~\eqref{fock}.
This remarkable feature will be the subject of discussion in section~\ref{sec:spurious}, where we illustrate that the time dependence of $\psi_0 (t)$ is associated with spurious degrees of freedom that decouple to all orders in $\lambda$ under the physical-state constraints.

\subsection{Removal of negative-norm states by physical-state conditions}
\label{sec:norm}

Here we illustrate that 
the proposed definition of physical states 
is sufficient to eliminate 
the negative-norm states present in the Fock space~\eqref{fock}.

Recall from eq.~\eqref{Psi_xi=1} that a one-particle state 
$\ket{1_{\Psi}} = \hat{\tilde{\mathcal{A}}}^{\dagger}[\Psi] \ket{0}$
defined in eq.~\eqref{1_particle_state}
is a physical state 
if the associated wave function $\Psi(t)$ 
is $2 \sigma$-periodic.
Now suppose that $\ket{1_{\Psi}}$ is 
an \textit{unphysical} one-particle state in the Fock space~\eqref{fock}
whose corresponding wavefunction $\Psi(t)$ 
has support only within a time interval $\sigma$. 
Such a state has zero norm:
\begin{align}
\inp{1_{\Psi}}{1_{\Psi}}
&=
\lim_{N \to \infty}
\frac{1}{(2 N \sigma)^2}
\int_{-N \sigma}^{N \sigma} dt 
\int_{-N \sigma}^{N \sigma} dt' \, 
\Psi^*(t) \, 
\Psi(t') \, 
\Theta(|t - t'| - \sigma) 
\\
&= 
\lim_{N \to \infty}
\frac{1}{(2 N \sigma)^2}
\int_{-N \sigma}^{N \sigma} dt \, 
\Psi^* (t) 
\left[ 
\int_{-N \sigma}^{t \, - \, \sigma} d t' \, 
\Psi(t')
+ 
\int_{t \, + \, \sigma}^{N \sigma} dt' \, 
\Psi(t')
\right]
= 
0
\, ,
\end{align}
since for values of $t$ where $\Psi^*(t) \neq 0$, 
the integration ranges $(-\infty, t - \sigma]$ 
and $[t + \sigma, \infty)$ in $t'$-space 
do not overlap with the support of $\Psi$.
If we consider 
the superposition of two such states 
$\ket{1_{\Psi}}$ and $\ket{1_{\Phi}}$, 
the norm becomes
\begin{align}
\bigl\lvert
\ket{1_{\Psi}} + \ket{1_{\Phi}}
\bigr\rvert^2
&= 
2 \Re \left\{ \inp{1_{\Psi}}{1_{\Phi}} \right\}
\label{norm_superpose} 
\nn \\
&= 
\lim_{N \to \infty}
\frac{2}{(2 N \sigma)^2}
\Re
\left\{ 
\int_{-N \sigma}^{N \sigma} dt
\int_{-N \sigma}^{N \sigma} dt' \, 
\Psi^* (t) \, 
\Phi (t') \, 
\Theta(|t - t'| - \sigma)
\right\}
\, ,
\end{align}
which could be negative
because the expression changes sign 
if the sign of either $\Psi$ or $\Phi$ is flipped.

However,
once we impose the physical-state constraints~\eqref{eom_a_xi=1} and~\eqref{eom_a_dag_xi=1},
the wave functions $\Psi(t)$ and $\Phi(t)$
are confined to being periodic with period $2 \sigma$:
\be 
\Psi(t) = \psi_0 (t) \, , 
\qquad 
\Phi(t) = \phi_0 (t) 
\qquad 
(\psi_0 \, , \phi_0
:
2 \sigma\text{-periodic})
\, .
\ee 
Meanwhile,
the dual wavefunction $\Psi^c (t)$ associated with 
a physical one-particle state $\bra{1_{\Psi}}$
was found in eq.~\eqref{psi_n^c}
to take the form\footnote{
Although the correction terms in $\displaystyle \Psi^c (t)$
largely dominate over the zeroth-order term due to $N$ being large, 
it is the combination $\displaystyle \Psi^{c}(t) / 2 N \sigma$ 
(finite in the $N\rightarrow\infty$ limit) 
that appears in the definition~\eqref{1_particle_dual} of states 
and ultimately in the calculations of inner products.
}
\begin{align}
\Psi^{c}(t)
&=
\psi_0^{\ast}(t)
+
\lim_{N \to \infty}
\left( 2 N \sigma \right)
\sum_{n = 1}^{\infty}
\psi_n^{c} (t)
\nn \\
&=
\psi_0^{\ast}(t)
+
\lim_{N \to \infty}
(2 N \sigma)
\left[ 
i \lambda \, b(t) \, \overbar{\psi}_0^{\ast} 
+ 
\frac{\lambda \, 
c'_1}{2 N \sigma}
+
\mathcal{O}(\lambda^2)
\right] 
.
\label{Psi^c_full}
\end{align}

In the free theory (order $\lambda^0$), 
we can express the inner product between two \emph{physical} one-particle states 
$\ket{1_{\Psi}}$ and $\ket{1_{\Phi}}$ as
\begin{align}
\inp{1_{\Psi}}{1_{\Phi}}
&=
\lim_{N \to \infty}
\frac{1}{(2 N \sigma)^2}
\int_{-N \sigma}^{N \sigma} dt
\int_{-N \sigma}^{N \sigma} dt' \, 
\psi_0^* (t) \, 
\phi_0 (t') \, 
\Theta(|t - t'| - \sigma)
\nn \\
&= 
\lim_{N \to \infty} 
\frac{1}{(2 N \sigma)^2}
\int_{-N \sigma}^{N \sigma} dt \, 
\psi_0^* (t)
\left[ 
\int_{-N \sigma}^{t \, - \, \sigma} 
dt' \, \phi_0 (t')
+ 
\int_{t \, + \, \sigma}^{N \sigma} 
dt' \, \phi_0 (t')
\right]
\nn \\
&= 
\overbar{\psi}_0^* \, \overbar{\phi}_0
\, ,
\end{align}
where we have again introduced the time-averaged wavefunctions 
$\overbar{\psi}_0^*$, $\overbar{\phi}_0$
defined in eqs.~\eqref{psi0_avg} and~\eqref{psi0_star_avg}.
Subsequently,
as opposed to eq.~\eqref{norm_superpose},
the norm of the superposition of 
two physical one-particle states is now given by 
\be 
\label{norm_superpose_phys}
\bigl\lvert
\ket{1_{\Psi}} + \ket{1_{\Phi}}
\bigr\rvert^2
=
\abs{\overbar{\psi}_0}^2 
+ 
\bigl| \overbar{\phi}_0 \bigr|^2
+ 
2 \Re \left\{ \overbar{\psi}_0^* \, \overbar{\phi}_0 \right\}
= 
\bigl| 
\overbar{\psi}_0 + \overbar{\phi}_0
\bigr|^2
\geq 
0
\, .
\ee 
Thus, the physical-state constraints
indeed decouple the negative-norm states from the system
at zeroth order in $\lambda$.
The extension of the above discussion 
to multi-particle states proceeds similarly.

In the interacting theory, 
the corrections induced by the background interaction 
modify the norm $\inp{1_{\Psi}}{1_{\Psi}}$ as
\begin{align}
\inp{1_{\Psi}}{1_{\Psi}}
&=
\abs{\overbar{\psi}_0}^2
+
\lim_{N \to \infty}
\int_{-N \sigma}^{N \sigma} dt
\int_{-N \sigma}^{N \sigma} dt' \, 
\psi_1^{c}(t) \, 
\frac{\psi_0(t')}{2 N \sigma} \, 
\Theta(|t - t'| - \sigma)
+
\mathcal{O}(\lambda^2)
\nn \\
&= 
\abs{\overbar{\psi}_0}^2
\left[ 
1 
+
i \lambda
\int_{-\infty}^{\infty}
b(t) \, dt
\right]
+ 
\lambda \, c'_1 \, \overbar{\psi}_0
+ 
\mathcal{O}(\lambda^2)
\, .
\label{norm_lambda}
\end{align}
The imaginary contribution to the norm~\eqref{norm_lambda} of a physical state 
arises from the mismatch between 
the complex conjugate $\Psi^{\ast} (t) = \psi_0^{\ast} (t)$ of the wavefunction 
and its dual $\Psi^c (t)$ given by eq.~\eqref{Psi^c_full}.
This is an inevitable consequence of implementing 
the involution symmetry~\eqref{involution} on the operator algebra,
while the equations of motion~\eqref{eom_a} and~\eqref{eom_a_dagger} 
are related by a different involution (the complex conjugation) 
under which $\sigma$ transforms as an imaginary number.
That said,
we can eliminate the imaginary piece at $\mathcal{O}(\lambda)$ 
by fixing the arbitrary constant
\be
c'_1 = 
i 
\left[\int_{-\infty}^{\infty} dt \, b(t)\right] 
\overbar{\psi}^{\ast}_0
\ee
in the definition~\eqref{psi_1_star} of $\psi_1^c$ 
so that the norm $\inp{1_{\Psi}}{1_{\Psi}} = \abs{\overbar{\psi}_0}^2$ 
is positive-definite up to first order in $\lambda$ 
as long as $\overbar{\psi}_0 \neq 0$.

Moreover,
based on eq.~\eqref{psi_n^c},
$\Psi^c (t)$ is a linear function of $\overbar{\psi}_0^{\ast}$,
and thus the $\mathcal{O}(\lambda^n)$ contribution to the norm 
has the form  
\be 
\label{norm_lambda_n}
\inp{1_{\Psi}}{1_{\Psi}}_n
=
\lambda^n 
\left[
\abs{\overbar{\psi}_0}^2
I_n [b]
+
\left( 
J_n[b]
+
c_n^{\prime} 
\right) 
\overbar{\psi}_0 
\right] 
,
\ee 
where 
\begin{align}
I_n [b]
&
=
i^n 
\int_{-\infty}^{\infty}
d t_0 \, b(t_0)
\prod_{j = 1}^{n - 1}
\int_{-\infty}^{\infty}
d t_j \, 
b(t_j) \, 
\Theta(t_j - t_{j - 1} - \sigma) 
\, , 
\\
J_n[b]
&=
\sum_{j = 0}^{n - 2}
i^{j + 1} \left( -1 \right)^j 
c_{n - j - 1}^{\prime}
\int_{-\infty}^{\infty}
d t_{j + 1} \, b(t_{j + 1})
\prod_{k = 1}^{j}
\int_{-\infty}^{\infty}
d t_k \, 
b(t_k) \, 
\Theta(t_{k + 1} - t_k - \sigma) 
\end{align}
are time-independent functionals of $b(t)$,
with $J_n [b]$ depending also on 
the arbitrary constants $\bigl\{ c_j^{\prime} \bigr\}_{j = 1}^{n - 1}$
in the lower-order terms $\bigl\{ \psi_j^c \bigr\}_{j = 1}^{n - 1}$~\eqref{psi_n^c} 
of the dual wavefunction.
By examining the second term 
$\left( J_n[b] + c_n^{\prime} \right) \overbar{\psi}_0$
in the square brackets of eq.~\eqref{norm_lambda_n},
it becomes clear that the norm of a physical state 
can be made \textit{positive-definite} to all orders in $\lambda$
if we set all the arbitrary constants $c_j^{\prime}$ 
to be proportional to $\overbar{\psi}_0^{\ast}$:
\be 
\label{cn_prime}
c_j^{\prime} \propto \overbar{\psi}_0^{\ast}
\qquad 
\forall \ j \geq 1
\, ,
\ee 
and then suitably selecting the 
$b$-dependent proportionality constant at each order.
As a matter of fact, 
one can even choose $\bigl\{ c_j^{\prime} \bigr\}_{j = 1}^n$
in a way that all higher-order corrections to the norm
are canceled out exactly up to order $\lambda^n$, leaving just 
$\inp{1_{\Psi}}{1_{\Psi}} = \abs{\overbar{\psi}_0}^2 + \mathcal{O}(\lambda^{n + 1})$.
Subtleties related to physical zero-norm states with $\bar{\psi}_0 = 0$ 
will be discussed below in section~\ref{sec:spurious}.

\subsection{Decoupling of zero-norm states}
\label{sec:spurious}

Even with the negative-norm states eliminated
in the physical Hilbert space~\eqref{H_phys}, 
the space of physical states in the nonlocal theory ($\sigma > 0$) 
remains significantly larger than the Fock space of the local theory with $\sigma = 0$. 
In the local theory, 
the zeroth-order wavefunctions are merely constants, 
whereas in the nonlocal case, 
the physical-state condition~\eqref{Psi_xi=1} 
allows for arbitrary $2 \sigma$-periodic functions.
The physical wavefunctions can be decomposed into a Fourier series as
\be 
\Psi(t) = \overbar{\Psi} 
+ 
\sum_{n \, \in \, \mathbb{Z} \setminus \{ 0 \}}
\alpha_n 
\exp \left( i \pi n \, t / \sigma \right)
,
\ee 
where $\overbar{\Psi}$ represents the time average of $\Psi(t)$, 
analogous to eq.~\eqref{psi0_avg}. 
In this section, 
we demonstrate that the \textit{zero-norm} physical states
---
characterized by periodic wavefunctions $e^{i \pi n \, t / \sigma}$
that average to zero over a cycle
---
decouple from the space of positive-norm physical states. 
As a result, 
the space of positive-norm physical states is ultimately 
equivalent to the Fock space of the local theory.

Notice from eqs.~\eqref{norm_lambda} and~\eqref{norm_lambda_n} that
only the ``zero mode'' $\overbar{\psi}_0$~\eqref{psi0_avg} 
of a one-particle wavefunction $\Psi(t) = \psi_0 (t)$
contributes to the norm of a physical state. 
This suggests that the system possesses a large redundancy, 
as there is an equivalence relation 
on the physical Hilbert space:
\be 
\label{gauge_symm}
\ket{1_{\Psi}}
=
\hat{\tilde{\mathcal{A}}}^{\dagger} [\Psi]
\ket{0}
\sim 
\hat{\tilde{\mathcal{A}}}^{\dagger} [\overbar{\Psi}]
\ket{0}
.
\ee 
This is analogous to what happens due to spurious states
in the covariant quantization of the string worldsheet theory.

It is clear from the equivalence relation~\eqref{gauge_symm}
that the \textit{spurious} physical states $\ket{\Psi^{\mathrm{sp}}}$ 
are those whose corresponding wavefunctions
are $2 \sigma$-periodic but average to zero over a cycle.
Let 
\be 
\label{spurious_mode}
\varphi^{\mathrm{sp}}_{0} (t)
=
\exp \left( i \pi n \,t / \sigma \right) 
\ee 
with $n \in \mathbb{Z} \setminus \{ 0 \}$
be a basis mode for the spurious wavefunctions in the free theory.
Due to the vanishing of the zero mode 
\be 
\overbar{\varphi}^{\mathrm{sp}}_{0}
\equiv 
\lim_{N \to \infty}
\frac{1}{2 N \sigma} 
\int_{-N \sigma}^{N \sigma} 
\varphi^{\mathrm{sp}}_{0} (t) \, dt 
=
\frac{1}{2 \sigma} 
\int_{-\sigma}^{\sigma} 
\varphi^{\mathrm{sp}}_{0} (t) \, dt 
=
0
\, ,
\ee 
the associated spurious basis state 
$\ket{\Phi^{\mathrm{sp}}} = \hat{\tilde{\mathcal{A}}}^{\dagger} [\Phi^{\mathrm{sp}}] \ket{0}$
decouples from all physical observables 
at zeroth order in $\lambda$. 

Remarkably,
the $\mathcal{O}(\lambda^n)$ contributions to 
the wavefunction $\Phi^{\mathrm{sp}}$ 
and its dual $(\Phi^{\mathrm{sp}})^c$
determined from 
eqs.~\eqref{Psi_xi=1} and~\eqref{psi_n^c}
turn out to be zero for all $n \geq 1$:\footnote{
The constants $c_n^{\prime}$ 
in the dual wavefunctions $\psi_n^c$~\eqref{psi_n^c} 
were previously chosen in eq.~\eqref{cn_prime} 
to ensure the positive-definiteness of the physical Hilbert space
$\mathcal{H}_{\mathrm{phys}}$. 
However, for the zero-norm states considered here, 
these constants are set to zero.}
\be 
\label{phi_sp_n}
\varphi^{\mathrm{sp}}_{n} (t) 
=
0
\, , 
\qquad 
\left( \varphi^{\mathrm{sp}}_{n} \right)^{c} (t) 
\propto 
\left( \overbar{\varphi}^{\mathrm{sp}}_{0} \right)^{\ast}
=
0
\, .
\ee 
This implies that the decoupling of the 
spurious basis state $\ket{\Phi^{\mathrm{sp}}}$
is not merely an artifact of the free theory,
but is a guaranteed feature 
to all orders in the perturbation theory.
Therefore, the mode functions~\eqref{spurious_mode} genuinely represent 
redundant degrees of freedom.

With eq.~\eqref{phi_sp_n}, 
one can further conclude that 
\be 
\inp{\Phi^{\mathrm{sp}}}{\Psi}
=
\inp{\Psi}{\Phi^{\mathrm{sp}}}
=
0
\qquad 
\text{for any} \ 
\ket{\Psi} \in \mathcal{H}_{\mathrm{phys}}
\ee 
in the full interacting theory.
This suggests that the \textit{physical} representation 
of the algebra~\eqref{comm0} is actually much smaller 
than the one that we have worked with so far by adopting 
eqs.~\eqref{n_particle_state} and~\eqref{A_dagger}.
Although the time dependence of the ladder operators 
$\hat{\tilde{a}}_0(t)$ and $\hat{\tilde{a}}_0^{\dagger}(t)$
seems to have introduced an infinite number of extra degrees of freedom
through the wavefunctions $\Psi (t)$,
it is sufficient to consider just 
\textit{constant} wavefunctions $\overbar{\Psi}$ 
for the creation operators 
$\hat{\tilde{\mathcal{A}}}^{\dagger} [\Psi]$~\eqref{1_particle_state}.

By excluding the spurious zero-norm states,
the space of physical states reduces to 
\be
\mathrm{span}
\bigl\{ (\hat{\tilde{\mathfrak{a}}}^{\dagger})^n \ket{0} \bigr\}
\, ,
\qquad 
\text{where}
\quad 
\hat{\tilde{\mathfrak{a}}}^{\dagger} \equiv 
\lim_{N\rightarrow\infty} \frac{1}{2N\sigma}
\int_{-N\sigma}^{N\sigma} dt \, \hat{\tilde{a}}_0^{\dagger}(t)
\, .
\label{A-def}
\ee
There is a one-to-one correspondence between states in this space
and the states in the Fock space of the local theory with $\sigma = 0$,
indicating that the space of physical states is in fact 
of the same dimension as the local theory,
in which the creation operators have no explicit time dependence.
Furthermore, all states in this space have positive norms, 
free from the pathologies typically associated with infinite-time-derivative theories~\cite{Ostrogradsky:1850fid, Woodard:2015zca}.

\subsection{Hamiltonian}
\label{sec:H}

In the Hamiltonian formalism, 
the Hamiltonian $\hat{H}(t)$ serves to generate 
the time evolution of operators $\hat{O}(t)$ in the Heisenberg picture 
through $\del_t \, \hat{O}(t) = -i \comm*{\hat{O}(t)}{\hat{H}(t)}$. 
Information about interactions is encoded in the Hamiltonian 
as an alternative approach to quantum mechanics 
alongside the path-integral formalism.
For the nonlocal 1D model under consideration, the time dependence of the operators $\hat{\tilde{a}}(t)$ and $\hat{\tilde{a}}^{\dag}(t)$, as well as the physical states $\ket{\Psi}$, has already been determined using information derived from the path-integral formalism.
Furthermore, with the dynamical equations imposed as physical constraints, 
the Hamiltonian no longer serves exactly the same role as the generator of time evolution 
as it does in local theories. 
Nevertheless, for the sake of completeness, we shall define and derive a Hamiltonian for this nonlocal model in this subsection.

Given the time dependencies of $\hat{\tilde{a}}(t)$, $\hat{\tilde{a}}^{\dagger}(t)$, and their commutator, a Hamiltonian $\hat{H}[\hat{\tilde{a}}, \hat{\tilde{a}}^{\dagger}](t)$ for the nonlocal 1D model~\eqref{S_a_tilde} can be constructed by reverse engineering it to reproduce the desired operator evolution 
through the Heisenberg equations
\be 
\label{heisenberg}
\comm*{\hat{\tilde{a}}(t)}{\hat{H}(t)} 
= 
i \del_t \, \hat{\tilde{a}}(t)
\, , 
\qquad 
\comm*{\hat{\tilde{a}}^{\dagger}(t)}{\hat{H}(t)} 
= 
i \del_t \, \hat{\tilde{a}}^{\dagger}(t)
\, .
\ee 
Assuming $\hat{H}(t)$ is quadratic in the ladder operators,
we proceed to construct it order by order in the coupling constant $\lambda$.

At zeroth order,
the Heisenberg equations~\eqref{heisenberg} take the forms
\be
\label{heisenberg0}
\comm*{\hat{\tilde{a}}_0(t)}{\hat{H}_0}
= i \del_t \, \hat{\tilde{a}}_0(t)
\, ,
\qquad
\comm*{\hat{\tilde{a}}_0^{\dagger}(t)}{\hat{H}_0}
= i \del_t \, \hat{\tilde{a}}_0^{\dagger}(t)
\, .
\ee 
By making the time-independent ansatz 
\be 
\label{H0}
\hat{H}_0 
= 
\int_{-\infty}^{\infty} d \tau 
\int_{-\infty}^{\infty} d \tau' \, 
h_0(\tau, \tau') \, 
\del_{\tau} \, \hat{\tilde{a}}_0^{\dagger}(\tau) \, 
\del_{\tau'} \, \hat{\tilde{a}}_0(\tau') 
\ee 
for the free Hamiltonian,
we find using the zeroth-order commutator~\eqref{comm0} that 
the function $h_0(\tau, \tau')$ has to satisfy 
the inhomogeneous difference equations 
\begin{align}
h_0(\tau \, , \tau' - \sigma) 
- 
h_0(\tau \, , \tau' + \sigma) 
&= 
i \, \delta(\tau - \tau')
\, ,
\\
h_0(\tau + \sigma \, , \tau') 
- 
h_0(\tau - \sigma \, , \tau') 
&= 
i \, \delta(\tau - \tau')
\end{align}
in order to reproduce~\eqref{heisenberg0}. 
The general solution to these algebraic equations 
can be written as 
\be 
\label{h0}
h_0 (\tau, \tau')
=
i \sum_{n = 0}^{\infty} 
\delta\left[ \tau - \tau' - (2n + 1) \, \sigma \right]
+
(\text{homogeneous term})
\, ,
\ee 
where the homogeneous solution can be any function 
that is periodic in both $\tau$ and $\tau'$ with period $2 \sigma$. 
The homogeneous solution will be dropped from now on,
as it does not affect how $\hat{H}_0$ acts on physical states
and therefore has no physical relevance.\footnote{
This is due to the fact that 
$\int_{-\infty}^{\infty} d \tau \, 
f(\tau) \, \del_{\tau} \, \hat{\tilde{a}}_0 (\tau)$
and 
$\int_{-\infty}^{\infty}
d \tau \, f(\tau) \, \del_{\tau} \, \hat{\tilde{a}}_0^{\dagger} (\tau)$
commute with all operators $\hat{O} [\hat{\tilde{a}}_0, \hat{\tilde{a}}_0^{\dagger}]$
if $f(\tau)$ is a periodic function with period $2 \sigma$. 
}
Leaving just the particular solution in eq.~\eqref{h0},
we get 
\be 
h_0 (\tau, \tau')
=
i \sum_{n = 0}^{\infty} 
\delta\left[ \tau - \tau' - (2n + 1) \, \sigma \right]
=
\frac{i}{2} 
\csch\left( \sigma \del_{\tau} \right) 
\delta(\tau - \tau')
\, .
\ee 
As a result, 
the zeroth-order Hamiltonian~\eqref{H0} can be expressed as 
\be 
\label{H0_2}
\hat{H}_0
=
\frac{i}{2} 
\int_{-\infty}^{\infty} d \tau \, 
\bigl[ \del_{\tau} \, \hat{\tilde{a}}_0^{\dagger} (\tau) \bigr] 
\csch\left( \sigma \del_{\tau} \right) 
\bigl[ \del_{\tau} \, \hat{\tilde{a}}_0 (\tau) \bigr] 
\, .
\ee 
which is inherently nonlocal as the nonlocality in the model~\eqref{S_a_tilde} is already encoded in the free-field action.
Note, however, that despite the presence of infinite time derivatives in the kinetic term of this model, the free Hamiltonian constructed above satisfies $\ev{\hat{H}_0}{\Psi} = 0$ in the physical Hilbert space with periodic wavefunctions~\eqref{periodic}.

The leading-order interaction Hamiltonian $\hat{H}_1 (t)$ 
is defined to generate the $\mathcal{O}(\lambda)$
corrections $\hat{\tilde{a}}_1 (t)$ and $\hat{\tilde{a}}_1^{\dagger} (t)$
consistent with the path integral, iteratively, through 
\be 
\comm*{\hat{\tilde{a}}_0(t)}{\hat{H}_1}
= 
i \del_t \, \hat{\tilde{a}}_1(t)
\, ,
\qquad
\comm*{\hat{\tilde{a}}_0^{\dagger}(t)}{\hat{H}_1}
= 
i \del_t \, \hat{\tilde{a}}_1^{\dagger}(t)
\, .
\ee 
Given that the time evolution of 
$\hat{\tilde{a}}_1 (t)$ and $\hat{\tilde{a}}_1^{\dagger} (t)$
in eqs.~\eqref{a1}--\eqref{a1_dagger} follow
\begin{align}
\del_t \, \hat{\tilde{a}}_1 (t)
&=
i \lambda \, 
\bigl[ 
\xi \, b(t - \sigma) \, \hat{\tilde{a}}_0 (t - \sigma) 
-
\left( \xi - 1 \right)
b(t + \sigma) \, \hat{\tilde{a}}_0 (t + \sigma)
\bigr]
\, , 
\\
\del_t \, \hat{\tilde{a}}_1^{\dagger} (t)
&=
- i \lambda \, 
\bigl[ 
\xi \, b(t - \sigma) \, \hat{\tilde{a}}_0^{\dagger} (t - \sigma) 
-
\left( \xi - 1 \right)
b(t + \sigma) \, \hat{\tilde{a}}_0^{\dagger} (t + \sigma)
\bigr]
\, ,
\end{align}
we propose an ansatz for $\hat{H}_1 (t)$ 
that meets the criteria:
\be 
\begin{aligned}
\hat{H}_1(t)
=
\ & 
\lambda \, \xi \, b(t - \sigma)
\int_{-\infty}^{\infty} d \tau 
\int_{-\infty}^{\infty} d \tau' \, 
h_1^- (\tau, \tau') \, \hat{\tilde{a}}_0^{\dagger}(t + \tau) \, \hat{\tilde{a}}_0(t + \tau')
\\
&+
\lambda \left( \xi - 1 \right) b(t + \sigma)
\int_{-\infty}^{\infty} d \tau 
\int_{-\infty}^{\infty} d \tau' \, 
h_1^+ (\tau, \tau') \, \hat{\tilde{a}}_0^{\dagger}(t + \tau) \, \hat{\tilde{a}}_0(t + \tau')
\, ,
\end{aligned}
\ee 
where the functions $h_1^- (\tau, \tau')$ and $h_1^+ (\tau, \tau')$
are required to satisfy 
\begin{align}
\int_{-\infty}^{\infty} d\tau \, 
\Theta(|\tau| - \sigma + \epsilon) \, 
h_1^{\pm}(\tau, \tau') 
&
= 
\pm \, \delta(\tau' \mp \sigma)
\, ,
\label{h1}
\\
\int_{-\infty}^{\infty} d\tau' \, 
\Theta(|\tau'| - \sigma + \epsilon) \, 
h_1^{\pm}(\tau, \tau')
&
= 
\pm \, \delta(\tau \mp \sigma)
\, . 
\label{h1_prime}
\end{align}
Notice that the infinitesimal term $\epsilon$ in the commutator~\eqref{comm0_epsilon}
is retained in this derivation.
Eqs.~\eqref{h1} and~\eqref{h1_prime} can be solved to give 
\be 
h_1^{\pm}(\tau , \tau') 
= 
\pm \, 
\delta(\tau \mp \sigma) \, 
\delta(\tau' \mp \sigma)
\, ,
\ee 
which then leads to
\be 
\begin{aligned}
\hat{H}_1 (t)
=
\ &
- \lambda \, \xi \, b(t - \sigma) \, 
\hat{\tilde{a}}_0^{\dagger}(t - \sigma) \, \hat{\tilde{a}}_0(t - \sigma)
\\
&
- \lambda \left( 1 - \xi \right) b(t + \sigma) \, 
\hat{\tilde{a}}_0^{\dagger}(t + \tau) \, \hat{\tilde{a}}_0(t + \tau')
\, .
\end{aligned}
\ee 
We observe that again the choice $\xi = 1$ 
for the parameter $\xi$ appearing in eq.~\eqref{g1} 
ensures causality, i.e.,
$\hat{H}_1(t)$ only depends on $b(t')$ for $t' < t$.

Extending the procedure to order $\lambda^n$,
we find that the interaction Hamiltonian must obey
\be 
\del_t \, \hat{\tilde{a}}_n (t)
=
- i 
\sum_{j = 0}^{n - 1} \, 
\bigl[ 
\hat{\tilde{a}}_j (t) \, ,
\hat{H}_{n - j} (t)
\bigr]
\, ,
\ee 
and it is shown explicitly in appendix~\ref{app:Ham} that for $\xi = 1$
the interaction Hamiltonian is given by 
\be 
\label{H_int}
\hat{H}_{\mathrm{int}} (t)
=
- \lambda \, b(t - \sigma) \, 
\hat{\tilde{a}}^{\dagger} (t - \sigma) \, 
\hat{\tilde{a}} (t - \sigma)
\ee 
to all orders in $\lambda$.

The full Hamiltonian $\hat{H}(t) = \hat{H}_0 + \hat{H}_{\mathrm{int}}(t)$ 
constructed above is invariant under the involution~\eqref{involution}. 
However, 
since it does not commute with either 
$\hat{\tilde{a}}(t)$ or $\hat{\tilde{a}}^\dagger(t)$, 
the physical Hilbert space $\mathcal{H}_{\mathrm{phys}}$ defined in~\eqref{H_phys} 
using the equation-of-motion constraints~\eqref{eom_a_full} and~\eqref{eom_a_dag_full}
is not preserved as an invariant subspace of $\hat{H}(t)$. 
Specifically, 
applying $\hat{H}(t)$ to a state $\ket{\Psi} \in \mathcal{H}_{\mathrm{phys}}$ results in  
$\hat{H}(t) \ket{\Psi} \notin \mathcal{H}_{\mathrm{phys}}$.
For example, 
acting the Hamiltonian $\hat{H}_0$~\eqref{H0_2} in the free theory
on a physical one-particle state $\hat{\tilde{\mathfrak{a}}}^\dagger \ket{0}$
defined via the creation operator in eq.~\eqref{A-def}
leads to 
\be 
\hat{H}_0 \, 
\hat{\tilde{\mathfrak{a}}}^\dagger \ket{0} 
=
- 
\lim_{N \rightarrow \infty}
\frac{i}{2 N \sigma} 
\left[ 
\hat{\tilde{a}}_0^{\dagger} (N \sigma) 
- 
\hat{\tilde{a}}_0^{\dagger} (- N \sigma) 
\right]
\ket{0}
\, ,
\ee 
which effectively \textit{annihilates} the state from $\mathcal{H}_{\mathrm{phys}}$
since 
\be 
\bra{0} \hat{\tilde{\mathfrak{a}}} \, 
\hat{H}_0 \, 
\hat{\tilde{\mathfrak{a}}}^\dagger \ket{0} 
= 
0
\, .
\ee 
Thus, the relationship between the constructed Hamiltonian 
in the Heisenberg picture and the usual notion of a Hamiltonian remains unclear. 
We leave this aspect for future investigation.

\subsection{Comments on related works}

In the spirit of Ostrogradski's framework~\cite{Ostrogradsky:1850fid}
of higher-derivative theories,
a general Hamiltonian formalism for nonlocal theories 
containing time derivatives of infinite order,
known as the \emph{$(1 + 1)$-dimensional formalism},
was developed in ref.~\cite{Llosa94}.
This formalism has since been further studied and 
applied to various examples~\cite{Gomis:2000gy, Bering:2000hc, Gomis:2000sp, Gomis:2003xv, Kolar:2020ezu}.
In this section, 
we comment on the similarities and differences 
between that formalism and the one introduced in this study.

To facilitate a comparison, we apply the $(1 + 1)$-dimensional formalism~\cite{Llosa94} to the free part of the nonlocal 1D model~\eqref{S_a_tilde}.
Their key idea is to extend the dynamical variables $\tilde{a}(t)$ and $\tilde{a}^{\dagger}(t)$ to fields $\tilde{A}(t \, , \tau)$ and $\tilde{A}^{\dagger}(t \, , \tau)$ in a space with one extra dimension $\tau \in \mathbb{R}$, satisfying the chirality conditions
\be 
\label{chirality}
\del_t \, \tilde{A}(t \, , \tau)
=
\del_{\tau} \, \tilde{A}(t \, , \tau)
\, , 
\qquad 
\del_t \, \tilde{A}^{\dagger}(t \, , \tau)
=
\del_{\tau} \, \tilde{A}^{\dagger}(t \, , \tau)
\, ,
\ee 
which lead to the following correspondences:
\be 
\label{A=a}
\tilde{A}(t \, , \tau) 
= 
\tilde{a}(t + \tau)
\, , 
\qquad
\tilde{A}^{\dagger}(t \, , \tau) 
= 
\tilde{a}^{\dagger}(t + \tau)
\, .
\ee 
Subsequently, the free-field action is rewritten as~\cite{Llosa94}
\be 
\label{S_A}
\begin{aligned}
S_A
=
\int dt \int d \tau \, 
\Bigl\{
\delta(\tau) \, 
\mathcal{L} [\tilde{A} \, , \tilde{A}^{\dagger}](t \, , \tau)
&+
P_A (t \, , \tau) 
\left[ 
\del_t \, \tilde{A} (t \, , \tau) 
- 
\del_{\tau} \, \tilde{A} (t \, , \tau) 
\right]
\\
&+
P_{A^{\dagger}} (t \, , \tau) \, 
\bigl[ 
\del_t \, \tilde{A}^{\dagger} (t \, , \tau) 
- 
\del_{\tau} \, \tilde{A}^{\dagger} (t \, , \tau) 
\bigr]
\Bigr\}
\, ,
\end{aligned}
\ee 
where $P_A (t \, , \tau)$ and $P_{A^{\dagger}} (t \, , \tau)$ 
are auxiliary fields that serve to 
enforce the desired conditions in eq.~\eqref{chirality},
whereas 
\be 
\mathcal{L}
[\tilde{A} \, , \tilde{A}^{\dagger}](t \, , \tau)
\equiv 
i \, \tilde{A}^{\dagger} (t \, , \tau - \sigma) \, 
\del_{\tau} \tilde{A} (t \, , \tau)
\ee 
is determined by the original free Lagrangian 
$L[\tilde{a} \, , \tilde{a}^{\dagger}] (t) 
\equiv  
i \, \tilde{a}^{\dagger}(t - \sigma) \, \del_t \, \tilde{a}(t)$ 
in eq.~\eqref{S_a_tilde} via the substitution
\be 
\tilde{a}(t) 
\to 
\tilde{A} (t \, , \tau)
\, , 
\qquad 
\tilde{a}^{\dagger} (t) 
\to 
\tilde{A}^{\dagger} (t \, , \tau)
\, ,
\qquad 
\del_t 
\to 
\del_{\tau}
\, .
\ee 
By replacing all the time derivatives $\del_t$ with 
derivatives $\del_{\tau}$ along the $\tau$-direction,
eq.~\eqref{S_A} defines a field theory 
which is local in the evolution time $t$,
with the nonlocality encoded in the internal parameter $\tau$.

In the Hamiltonian formalism for the field theory~\eqref{S_A},
$P_A (t \, , \tau)$ and $P_{A^{\dagger}} (t \, , \tau)$
act as nondynamical fields that are identified as the conjugate momenta of 
$\tilde{A}(t \, , \tau)$ and $\tilde{A}^{\dagger}(t \, , \tau)$,
respectively.
The reduced Hamiltonian can thus be written as 
\be 
\label{H_A}
H_A (t)
=
\int_{-\infty}^{\infty} d \tau \, 
\Bigl\{
P_A(t \, , \tau) \, \del_{\tau} \, \tilde{A}(t \, , \tau)
+
P_{A^{\dagger}}(t \, , \tau) \, \del_{\tau} \, \tilde{A}^{\dagger}(t \, , \tau)
-
\delta(\tau) \, 
\mathcal{L}[\tilde{A}, \tilde{A}^{\dagger}](t \, , \tau)
\Bigr\}
\, ,
\ee 
from which we see that the conditions~\eqref{chirality}
are realized as the equations of motion for 
$\tilde{A}(t \, , \tau)$ and $\tilde{A}^{\dagger}(t \, , \tau)$. 
In addition,
the conjugate momenta are subject to 
the constraints~\cite{Llosa94, Gomis:2000gy}
\begin{align}
P_A (t \, , \tau) 
&
\approx 
\int_{-\infty}^{\infty} d \tau' 
\left[ \Theta(\tau) - \Theta(\tau') \right]
\frac{\delta \mathcal{L}[\tilde{A} \, , \tilde{A}^{\dagger}](t \, , \tau')}
{\delta \tilde{A} (t \, , \tau)}
\, ,
\\
P_{A^{\dagger}} (t \, , \tau) 
&
\approx 
\int_{-\infty}^{\infty} d \tau' 
\left[ \Theta(\tau) - \Theta(\tau') \right]
\frac{\delta \mathcal{L}[\tilde{A} \, , \tilde{A}^{\dagger}](t \, , \tau')}
{\delta \tilde{A}^{\dagger} (t \, , \tau)} \, ,
\end{align}
whose corresponding secondary constraints recover 
the original equations of motion for 
$\tilde{a}(t)$ and $\tilde{a}^{\dagger}(t)$~\cite{Llosa94, Gomis:2000gy}.\footnote{
The weak equality symbol ``$\approx$''
denotes equality modulo first-class constraints;
namely, the equation holds only on the constrained surface.
}
Here we make a few important remarks:
\begin{itemize}[leftmargin=*]
\item
The newly introduced field variables 
$\tilde{A}(t \, , \tau)$ and $\tilde{A}^{\dagger}(t \, , \tau)$ 
are only used in intermediate steps.
As the Dirac brackets of this constrained system 
respect the time evolution generated by the Hamiltonian~\eqref{H_A},
the relation~\eqref{A=a} allows us to eventually express the Dirac brackets 
in terms of the original variables 
$\tilde{a}(t)$ and $\tilde{a}^{\dagger}(t)$.
The Dirac Bracket 
$\left\{ 
\tilde{a} (t) \, , 
\tilde{a}^{\dagger} (t') 
\right\}_{\mathrm{D}}$
produced in this setup is provided in appendix~\ref{app:Llosa}
and is shown to be different from the algebra~\eqref{comm0} 
we have obtained through the path-integral correlation function.  
Consequently, 
the $(1 + 1)$-dimensional formalism does not share the desirable properties of our approach, including the absence of negative-norm states and the decoupling of zero-norm states.

\item
In this formalism,
the point $\tau = 0$ holds a special role in the action~\eqref{S_A}.\footnote{
The delta function $\delta(\tau)$ in the action~\eqref{S_A} 
cannot simply be replaced by 
a constant term of mass dimension one, 
as doing so would effectively rescale the Planck constant $\hbar$
in the weight $\exp \left( i S_A / \hbar \right)$
by an infinite factor.
}
However,
this choice is artificial and 
spoils the translation symmetry along the $\tau$-direction,
leading to a discrepancy with our results, 
as illustrated in appendix~\ref{app:Llosa}. 

\item
The original equations of motion for $\tilde{a}(t)$ and $\tilde{a}^{\dagger}(t)$
appear as secondary constraints in this Hamiltonian formalism.
As constraints, 
they can be either used to derive the Dirac brackets 
before quantization (as in ref.~\cite{Llosa94})
or imposed on physical states after quantization, 
as we do in this work.
In either case,
the Hamiltonian no longer plays the role of 
determining the time evolution of the system,
as discussed in section~\ref{sec:H}.

\end{itemize}

\section{Hamiltonian Formalism for 2D Toy Model}
\label{sec:stringy}

We are now ready to apply the lessons learned from the 1D nonlocal theory discussed in section~\ref{sec:toy_Ham} to the 2D toy model~\eqref{S_2d_phi_tilde} and examine the physical implications.
We leave further generalizations to string field theories~\eqref{SFT_action2}, including infinitely many fields and higher-order interactions for future works.

\subsection{From 1D to 2D}
\label{sec:toy_to_stringy}

As made clear in section~\ref{sec:toy} through the correspondence~\eqref{corres_a_tilde}--\eqref{corres_B_b}, the 2D toy model~\eqref{S_2d_phi_tilde} consists of an infinite set of independent Fourier modes $\tilde{a}_{\Omega}(V)$ and $\tilde{a}^{\dagger}_{\Omega}(V)$ labelled by the outgoing light-cone frequency $\Omega$, each of which can be seen as 
dynamical fields in the 1D theory~\eqref{S_a_tilde} 
characterized by the nonlocal length scale $\sigma_{\Omega} \equiv 4 \ell_E^2 \, \Omega$.
As a result, making use of eqs.~\eqref{corres_a_tilde}--\eqref{corres_B_b}, the defining formulas that we have obtained in the Hamiltonian formalism for the 1D model can be directly transferred to apply to the 2D model.

For instance, 
according to eqs.~\eqref{D} and~\eqref{Dn},
the correlation function for the action~\eqref{S_2d_phi_tilde}
has the perturbative expansion
\begin{alignat}{2}
&
\bigl\langle 
\tilde{a}_{\Omega}(V) \, \tilde{a}_{\Omega'}^{\dagger} (V')
\bigr\rangle
&&
\nn \\
&
=
\Biggl\{ 
\Theta( V - V' -\sigma_{\Omega} )
+ 
\sum_{n = 1}^{\infty} 
\left( \frac{i \lambda}{\Omega} \right)^n 
\bigintssss_{-\infty}^{\infty} 
\Biggl[
&&
\prod_{j = 1}^n \, 
d V_j \, B(V_j)
\Biggr]
\Theta(V - V_n - \sigma) \, 
\Theta(V_1 - V' - \sigma)
\nn \\ 
& &&
\times 
\Theta(V_n - V_{n - 1} - \sigma) 
\cdots 
\Theta(V_2 - V_1 - \sigma) 
\Biggr\} \,
\delta(\Omega - \Omega')
\, .
\label{D_stringy}
\end{alignat}
Similar to the 1D model, due to the vanishing of the step functions for $\sigma_{\Omega} \gtrsim V - V'$ in the above expression, the corrections to the correlation function induced by the interaction with the background field $B(V)$ are largely suppressed for ultra-high frequency modes.

With 
$\bigl\langle 
\tilde{a}_{\Omega}(V) \, \tilde{a}_{\Omega'}^{\dagger} (V')
\bigr\rangle$ 
at hand,
the two-point correlation function of $\tilde{\phi}$
can be inferred based on the Fourier decomposition~\eqref{phi_tilde_Fourier} as
\be 
\begin{aligned}
\ev{\tilde{\phi}(U, V) \, \tilde{\phi}(U' , V')}
=
\int_0^{\infty} \int_0^{\infty}
\frac{d \Omega \, d \Omega'}{4 \pi \sqrt{\Omega \, \Omega'}} \, 
\Bigl[
&
e^{- i \Omega ( U - U' )} \, 
\bigl\langle 
\tilde{a}_{\Omega}(V) \, \tilde{a}_{\Omega'}^{\dagger} (V')
\bigr\rangle
\\
&+
e^{i \Omega ( U - U' )} \, 
\bigl\langle 
\tilde{a}_{\Omega'}(V') \, \tilde{a}_{\Omega}^{\dagger} (V)
\bigr\rangle
\Bigr]
\, .
\end{aligned}
\ee 
In the free theory, this results in 
\be 
\begin{aligned}
\label{G0_stringy}
\ev{\tilde{\phi}(U, V) \, \tilde{\phi}(U' , V')}_0
=
\int_0^{\abs{V - V'} / 4 \ell_E^2}
\frac{d \Omega}{4 \pi \Omega} \, 
\Bigl[
&
\Theta(V - V') \, 
e^{- i \Omega ( U - U' )}
\\ 
&
+
\Theta(V' - V) \, 
e^{i \Omega ( U - U' )}
\Bigr]
\, .
\end{aligned}
\ee 
Moreover, 
by defining the vacuum state $\ket{0}$ 
using the zeroth-order lowering operator $\hat{\tilde{a}}_{\Omega,\, 0} (V)$ as 
\be 
\label{vac_2D}
\hat{\tilde{a}}_{\Omega, \, 0} (V) \ket{0}
=
0
\qquad 
\forall \ 
\Omega > 0 
\ \text{and} \ 
V
\, ,
\ee 
it follows from the zeroth-order commutator~\eqref{comm0} in the 1D model that 
\be 
\label{comm0_stringy}
\bigl[
\hat{\tilde{a}}_{\Omega}(V) \, ,
\hat{\tilde{a}}_{\Omega'}^{\dagger}(V')
\bigr]_0
=
\delta(\Omega - \Omega') \, 
\Theta( |V - V'| - \sigma_{\Omega})
\ee 
in the operator formalism for the 2D toy model. 
We have thus reproduced both the propagator and the operator algebra proposed in ref.~\cite{Ho:2023tdq}, albeit this time providing the theoretical basis in support of the proposal.

Following the scheme of operator perturbation theory 
laid out in section~\ref{sec:2pt_function},
the Heisenberg evolution of the ladder operators 
\begin{align}
\hat{\tilde{a}}_{\Omega}(V)
&=
\hat{\tilde{a}}_{\Omega, \, 0}(V)
+
\hat{\tilde{a}}_{\Omega, \, 1}(V)
+
\mathcal{O}(\lambda^2)
\, , 
\\
\hat{\tilde{a}}_{\Omega}^{\dagger}(V)
&=
\hat{\tilde{a}}_{\Omega, \, 0}^{\dagger}(V)
+
\hat{\tilde{a}}_{\Omega, \, 1}^{\dagger}(V)
+
\mathcal{O}(\lambda^2)
\end{align}
are determined perturbatively in $\lambda$
through the Schwinger-Dyson equations (cf. eqs.~\eqref{SD1}--\eqref{SD2}):
\begin{align}
\del_V
\ev{\mathcal{T} \, 
\bigl\{ \hat{\tilde{a}}_{\Omega}(V) \, 
\hat{\tilde{a}}_{\Omega'}^{\dagger}(V') \bigr\}}{0} 
&=
\delta( \Omega - \Omega') \, 
\delta(V - V' - \sigma_{\Omega})
\nn \\
& \quad \,
+
\frac{i \lambda}{\Omega} \, 
\delta(\Omega - \Omega') \, 
B(V - \sigma_{\Omega})
\ev{\mathcal{T} \, 
\bigl\{ 
\tilde{a}_{\Omega}(V - \sigma_{\Omega}) \, 
\tilde{a}_{\Omega'}^{\dagger}(V') \bigr\}}{0} 
\, ,
\\ 
\del_{V'}
\ev{\mathcal{T} \, 
\bigl\{ \hat{\tilde{a}}_{\Omega}(V) \, 
\hat{\tilde{a}}_{\Omega'}^{\dagger}(V') \bigr\}}{0} 
&=
- \, \delta( \Omega - \Omega') \, 
\delta(V - V' - \sigma_{\Omega})
\nn \\
& \quad \,
-
\frac{i \lambda}{\Omega} \, 
\delta( \Omega - \Omega') \, 
B(V' + \sigma_{\Omega}) 
\ev{\mathcal{T} \, 
\bigl\{ 
\tilde{a}_{\Omega}(V) \, 
\tilde{a}_{\Omega'}^{\dagger}(V' + \sigma_{\Omega}) \bigr\}}{0} 
\, ,
\end{align}
which are necessary and sufficient conditions 
for the vacuum expectation value of the time-ordered product 
to maintain consistency with the known path-integral correlation functions,
i.e.,
\be
\ev{\mathcal{T} \, 
\bigl\{ \hat{\tilde{a}}_{\Omega}(V) \, 
\hat{\tilde{a}}_{\Omega}^{\dagger}(V') \bigr\}}{0} 
=
\bigl\langle 
\tilde{a}_{\Omega}(V) \, 
\tilde{a}_{\Omega}^{\dagger}(V')
\bigr\rangle
\, .
\ee 
Specifically, 
by establishing this correspondence, 
the leading-order correction to the operators 
can be inferred from eqs.~\eqref{an} and~\eqref{an_dag} as
\begin{align}
\hat{\tilde{a}}_{\Omega, \, 1} (V)
&=
\frac{i \lambda}{\Omega}
\int_{-\infty}^{V \, - \, \sigma_{\Omega}} dV'' \, 
B(V'') \, \hat{\tilde{a}}_{\Omega, \, 0} (V'')
\, ,  
\\
\hat{\tilde{a}}_{\Omega, \, 1}^{\dagger} (V)
&=
- \frac{i \lambda}{\Omega}
\int_{-\infty}^{V \, - \, \sigma_{\Omega}} dV'' \, 
B(V'') \, \hat{\tilde{a}}_{\Omega, \, 0}^{\dagger} (V'')
\, ,
\end{align}
and the commutator algebra is deformed up to $\mathcal{O}(\lambda)$ as 
\be 
\label{comm1_stringy}
\begin{aligned}
\bigl[
\hat{\tilde{a}}_{\Omega}(V) \, ,
\hat{\tilde{a}}_{\Omega'}^{\dagger} (V')
\bigr]
=
\delta(\Omega - \Omega') \, 
\Biggl[
&
\Theta( |V - V'| - \sigma_{\Omega})
\\
&+
\frac{i \lambda}{\Omega} \, 
\Theta( |V - V'| - 2 \sigma_{\Omega}) 
\int_{V' \, + \, \sign(V - V') \, \sigma_{\Omega}}^{V \, - \, \sign(V - V') \, \sigma_{\Omega}}
B(V'') \, dV''
\Biggr]
\, ,
\end{aligned}
\ee 
in which the correction term is related to eq.~\eqref{comm1} 
via the mapping~\eqref{corres_a_tilde}--\eqref{corres_B_b}. 

With regard to the quantum states in the 2D toy model, the Fock space is spanned by acting on the vacuum state $\ket{0}$~\eqref{vac_2D} 
with all possible combinations of creation operators of the form 
\be 
\hat{\tilde{\mathcal{A}}}_{\Omega}^{\dagger} [\Psi_{\Omega}]
\equiv 
\lim_{N \to \infty} \frac{1}{2N \sigma_{\Omega}}
\int_{-N \sigma_{\Omega}}^{N \sigma_{\Omega}} 
d V \, 
\Psi_{\Omega} (V) \, 
\hat{\tilde{a}}_{\Omega, \, 0}^{\dagger} (V)
\, ,
\ee 
analogous to eq.~\eqref{n_particle_state}. 
On the other hand, as discussed in section~\ref{sec:phys},
in this formalism the field equation for $\tilde{\phi}$
should be imposed as a physical-state constraint on the Hilbert space.
More explicitly, 
let us make the decomposition 
$\hat{\tilde{\phi}} = \hat{\tilde{\phi}}^{(+)} + \hat{\tilde{\phi}}^{(-)}$,
where 
\begin{align}
\hat{\tilde{\phi}}^{(+)} (U, V) 
&\equiv
\int_0^{\infty} 
\frac{d \Omega}{\sqrt{4 \pi \Omega}} \, 
\hat{\tilde{a}}_{\Omega}(V) \, e^{-i \Omega U}
\, ,
\\
\hat{\tilde{\phi}}^{(-)} (U, V) 
&\equiv
\int_0^{\infty} 
\frac{d \Omega}{\sqrt{4 \pi \Omega}} \, 
\hat{\tilde{a}}_{\Omega}^{\dagger}(V) \, e^{i \Omega U}
\, .
\end{align}
Then for any Heisenberg state $\ket{\Psi}$ 
in the physical Hilbert space,
it satisfies the constraint 
\be 
\label{eom_stringy}
\left[ 
\Box \, e^{- i \, \ell_E^2 \, \Box} \, 
\hat{\tilde{\phi}}^{(+)} (U, V) 
+
4 \lambda \, B(V) \, 
\hat{\tilde{\phi}}^{(+)} (U, V) 
\right]
\ket{\Psi}
=
0
\, .
\ee 
Furthermore,
the corresponding dual $\bra{\Psi}$ of the physical state is defined by 
the conjugate constraint
\be 
\label{eom_conj_stringy}
\bra{\Psi}
\left[ 
\Box \, e^{- i \, \ell_E^2 \, \Box} \, 
\hat{\tilde{\phi}}^{(-)} (U, V) 
+
4 \lambda \, B(V) \, 
\hat{\tilde{\phi}}^{(-)} (U, V) 
\right]
=
0
\, .
\ee 
These constraints suffice to guarantee that the equation-of-motion operator 
has vanishing matrix elements between the physical states:
\be
\ev{
\left[ 
\Box \, e^{- i \, \ell_E^2 \, \Box} \, 
\hat{\tilde{\phi}}
+
4 \lambda \, B(V) \, 
\hat{\tilde{\phi}}
\right]
}{\Psi}
=
0
\, ,
\ee 
thus ensuring that the quantum theory has the correct low-energy limit.

As further illustrated in section~\ref{sec:phys}, the equation-of-motion constraints~\eqref{eom_stringy} and~\eqref{eom_conj_stringy} translate into conditions on the temporal profiles of the wavefunction 
\be 
\Psi_{\Omega} (V)
=
\psi_{\Omega, \, 0} (V)
+
\lim_{N \to \infty}
\left( 2 N \sigma_{\Omega} \right)
\sum_{n = 1}^{\infty}
\psi_{\Omega, \, n} (V)
\ee 
and its dual
\be 
\Psi_{\Omega}^{c} (V)
=
\psi_{\Omega, \, 0}^{\ast} (V)
+
\lim_{N \to \infty}
\left( 2 N \sigma_{\Omega} \right)
\sum_{n = 1}^{\infty}
\psi_{\Omega, \, n}^c (V)
\ee 
associated with physical states.
For example, according to eqs.~\eqref{Psi_xi=1} and~\eqref{Psi^c_full} derived in the 1D model, a physical one-particle state $\ket{1_{\Omega}}$ in the 2D toy model and its counterpart $\bra{1_{\Omega}}$ in the dual space has the respective form
\begin{align}
\ket{1_{\Omega}}
&=
\lim_{N \to \infty}
\int_{-N \sigma}^{N \sigma} 
d V \, 
\frac{\psi_{\Omega, \, 0} (V)}{2N \sigma_{\Omega}} \, 
\hat{\tilde{a}}_{\Omega, \, 0}^{\dagger} (V)
\ket{0}
\, , 
\\
\bra{1_{\Omega}}
&=
\lim_{N \to \infty}
\int_{-N \sigma}^{N \sigma} 
d V \, 
\biggl[ 
\frac{\psi_{\Omega, \, 0}^{\ast} (V)}{2N \sigma_{\Omega}}
+
\frac{i \lambda}{\Omega} \, 
\overbar{\psi}_{\Omega, \, 0}^{\ast} \, 
B(V)
+
\mathcal{O}(\lambda^2)
\biggr]
\bra{0}
\hat{\tilde{a}}_{\Omega, \, 0} (V)
\, ,
\end{align}
where the zeroth-order wavefunction $\psi_{\Omega, 0} (V)$
is periodic in time with period $2 \sigma_{\Omega}$
as shown in eq.~\eqref{periodic},
and 
\be 
\label{zero_mode_stringy}
\overbar{\psi}_{\Omega, \, 0}
\equiv 
\frac{1}{2 \sigma_{\Omega}} 
\int_{-\sigma_{\Omega}}^{\sigma_{\Omega}} 
\psi_{\Omega, \, 0}(V) \, dV
=
\frac{1}{2 N \sigma_{\Omega}} 
\int_{-N \sigma_{\Omega}}^{N \sigma_{\Omega}} 
\psi_{\Omega, \, 0}(V) \, dV
\ee 
is its average over a cycle. 

As illustrated in section~\ref{sec:norm}, 
the physical Hilbert space constructed in this way is free of negative-norm states.
Secondly, 
the only physical degree of freedom 
in the wavefunction $\Psi_{\Omega} (V) = \psi_{\Omega, \, 0}(V)$ 
characterizing single-frequency physical states
is its zero mode $\overbar{\psi}_{\Omega, \, 0}$~\eqref{zero_mode_stringy}.
In fact, it follows from the analysis in section~\ref{sec:spurious} that 
physical wavefunctions $\Psi_{\Omega} (V)$ differing by the addition of any 
complex periodic function that averages to zero over a cycle $2 \sigma_{\Omega}$ 
are physically equivalent.
In other words, 
the physical Hilbert space defined by 
the equation-of-motion constraints~\eqref{eom_stringy} and~\eqref{eom_conj_stringy} 
exhibits the following equivalence relation within each subspace of fixed $\Omega$:
\be 
\prod_{i = 1}^n
\hat{\tilde{\mathcal{A}}}_{\Omega}^{\dagger} [\Psi_{\Omega}^{(i)}]
\ket{0}
\sim 
\prod_{i = 1}^n
\hat{\tilde{\mathcal{A}}}_{\Omega}^{\dagger} [ \, \overbar{\Psi}_{\Omega}^{(i)} ]
\ket{0}
\, .
\ee 
This is established in section~\ref{sec:spurious} by demonstrating that the mode functions $e^{i \pi n V / \sigma_{\Omega}}$ ($n \in \mathbb{Z} \setminus \{ 0 \}$) are associated with \textit{spurious} zero-norm states that decouple from physical observables to all orders in the interacting theory.
Therefore, we conclude that the physical state space of the 2D toy model is given by 
\be
\mathrm{span}
\left\{
\Pi_{i = 1}^n \, 
\hat{\tilde{\mathfrak{a}}}_{\Omega_i}^{\dagger}
\ket{0}
\right\}
\, ,
\qquad 
\text{where}
\quad 
\hat{\tilde{\mathfrak{a}}}_{\Omega}^{\dagger} 
\equiv 
\lim_{N\rightarrow\infty} \frac{1}{2N \sigma_{\Omega}}
\int_{-N \sigma_{\Omega}}^{N \sigma_{\Omega}} dV \, 
\hat{\tilde{a}}_{\Omega, \, 0}^{\dagger}(V)
\, .
\ee

\subsection{Comments on analytic continuation}
\label{sec:continuation}

Notice that eqs.~\eqref{G0_stringy} and~\eqref{comm0_stringy} 
differ from their counterparts in the ordinary low-energy theory
merely by a time-dependent UV cutoff 
\be 
\label{uv_ir}
\Omega 
\, \leq \, 
\frac{\abs{V - V'}}{4 \ell_E^2}
\ee  
that arises from the vanishing of the step function 
$\Theta( \abs{V - V'} - \sigma_{\Omega})$.
As pointed out in ref.~\cite{Ho:2023tdq}, this reflects a UV/IR relation in the 2D toy model, where correlators are nonvanishing only when the separation in time $V$ extends greater than $\sigma_{\Omega} \equiv 4 \ell_E^2 \, \Omega$, which becomes a macroscopic time scale for large $\Omega$. 
This property very likely extends to string field theories.
In the interacting theory, we have also seen that the factor 
$\Omega^{-n} \, 
\Theta \bigl( V - V' - \left( n + 1 \right) \sigma_{\Omega} \bigr)$
in the $\mathcal{O}(\lambda^n)$ correction~\eqref{D_stringy} 
to the correlation function suppresses the effects of UV interactions. 

As mentioned at the end of section~\ref{sec:toy}, 
all results obtained from the complexified string length parameter 
$\ell^2 = i \ell_E^2$~\eqref{ell_to_ellE (sec2)}
can be reinterpreted in terms of the Euclidean time $V_E = i V$ 
with the original \textit{real} string length parameter $\ell \in \mathbb{R}$. 
Therefore, the UV/IR relation~\eqref{uv_ir} also implies that 
a Wick-rotated background field with a characteristic Euclidean time scale $\Delta V_E$ 
is invisible to very high-frequency modes 
whose frequencies exceed the bound
\be
\label{uv_ir_E}
\Omega \leq \frac{\Delta V_E}{4\ell^2} 
\, .
\ee

That said, at this stage, 
the UV suppression is only demonstrated 
for purely imaginary string length squared $\ell^2 = i \ell_E^2$ ($\ell_E^2 > 0$) 
or purely imaginary light-cone time $V = - i V_E$ ($V_E \in \mathbb{R}$). 
How it translates to UV suppression 
in real physical applications is still unclear. 
In section~\ref{sec:THR}, 
we consider Hawking radiation as an example where 
the UV/IR connection yields a UV suppression effect in real spacetime.

It is important to note that 
the ubiquitous nonlocal factor $\mathrm{exp}(\ell^2 k^2)$ in string field theories 
makes it impossible to perform general calculations 
in either the path-integral or Hamiltonian formalism 
without resorting to analytic continuation. 
Even in standard calculations carried out in Euclidean space, 
where the Minkowski time $t$ and energy $k^0$ are Wick-rotated, 
whether the theory maintains genuine UV suppression 
after analytically continuing back to Lorentzian signature 
remains an open question.

On the other hand, 
it is a well-established feature of string theory 
that interactions are universally suppressed in the UV limit. 
It is therefore reasonable to assume that 
physical observables in string field theories~\eqref{SFT_action} 
should \textit{always} reflect 
the suppression of interactions in the UV regime. 
Under this assumption, the step function $\Theta(\abs{V - V'} - 4 \ell_E^2 \, \Omega)$ 
in the analytically continued space 
offers an intuitive explanation for the UV suppression in the 2D toy model.

In this context, 
eq.~\eqref{uv_ir_E} implies a space-time uncertainty relation~\cite{Ho:2023tdq}
\be
\label{stur_UV}
\Delta U \Delta V \gtrsim \ell^2
\ee
between the light-cone coordinates, 
as the frequency $\Omega$ imposes a bound on the precision of $U$ 
such that $\Delta U \gtrsim 1/\Omega$. 
As shown in appendix~\ref{app:STUR}, 
the light-cone uncertainty relation~\eqref{stur_UV} 
implies the original space-time uncertainty relation 
$\Delta t \, \Delta x \gtrsim \ell^2$
proposed by Yoneya~\cite{Yoneya:1987gb, Yoneya:1989ai, Yoneya:1997gs, Yoneya:2000bt}.

\subsection{Termination of Hawking radiation}
\label{sec:THR}

The origin of Hawking radiation~\cite{Hawking:1975vcx} 
lies in the exponential relationship between 
the proper times of freely falling observers 
and fiducial observers in a black-hole spacetime. 
The corresponding exponential blueshift suggests that 
low-energy Hawking radiation originates from 
ultra-energetic quantum fluctuations 
near the black-hole horizon in the past~\cite{Jacobson:1991gr}.
With the event horizon acting as the surface of infinite blueshift, 
the detection of late-time Hawking radiation by a distant observer
is in fact a probe of 
the short-distance Wightman function~\cite{Fredenhagen:1989kr}.

Perturbative approaches to nonlocality have argued that 
Hawking radiation remains unmodified 
in nonlocal UV theories of the radiation field~\cite{Kajuri:2018myh, Boos:2019vcz}. 
However, this section provides a concise recap of how 
a proper nonperturbative treatment of 
the nonlocality in string field theories 
may indicate significant modifications to 
the amplitude of late-time Hawking radiation.\footnote{
Conversely, the Unruh effect may remain unchanged 
in open string field theory~\cite{Hata:1995di}.
This is because Lorentz symmetry is preserved in Minkowski space, 
but effectively broken by the collapsing matter 
in a dynamic black-hole background. 
The role of collapsing matter in producing 
a trans-Planckian, Lorentz-invariant energy scale 
in the Hawking process was highlighted in ref.~\cite{Ho:2021sbi}.
}
A more comprehensive discussion 
can be found in the original work~\cite{Ho:2023tdq} 
(as well as in a recent review~\cite{Ho:2024tby}).
Here, we revisit the derivation to illustrate 
how the analytic continuation of the string tension parameter $\ell^{-2}$
is carried out in the calculation of a physical quantity.

Let $\hat{\tilde{b}}_{\omega}$ and $\hat{\tilde{b}}^{\dagger}_{\omega}$ 
be the ladder operators associated to an outgoing mode $e^{-i \omega u}$ 
with positive frequency $\omega > 0$, 
defined with respect to the 
Eddington retarded coordinate $u \simeq t - r$ 
at large distances from the black hole.
A Hawking particle with frequency $\omega = \omega_0$ 
detected by a distant observer at retarded time $u = u_0$ 
can be described by a wave packet
\be 
\label{hawking_psi}
\Psi_{\omega_0, u_0}(u)
=
\int_0^{\infty} 
\frac{d \omega}{{\sqrt{4 \pi \omega}}} \, 
\psi_{\omega_0}(\omega) \, 
e^{-i \omega (u - u_0)}
\, ,
\ee 
which is a superposition of purely positive-frequency plane waves 
characterized by a profile function $\psi_{\omega_0}(\omega)$
that is sharply peaked around $\omega = \omega_0$ in the frequency domain.
Here, we focus on large black holes
with Schwarzschild radius $r_s \gg \ell$,
where the emitted Hawking particles
have characteristic frequencies $\omega_0 \sim 1 / r_s \ll \ell^{-1}$
that lie well within the infrared regime.
The annihilation operator $\hat{\tilde{b}}_{\Psi}$ 
corresponding to the wave packet~\eqref{hawking_psi}
can be defined as 
\be 
\label{b_psi}
\hat{\tilde{b}}_{\Psi} (v)
\equiv 
\int_0^{\infty} d\omega \, 
\psi_{\omega_0}(\omega) \, 
e^{-i \omega u_0} \, 
\hat{\tilde{b}}_{\omega} (v)
\, ,
\ee 
where $v \simeq t + r$ is the Eddington advanced time.
Notably, recall that the ladder operators constructed in the light-cone 2D toy model acquire dependence on the advanced time $v$ due to nonlocality.

In the reference frame of a freely falling observer 
near the horizon of the black hole,
the outgoing sector of the field $\hat{\tilde{\phi}}$ 
can be decomposed as in eq.~\eqref{phi_tilde_Fourier}:
\be 
\hat{\tilde{\phi}} (U, V)
= 
\int_0^{\infty} 
\frac{d\Omega}{\sqrt{4\pi\Omega}} 
\left[ 
\hat{\tilde{a}}_{\Omega}(V) \, e^{- i \Omega U} 
+ 
\hat{\tilde{a}}^{\dagger}_{\Omega}(V) \, e^{i \Omega U} 
\right]
, 
\ee 
except here $(U, V)$ represent the Kruskal coordinates,
which are related to $(u, v)$ via 
\begin{align}
U(u)
&=
- 2 \, r_s \, e^{-u / 2 \, r_s}
\, , 
\label{U(u)} 
\\
V(v)
&=
2 \, r_s \, e^{v / 2 \, r_s}
\, .
\label{V(v)}
\end{align}
The key observable characterizing Hawking radiation is 
the expectation value of the number operator
$\hat{\mathcal{N}}_{\Psi}(v, v') 
\equiv 
\hat{\tilde{b}}_{\Psi}^{\dagger} (v) \, \hat{\tilde{b}}_{\Psi} (v')$
in the free-fall vacuum state $\ket{0}_a$ defined by 
\be 
\hat{\tilde{a}}_{\Omega}(V) \ket{0}_a = 0
\qquad 
\forall \ \Omega > 0 
\, .
\ee 
The exponential mapping~\eqref{U(u)}
between the light-cone coordinates $U$ and $u$
gives rise to a nonzero $b$-particle number 
in the $a$-vacuum $\ket{0}_a$~\cite{Hawking:1975vcx}:
\be 
\label{b_particle}
\prescript{}{a}\langle 0 | \, 
\hat{b}_{\omega}^{\dagger} \bigl( v(V) \bigr) \, 
\hat{b}_{\omega'} \bigl( v'(V') \bigr) 
\ket{0}_a
=
\int_{0}^{\infty} d \Omega 
\int_{0}^{\infty} d \Omega' \, 
\beta_{\omega \Omega}^{\ast} \, 
\beta_{\omega'\Omega'} 
\prescript{}{a}\langle 0 | \, 
[ \hat{\tilde{a}}_{\Omega}(V) \, , \hat{\tilde{a}}_{\Omega'}^{\dagger}(V') ]
\ket{0}_a
\, ,
\ee 
where the Bogoliubov coefficients $\beta_{\omega \Omega}$
are given by 
\be 
\beta_{\omega \Omega} 
=
\frac{1}{2 \pi} 
\sqrt{\frac{\omega}{\Omega}} \, 
\int_{-\infty}^{\infty} du \, e^{i \omega u} \, e^{i \Omega U(u)}
=
\frac{r_s}{\pi} 
\sqrt{\frac{\omega}{\Omega}} 
\left( 2 \, r_s \, \Omega \right)^{2 i \, r_s \, \omega}
e^{- \pi r_s \, \omega} \, \Gamma(-2 i \, r_s \, \omega)
\ee 
for $\omega, \Omega > 0$,
and $v(V)$ denotes the inverse of~\eqref{V(v)}.
Note that $\beta_{\omega \Omega}$ is completely determined by
the coordinate transformation $U(u)$~\eqref{U(u)},
and thus its expression remains unchanged 
regardless of whether we work with 
the field $\phi$ or $\tilde{\phi} \equiv e^{\ell^2 \Box / 2} \, \phi$.
However, as emphasized at the beginning of section~\ref{sec:toy}, since physical measurements rely on interactions, it is the correlation functions of $\tilde{\phi}$ that is directly probed in string field theories.

It can be justified using eq.~\eqref{D_stringy} that the effects of the curved background, including the collapsing matter, on vacuum fluctuations in a freely falling frame are highly suppressed if the outgoing mode frequency $\Omega$ is much greater than the characteristic energy scale $1 / r_s$ of the black hole.
This allows us to employ the commutator 
\be 
[\hat{\tilde{a}}_{\Omega}(V) \, , \hat{\tilde{a}}_{\Omega'}^{\dagger}(V')]
\simeq 
\delta(\Omega - \Omega') \, 
\Theta( |V - V'| - 4 \ell_E^2 \Omega )
\ee 
as a good approximation when discussing Hawking radiation 
at sufficiently late times.
Inserting this into eq.~\eqref{b_particle}
leads to the number expectation value~\cite{Ho:2023tdq}
\begin{align}
&
\prescript{}{a}\langle 0 | \, 
\hat{\mathcal{N}}_{\Psi} (V, V')
\ket{0}_a
\nn \\
&\simeq 
\int_0^{\infty} d \omega 
\int_0^{\infty} d \omega' \, 
\psi_{\omega_0}^{\ast}(\omega) \, 
\psi_{\omega_0}(\omega') \, 
e^{i \left( \omega - \omega' \right) \, u_0} 
\prescript{}{a}\langle 0 | \, 
\hat{b}_{\omega}^{\dagger} \bigl( v(V) \bigr) \, 
\hat{b}_{\omega'} \bigl( v'(V') \bigr) 
\ket{0}_a
\nn \\
&=
\frac{r_s / \pi}{e^{4 \pi r_s \, \omega_0} - 1}
\int_0^{\infty} d \omega 
\int_0^{\infty} d \omega' \, 
\psi_{\omega_0}^{\ast}(\omega) \, 
\psi_{\omega_0}(\omega') \, 
e^{i \left( \omega - \omega' \right) \, u_0} 
\int_0^{\abs{V - V'} / 4 \ell_E^2} 
\frac{d \Omega}{\Omega} 
\left( 2 r_s \, \Omega \right)^{-2 i \, r_s \left( \omega - \omega' \right)}
\, .
\label{N_vev}
\end{align}
With a change of variable 
$\Omega \mapsto u(\Omega) = 2 r_s \ln \left( 2 r_s \, \Omega \right)$,
the $\Omega$-integral can be further written as 
\be 
\label{Omega_to_u}
\int_0^{\abs{V - V'} / 4 \ell_E^2} 
\frac{d \Omega}{\Omega} 
\left( 2 r_s \, \Omega \right)^{-2 i r_s \left( \omega - \omega' \right)}
=
\int_{-\infty}^{u_{\Lambda} (\abs{V - V'} \, , \, \ell_E^2)}
\frac{du}{2 r_s} \,
e^{-i \left( \omega - \omega' \right) u}
\, ,
\ee 
where 
\be 
u_{\Lambda} (|V - V'| \, , \ell_E^2)
\equiv 
2 r_s
\ln 
\left( \frac{r_s \abs{V - V'}}{2 \ell_E^2} \right)
.
\ee 
Notice that, 
after performing the analytic continuation $\ell_E^2 \to - i \ell^2$ 
back from the complexified string length, 
$u_{\Lambda}$ as a function of $\ell_E^2$ gets turned into 
\be 
u_{\Lambda} (|V - V'| \, , \ell_E^2)
\quad \rightarrow \quad 
u_{\Lambda} (|V - V'| \, , \ell^2)
+
i \pi r_s
\, ,
\ee 
and thus eq.~\eqref{N_vev} becomes 
\be 
\label{N_vev2}
\prescript{}{a}\langle 0 | \, 
\hat{\mathcal{N}}_{\Psi} (V, V')
\ket{0}_a
\simeq 
\frac{2 \omega_0}{e^{4 \pi r_s \, \omega_0} - 1}
\int_{-\infty}^{u_{\Lambda} (\abs{V - V'} \, , \, \ell^2)} 
\abs{\Psi_{\omega_0 , u_0} (u)}^2 \, 
du 
\, ,
\ee 
where the approximation 
$e^{\pi r_s \left( \omega - \omega' \right)} 
\simeq 
e^{\pi r_s \left( \omega_0 - \omega_0 \right)}
= 1$
has been made.
This simplification is valid based on the assumption that 
the profile function $\psi_{\omega_0}(\omega)$ 
has a narrow width $\Delta \omega \ll \omega_0 \sim 1 / r_s$ 
in order for the wave packet $\Psi_{\omega_0 , u_0} (u)$
in eq.~\eqref{hawking_psi} 
to have a well-defined frequency $\omega_0$.
We see that the analytic continuation $\ell_E^2 \to -i \ell^2$ 
has a negligible impact on the outcome, 
and thus the UV/IR connection~\eqref{uv_ir} established for real $\ell_E^2$ 
remains valid for real $\ell^2$ in this scenario.

In the ordinary low-energy theory where $\ell = 0$, the upper bound of the integral in eq.~\eqref{N_vev2} goes to infinity, yielding a Planck distribution 
at the Hawking temperature $1 / 4 \pi r_s$ with a time-independent amplitude.
In the 2D toy model, the finite upper bound $u_{\Lambda}$ stems from the UV cutoff $\Omega \leq \abs{V - V'} / 4 \ell^2$, as shown in eq.~\eqref{Omega_to_u}. 
This cutoff causes the amplitude of Hawking radiation 
to be dependent on the measuring time $u_0$,
reflecting a time-dependent characteristic of Hawking radiation 
that has been observed in previous studies of 
Hawking radiation in UV theories~\cite{Jacobson:1993hn, Barcelo:2008qe, Barman:2017vqx, Ho:2022gpg, Akhmedov:2023gqf, Chau:2023zxb}.

In particular, 
since the wave function $\Psi_{\omega_0 , u_0} (u)$~\eqref{hawking_psi} 
of a Hawking particle is localized 
around the retarded time $u = u_0$ with a width of $\mathcal{O}(r_s)$,
eq.~\eqref{N_vev2} indicates that 
for a finite detection duration $T_V$, 
the mean number of Hawking particles detected 
would essentially vanish for sufficiently large values of $u_0$ when 
\be 
u_0 
\, \gg \, 
u_{\Lambda} (T_V , \ell^2)
+
\mathcal{O}(r_s)
\, \geq \, 
u_{\Lambda} (|V - V'| \, , \ell^2)
+
\mathcal{O}(r_s)
\, .
\ee 
Furthermore, 
as argued in ref.~\cite{Ho:2023tdq},
late-time Hawking particles that require detection ranges
\be 
T_V
\, \gg \, 
\frac{\ell^2}{r_s} \, e^{u_0 / 2 r_s} 
\, \gg \, 
\mathcal{O}(r_s)
\ee 
vastly exceeding the size $r_s$ of the black hole would not be produced by their stringy ancestors in the first place. 
It was therefore concluded~\cite{Ho:2023tdq} that Hawking radiation could potentially terminate around the scrambling time $u_0 \sim \mathcal{O} ( r_s \ln(r_s / \ell) )$~\cite{Sekino:2008he} within a framework of string theory where the space-time uncertainty relation~\eqref{stur_UV} is manifest. 

\section{Summary and Outlook}
\label{sec:summary}

In this work, we take initial steps toward developing the long-sought Hamiltonian formalism for a class of nonlocal field theories relevant to string field theories~\eqref{SFT_action}. 
The primary challenge lies in addressing the nonlocality in time 
introduced by the exponential factor $e^{\ell^2 \Box / 2}$ 
appearing in all interaction vertices. 
To account for the effects of this nonlocal operator 
in a fully nonperturbative manner 
(i.e., without expanding it in powers of the derivatives $\partial_{\mu}$), 
some form of analytic continuation is necessary. 
To this end, we propose complexifying the string length parameter $\ell$ and working with $\ell^2 = i \ell_E^2$ ($\ell_E^2 > 0$). 
As demonstrated explicitly in this study for a specific toy model~\eqref{S_phi}, this prescription enables the construction of a Hamiltonian formalism.
The Hamiltonian formalism is formulated in the light-cone frame $(U, V)$ of Minskowski space,
where the nonlocality manifests as a finite shift in the advanced time coordinate $V$. 
Equivalently, this formalism can be reinterpreted as one in which the advanced time $V_E = i V \in \mathbb{R}$ is Euclidean, while the string length $\ell$ remains real.

Our Hamiltonian formalism is constructed from the information about 
the correlation functions derived in the path-integral formalism. 
As is common in proposals~\cite{Llosa94, Woodard:2000bt} for Hamiltonian formalisms for theories containing infinite time derivatives, the field operators in the Heisenberg picture do not satisfy the equations of motion.
Instead, the equations of motion are imposed as physical-state constraints on the Fock space.

Typically, 
infinite-derivative theories such as~\eqref{S_phi} possess 
a much larger phase space than their local counterparts, 
leading to the existence of negative-norm states that violate unitarity 
or Hamiltonians that are unbounded from below~\cite{Eliezer:1989cr, Woodard:2015zca}. 
However, in our case, the physical-state constraints 
eliminate all negative-norm states from the Fock space. 
Furthermore, these constraints render the zero-norm states spurious, 
decoupling them entirely from the positive-norm physical Hilbert space. 
Consequently, the remaining physical degrees of freedom 
align precisely with those of standard local quantum field theory. 
As a result, the pathological issues commonly associated with the quantization of time-nonlocal theories are avoided.

The decoupling of zero-norm states is perhaps the most remarkable and critical feature of our Hamiltonian formalism for the nonlocal theory~\eqref{S_phi}.
Without this decoupling, zero-norm states would typically lead to violations of unitarity. 
While nonlocal theories of this type have been considered previously~\cite{Barci:1995ad, Gomis:2000gy, Woodard:2000bt, Gomis:2003xv, Calcagni:2007ef, Kolar:2020ezu}, this feature has not been obtained before.
Our approach provides a starting point for extending the Hamiltonian formalism from the toy model~\eqref{S_phi} to the more general action~\eqref{SFT_action} involving multiple nonlocal fields and higher-point interaction vertices. 
An important open question is whether the absence of instabilities persists in this broader setting.

A key contribution of this work is the introduction of a new paradigm for quantizing nonlocal theories.
At the moment, the applicability of our approach is limited to nonlocality of the exponential type $\exp \left[ \alpha' \mathcal{O}(1) \, \del_{\mu} \, \del^{\mu} \right]$.
However, infinite-derivative structures may appear in other forms in the effective action for the component fields in string field theory (e.g., as derivative couplings), and their implications for the well-posedness of our Hamiltonian framework are not addressed in the current work.
More broadly, different types of nonlocal theories exhibit widely varying characteristics, and it may not always be possible to interpret them coherently within a single framework.\footnote{
Additional examples of string-inspired nonlocal theories that have been shown to be unitary and possess a spectrum bounded from below are discussed in ref.~\cite{Cheng:2008qz}.
}

Last but not least, our formalism provides a quantitative understanding of 
the spacetime nonlocality in string field theories at the quantum level.
In particular, we illustrated that the nonlocality is characterized by a stringy uncertainty relation 
$\Delta U \Delta V \gtrsim \ell^2$ between the light-cone coordinates, offering an explicit realization of the \textit{space-time uncertainty principle} proposed by Yoneya~\cite{Yoneya:1987gb, Yoneya:1989ai, Yoneya:1997gs, Yoneya:2000bt}.
This relation can be interpreted as the physical principle behind the strong suppression of background effects on high-energy quantum modes, which are highly nonlocal in time $V$.
As was pointed out in section~\ref{sec:THR}, the suppression of UV interactions with the background implies the shutdown of Hawking radiation after the scrambling time of a black hole --- a result initially suggested in ref.~\cite{Ho:2023tdq} based on an incomplete Hamiltonian formalism for the theory.
It would also be interesting to explore how this formalism might further illuminate 
nonperturbative aspects of string theory in cosmological backgrounds, 
especially the implications of the space-time uncertainty relation 
$\Delta t \, \Delta x \gtrsim \ell^2$ in the early universe~\cite{Brandenberger:2002nq, Brandenberger:2024vgt}.

\section*{Acknowledgement}

We thank Chong-Sun Chu, Sumit Das, Yosuke Imamura, Takaaki Ishii, 
Puttarak Jai-akson, Hikaru Kawai, Yoichi Kazama, 
Christy Kelly, Bum-Hoon Lee, Ryo Namba, 
Toshifumi Noumi, Nobuyoshi Ohta, Tsukasa Tada, and Yuki Yokokura 
for valuable discussions. 
P.M.H. and W.H.S. are supported in part 
by the Ministry of Science and Technology, R.O.C.
(MOST 110-2112-M-002-016-MY3),
and by National Taiwan University.

\appendix

\section{Analytic Continuation of the String Tension}
\label{app:AC}

In string theory, 
scattering amplitudes are known to be analytic functions of the external momenta, 
in a manner consistent with unitarity~\cite{DHoker:1993hvl, DHoker:1993vpp}. 
Specifically, for a given set of external momenta $p_i$,
the amplitudes are analytic in the combinations $\alpha' p_i \cdot p_j$,
where $\alpha' = \ell^2$ is the inverse of the string tension.
For example, 
four-point amplitudes $\mathcal{A}_4 (\alpha' s, \alpha' t, \alpha' u)$ are analytic in the Mandelstam variables $s, t, u$ (with only two of them independent due to energy-momentum conservation), as well as the parameter $\alpha' = \ell^2$. 
Since any consistent string field theory must reproduce 
the perturbative worldsheet amplitudes~\cite{Sen:2024nfd}, 
its scattering amplitudes must likewise be analytic in $\ell^2$.

This leads to the follow-up question: 
is the analytic continuation $\ell^2 \rightarrow i \ell_E^2$~\eqref{ell_to_ellE (sec2)}
employed in this work physical?
In particular, 
is it equivalent to the conventional analytic continuation (Wick rotation) procedure 
$k^{\mu} \rightarrow k_E^{\mu}$
that preserves unitarity and causality?
For the calculation of Feynman diagrams 
in a perturbation theory with local interactions,
consistency with unitarity (or causality) is maintained by simply applying
the standard $i \epsilon$-prescription to the internal propagators.
In the case of the stringy model~\eqref{SFT_action2}, 
which involves only local interactions among the fields $\tilde{\phi}_j$,
we shall therefore also examine just the analytic continuation of its free propagator~\eqref{prop_p}.

Let us begin by recalling Feynman’s $i\epsilon$-prescription,
which specifies that the propagator in ordinary relativistic quantum field theory 
takes the form
\be
\frac{1}{k^2 + m^2 - i\epsilon} 
= 
i \int_0^{\infty} d\tau \, e^{-i\tau \, (k^2 \, + \, m^2 \, - \, i \epsilon)}
\, ,
\label{prop0_L (app1)}
\ee
where the Schwinger parameter $\tau$ represents the proper time in Lorentzian signature.
Convergence of the integral is ensured by 
the damping factor $e^{- \epsilon \tau}$ in the integrand. 
The standard analytic continuation of eq.~\eqref{prop0_L (app1)} 
to Euclidean space with Wick-rotated momentum $k^{\mu} \rightarrow k_E^{\mu}$ is given by  
\be
\frac{1}{k^2_E + m^2} 
= 
\int_0^{\infty} d\tau_E \, e^{- \tau_E \, (k_E^2 \, + \, m^2)}
\, ,
\label{prop0_E (app1)}
\ee
where $\tau_E = i\tau$ is the Euclidean (imaginary) Schwinger proper time.
As in the Lorentzian case,
the contributions from large values of $\tau_E$ are suppressed
as long as $k_E^2 + m^2 > 0$.
This shows that the Wick rotation $k^{\mu} \rightarrow k_E^{\mu}$ in momentum space 
can be equivalently understood as a rotation 
$\tau \rightarrow -i \tau_E$ of the Schwinger parameter.
Since the Schwinger proper time $\tau$ 
can be interpreted as the modular parameter of a particle's worldline, 
the Wick rotation $\tau_E = i\tau$ corresponds to a specific contour 
in the complexified modular parameter space. 
This perspective naturally generalizes to 
the moduli space of string worldsheets in perturbative string theory~\cite{Berera:1992tm, Witten:2013pra}, 
which underlies the analytic continuation~\eqref{ell_to_ellE (sec2)} considered in this work.

By rescaling the Schwinger parameter as $\tau = \ell_E^2 \, \tau'$, 
the Lorentzian Feynman propagator~\eqref{prop0_L (app1)} becomes
\be
\frac{1}{k^2 + m^2 - i\epsilon} 
= 
i \, \ell_E^2 
\int_0^{\infty} d\tau' \, e^{- i \, \ell_E^2 \, \tau'(k^2 \, + \, m^2 \, - \, i \epsilon)}
\label{prop0_L2 (app1)}
\ee
for an arbitrary positive constant $\ell_E^2 > 0$.
It is then clear that besides Wick-rotating $k^{\mu} \rightarrow k_E^{\mu}$ and 
$\tau' \rightarrow -i \tau_E^{\prime}$,
one can equivalently perform the analytic continuations
$k^{\mu} \rightarrow k_E^{\mu}$ and $\ell_E^2 \rightarrow - i \ell^2$,
which result in the Euclidean-space propagator
\be 
\frac{1}{k^2_E + m^2} 
= 
\ell^2 
\int_0^{\infty} d\tau' \, 
e^{- \ell^2 \tau'(k_E^2 \, + \, m^2)}
\, .
\label{prop0_E2 (app1)}
\ee 
Note that this matches the earlier Euclidean expression~\eqref{prop0_E (app1)}  
after the change of variable $\tau_E = \ell^2 \tau'$.\footnote{
For readers that are not expecting a real $\ell^2$ 
to appear in the Euclidean propagator~\eqref{prop0_E2 (app1)}, 
recall that the string worldsheet action (including the string tension $\ell^{-2}$) 
is naturally defined on complex manifolds with Euclidean signature.}
Hence, 
standard calculations performed in Euclidean momentum space with $\ell^2 > 0$ 
(as in eq.~\eqref{prop0_E2 (app1)})
can be viewed as analytic continuations of calculations performed in 
Lorentzian momentum space with $\ell_E^2 = - i \ell^2 > 0$
(as in eq.~\eqref{prop0_L2 (app1)}).

The equivalence between the two analytic continuation schemes 
$(k, \tau) \rightarrow (k_E, - i \tau_E)$ and $(k, \ell_E^2) \rightarrow (k_E, - i \ell^2)$ 
remains valid even with the introduction of the exponential factor $\mathrm{exp}(- \ell^2 k^2)$ 
in the propagator~\eqref{prop_p} of the stringy model~\eqref{SFT_action2}.
Analogous to the expressions~\eqref{prop0_L (app1)}--\eqref{prop0_E2 (app1)} 
for an ordinary propagator discussed earlier,
the propagators for a field $\tilde{\phi}_i$ with mass $m_i = m$ 
in the stringy model can be written as follows:
\begin{align}
\frac{e^{- i \, \ell_E^2 \, k^2}}{k^2 + m^2 - i\epsilon} 
&= 
i \, e^{i \, \ell_E^2 \, m^2} 
\int_{\ell_E^2}^{\infty} d\tau \, 
e^{-i\tau \, (k^2 \, + \, m^2 \, - \, i \epsilon)}
= 
i \, \ell_E^2 \, e^{i \, \ell_E^2 \, m^2} 
\int_{1}^{\infty} d\tau' \, 
e^{- i \, \ell_E^2 \tau' (k^2 \, + \, m^2 \, - \, i \epsilon)}
\, ,
\label{prop_L (app1)}
\\
\frac{e^{- \ell^2 k_E^2}}{k^2_E + m^2} 
&= 
e^{\ell^2 m^2}
\int_{\ell^2}^{\infty} d\tau_E \, 
e^{- \tau_E \, (k_E^2 \, + \, m^2)}
= 
\ell^2 \, e^{\ell^2 m^2}
\int_{1}^{\infty} d\tau' \, 
e^{- \ell^2 \tau' (k_E^2 \, + \, m^2)}
\, ,
\label{prop_E (app1)}
\end{align}
where the first line~\eqref{prop_L (app1)} represents the Lorentzian propagator~\eqref{prop_p} 
with complexified string tension, i.e., $\ell_E^2 = - i \ell^2 > 0$,
whereas the second line~\eqref{prop_E (app1)} represents its Euclidean counterpart with the string length $\ell$ kept real.
The exponential suppression factors $e^{- i \, \ell_E^2 \, k^2}$ and $e^{-\ell^2 k_E^2}$ 
regularize the UV behavior of both propagators, 
effectively lifting the lower limit of integration in the Schwinger representation 
from 0 to $\ell_E^2$ (or to $\ell^2$ in the Euclidean case).
Thus, the right-hand sides of eqs.~\eqref{prop_L (app1)} and~\eqref{prop_E (app1)} 
are simply modified versions of eqs.~\eqref{prop0_L2 (app1)} and~\eqref{prop0_E2 (app1)} 
with a lifted lower bound of the $\tau'$-integral. 

The conventional analytic continuation 
proceeds from the Lorentzian propagator in eq.~\eqref{prop_p}
to its Euclidean counterpart on the left-hand side of eq.~\eqref{prop_E (app1)}.
On the other hand, 
the string worldsheet action --- including the string tension ---
is typically defined on a complex manifold with Euclidean signature, 
and Feynman diagram calculations are carried out in the Euclideanized phase space,
where the corresponding expression for the propagator 
is given by the middle expression in eq.~\eqref{prop_E (app1)}.
Crucially, the right-hand sides of eqs.~\eqref{prop_L (app1)} and~\eqref{prop_E (app1)} 
make it manifest that the Wick rotation $k^\mu \rightarrow k_E^\mu$ 
is equivalent to the analytic continuation $(k, \ell_E^2) \rightarrow (k_E, -i \ell^2)$. 
In particular, the Lorentzian propagator can be recovered from the Euclidean one 
via the continuation $\ell^2 \rightarrow i \ell_E^2$.

\section{Consistency Between Different Continuation Schemes}
\label{app:continuation_schemes}

According to the action~\eqref{SFT_action2}, 
the Feynman propagator for a massless scalar field in $(D + 1)$ dimensions is given by
\be 
- i \, \frac{e^{-\ell^2 k^2}}{k^2 - i\epsilon}
\, .
\label{D(k)-def}
\ee  
Since the factor $e^{-\ell^2 k^2}$ diverges in timelike directions, 
deriving the Feynman propagator 
\be 
\ev{\tilde{\phi}(X) \, \tilde{\phi}(0)}_0
= 
- i \int \frac{d^{D + 1} k}{(2\pi)^{D + 1}} \, 
\frac{e^{-\ell^2 k^2}}{k^2 - i\epsilon} \, e^{i k\cdot X}
\ee 
in position space requires the use of analytic continuation.  
In this appendix, 
we demonstrate that the same position-space propagator 
can be obtained through three different methods of analytic continuation:  
(1) Wick rotation of the Minkowski time $t$,
(2) Wick rotation of the light-cone time $V$, 
(3) complexification of the string length parameter $\ell$.

\subsection{Wick-rotate $t$}

We evaluate the propagator in four dimensions ($D = 3$) for simplicity.
We analytically continue the Lorentzian momentum $k_{\mu} = (k_0 \, , \vec{\vb{k}})$
into its Euclidean counterpart $(k_E)_{\mu} = \bigl( (k_E)_0 \, , \vec{\vb{k}} \bigr)$ 
according to
\be
k_0 \rightarrow k_0 = - i (k_E)_0 \, .
\ee 
Similarly, the spacetime coordinates 
$X^{\mu} = (t \, , \vec{\vb{x}})$ are Euclideanized to 
$X_E^{\mu} = (t_E \, , \vec{\vb{x}})$ via
\be 
t \rightarrow t = i t_E \, .
\ee
By introducing the Euclidean Schwinger proper time $\tau_E$,
the propagator in Euclidean space becomes 
\begin{align}
\ev{\tilde{\phi}(X_E) \, \tilde{\phi}(0)}_0
&=
\int\frac{d^4 k_E}{(2\pi)^4} \, 
\frac{e^{- \ell^2 k_E^2}}{k_E^2} \, 
e^{i k_E \cdot X_E}
\nn \\
&
=
\int\frac{d^4 k_E}{(2\pi)^4} 
\int_{\ell^2}^{\infty} d \tau_E \,
e^{- \tau_E \, k_E^2} \, e^{i k_E \cdot X_E}
\nn \\
&=
\frac{1}{16 \pi^2}
\int_{\ell^2}^{\infty} 
\frac{d \tau_E}{\tau_E^2} \,
e^{- X_E^2 / 4 \tau_E}
\, .
\end{align}
Subsequently, 
performing the change of variables $z = 1 / \tau_E$
leads to 
\be 
\ev{\tilde{\phi}(X_E) \, \tilde{\phi}(0)}_0
=
\frac{1}{16\pi^2} 
\int_0^{1/\ell^2} d z \,
e^{- X_E^2 \, z / 4}
=
\frac{1}{4\pi^2} \,
\frac{1 - e^{- X_E^2 / 4\ell^2}}{X_E^2}
\quad \rightarrow \quad 
\frac{1}{4\pi^2} \,
\frac{1 - e^{- X^2 / 4 \ell^2}}{X^2}
\, ,
\ee 
where we Wick-rotated back to the Minkowski spacetime coordinates $X^{\mu}$
in the last step.

\subsection{Wick-rotate $V$}
\label{sec:WickRotateV}

In the light-cone frame 
\be 
X^{\mu} = (U \equiv t - x \, , V \equiv t + x \, , \vec{\vb{x}}_{\perp})
\ee 
with corresponding light-cone frequencies 
$\Omega_U \equiv ( k^0 + k^1 ) / 2$
and $\Omega_V \equiv ( k^0 - k^1 ) / 2$
defined in eq.~\eqref{lightcone_freq},
one can define the correlation functions
with Wick-rotated light-cone time $V$ and frequency $\Omega_V$:
\be 
V \rightarrow V = - i V_E
\, , 
\qquad 
\Omega_V \rightarrow \Omega_V = i \Omega_V^E
\, .
\ee 
In this case, the momentum-space propagator has the form
\be 
\frac{i \, 
\mathrm{exp} \bigl[ \ell^2 ( 4 i \Omega_U \Omega_V^E - \vec{\vb{k}}_{\perp}^2 ) \bigr]}
{4 i \Omega_U \Omega_V^E - \vec{\vb{k}}_{\perp}^2}
=
- i
\int_{\ell^2}^{\infty} d \tau_E \, 
\mathrm{exp} \bigl[ - \tau_E \, ( \vec{\vb{k}}_{\perp}^2 - 4 i \Omega_U \Omega_V^E ) \bigr]
\, ,
\ee 
and the position-space representation of the propagator 
can be expressed as
\begin{align}
\ev{\tilde{\phi}(U, V_E) \, \tilde{\phi}(0)}_0
&=
2 \int_{-\infty}^{\infty} 
\frac{d \Omega_U \, d \Omega_V^E \, d^2 \vec{\vb{k}}_{\perp}}{(2 \pi)^4} \, 
e^{- i \Omega_U U} \, e^{- i \Omega_V^E V_E} \, 
e^{i \vec{\vb{k}}_{\perp} \cdot \, \vec{\vb{x}}_{\perp}}
\int_{\ell^2}^{\infty} d \tau_E \, 
e^{- \tau_E \, ( \vec{\vb{k}}_{\perp}^2 - 4 i \Omega_U \Omega_V^E )}
\nn \\
&=
2 \int_{-\infty}^{\infty} 
\frac{d \Omega_U \, d^2 \vec{\vb{k}}_{\perp}}{(2 \pi)^3} \, 
e^{- i \Omega_U U} \, 
e^{i \vec{\vb{k}}_{\perp} \cdot \, \vec{\vb{x}}_{\perp}}
\int_{\ell^2}^{\infty} d \tau_E \, 
\delta(4 \tau_E \, \Omega_U - V_E) \, 
e^{- \tau_E \, \vec{\vb{k}}_{\perp}^2}
\nn \\
&=
\int_{-\infty}^{\infty} 
\frac{d \Omega_U}{4 \pi \abs{\Omega_U}} \, 
e^{- i \Omega_U U} \,
\Theta \left( \frac{V_E}{4 \Omega_U} - \ell^2 \right) 
\int \frac{d^2 \vec{\vb{k}}_{\perp}}{(2 \pi)^2} \, 
e^{i \vec{\vb{k}}_{\perp} \cdot \, \vec{\vb{x}}_{\perp}} \, 
e^{- V_E \, \vec{\vb{k}}_{\perp}^2 / 4 \Omega_U}
\, .
\end{align}
Note that the ratio $V_E / \Omega_U$ in the integrand above
is constrained to be positive,
and thus the Gaussian integral over transverse momenta $\vec{\vb{k}}_{\perp}$
is well-defined.
In fact, we can further write 
\be 
\begin{aligned}
\ev{\tilde{\phi}(U, V_E) \, \tilde{\phi}(0)}_0
=
\int_{0}^{\abs{V_E} / 4 \ell^2} 
\frac{d \Omega_U}{4 \pi \Omega_U} 
\int \frac{d^2 \vec{\vb{k}}_{\perp}}{(2 \pi)^2} \, 
e^{i \vec{\vb{k}}_{\perp} \cdot \, \vec{\vb{x}}_{\perp}} 
\Bigl[
&
\Theta(V_E) \, e^{- i \Omega_U U} \, e^{- V_E \, \vec{\vb{k}}_{\perp}^2 / 4 \Omega_U}
\\
&
+
\Theta(-V_E) \, e^{i \Omega_U U} \, e^{- \abs{V_E} \, \vec{\vb{k}}_{\perp}^2 / 4 \Omega_U}
\Bigr] 
\, ,
\end{aligned}
\ee 
which is evaluated to be 
\be 
\ev{\tilde{\phi}(U, V_E) \, \tilde{\phi}(0)}_0
=
\frac{1}{4 \pi^2} \, 
\frac{1}{- i U V_E - \vec{\vb{x}}_{\perp}^2}
\left[ 
e^{(-i U V_E - \vec{\vb{x}}_{\perp}^2) / 4 \ell^2}
- 1
\right]
.
\ee 
Subsequently, performing the analytic continuation 
$V_E \to V_E = i V$ results in 
\be 
\label{G_VE}
\ev{\tilde{\phi}(U, V_E) \, \tilde{\phi}(0)}_0
\quad \rightarrow \quad 
\ev{\tilde{\phi}(U, V) \, \tilde{\phi}(0)}_0
=
\frac{1}{4\pi^2} \,
\frac{1 - e^{- X^2 / 4 \ell^2}}{X^2}
\, ,
\ee 
where $X^2 = - UV + \vec{\vb{x}}_{\perp}^2$ 
is the Lorentzian interval between 
the two spacetime points 
$X^{\mu} = (U, V, \vec{\vb{x}}_{\perp})$ and $X^{\prime \mu} = 0$.

\subsection{Anti-Wick-rotate $\ell^2$}
\label{sec:AntiWickRotateell}

In the following,
we consider the analytic continuation of the string length parameter:
\be
\ell^2 \rightarrow \ell^2 = \pm i \, \ell_E^2
\qquad 
\text{with}
\quad 
\ell_E^2 > 0
\, .
\label{ell-ellE}
\ee
The momentum-space Feynman propagator 
can then be expressed as an integral 
over Lorentzian Schwinger proper time $\tau$:
\be 
-i \, 
\frac{\mathrm{exp} ( \mp i \, \ell_E^2 \, k^2 )}{k^2 - i \epsilon}
=
\int_{\pm \ell_E^2}^{\infty}
d \tau \, 
e^{- i \tau (k^2 - i \epsilon)}
\, .
\ee 
From the viewpoint of the string worldsheet,
$\tau$ is part of the moduli space coordinates of Riemann surfaces.
Therefore, the continuation~\eqref{ell-ellE}
can be interpreted as complexifying the modular parameters~\cite{Berera:1992tm, Witten:2013pra}.

In the space of imaginary $\ell^2$, 
the spacetime propagator can be written as
\begin{align}
\ev{\tilde{\phi}(X) \, \tilde{\phi}(0)}_0
(\ell_E^2)
&=
\int \frac{d^4 k}{(2 \pi)^4} \, e^{i k \cdot X}
\int_{\pm \ell_E^2}^{\infty} d \tau \, 
e^{- i \tau (k^2 - i \epsilon)}
\nn \\
&=
\int_{\pm \ell_E^2}^{\infty} d \tau \, 
e^{i X^2 / 4 \tau} 
\int \frac{d^4 k}{(2 \pi)^4} \, 
e^{- i \tau (k - X / 2 \tau)^2}
\nn \\
&=
- \frac{i}{16 \pi^2} 
\int_{\pm \ell_E^2}^{\infty} \frac{d \tau}{\tau^2} \, 
e^{i X^2 / 4 \tau} 
\, .
\label{G_ellE}
\end{align}
Due to the essential singularity at $\tau = 0$,
the expression~\eqref{G_ellE} is well defined only for the upper sign case, 
i.e. $\ell^2 = + i \, \ell_E^2$,
which is the prescription that is adopted in this work 
(see eq.~\eqref{ell_to_ellE (sec2)}).
For the upper sign case, making the variable change $z = 1 / \tau$,
eq.~\eqref{G_ellE} gives
\be 
\ev{\tilde{\phi}(X) \, \tilde{\phi}(0)}_0
(\ell_E^2)
=
- \frac{i}{16 \pi^2} 
\int_0^{1 / \ell_E^2} 
dz \, e^{i X^2 z / 4}
=
\frac{1}{4 \pi^2} \, 
\frac{1 - e^{i X^2 / 4 \ell_E^2}}{X^2}
\, .
\ee 
Finally, after performing the continuation $\ell_E^2 \to - i \ell^2$ back to $\ell^2$, the propagator becomes 
\be 
\ev{\tilde{\phi}(X) \, \tilde{\phi}(0)}_0
(\ell_E^2)
\quad \rightarrow \quad 
\ev{\tilde{\phi}(X) \, \tilde{\phi}(0)}_0
(\ell^2)
=
\frac{1}{4\pi^2} \, \frac{1 - e^{-X^2 / 4 \ell^2}}{X^2}
\, .
\ee 
Thus, we have demonstrated that all three approaches to analytic continuation 
yield the same propagator in spacetime.

\section{Perturbative Treatment of Nonlocality}
\label{app:pert}

In this appendix, we show that applying the naive Hamiltonian formalism to the variables $(a, a^{\dagger})$ of the nonlocal toy model defined by the action~\eqref{S_a}
\be
\label{S_a_1}
S[a, a^{\dagger}]
=
\int dt 
\left[
i \, a^{\dagger}(t) \, \del_t \, a(t)
+
\lambda \, b(t) \, 
a(t - \sigma / 2) \, a^{\dagger}(t + \sigma / 2)
\right]
\ee
cannot reproduce the path-integral correlation function
in the interacting theory.

In the path-integral formalism of~\eqref{S_a_1},
the two-point correlation function 
at zeroth order in $\lambda$ is given by 
\be
\label{aa0}
\langle a(t) \, a^{\dagger}(t') \rangle_0 
= 
\Theta(t - t')
\, .
\ee
On the other hand,
in the usual Hamiltonian approach to canonical quantization,
the unperturbed equations of motion $\del_t \, \hat{a}_0 (t) = 0$
and $\del_t \, \hat{a}_0^{\dagger}(t) = 0$
for the operators imply that they are time-independent:
\be
\hat{a}_0(t) = \hat{a}_0
\, ,
\qquad
\hat{a}_0^{\dagger}(t) = \hat{a}_0^{\dagger}
\, .
\ee
Furthermore, 
the equal-time canonical commutation relation 
$[ \hat{a}_0 (t) \, , \delta S / \delta \dot{\hat{a}}_0 (t)] = i$
leads to 
\be
[\hat{a}_0 \, , \hat{a}_0^{\dagger}] = 1
\, .
\ee
Keeping in mind the correspondence~\eqref{corres_a} with the 2D toy model in the light-cone frame, we identify $(\hat{a}_0, \hat{a}_0^{\dagger})$ as the creation and annihilation operators, and define the vacuum state $\ket{0}$ as 
\be
\hat{a}_0 \ket{0}
=
0 \, ,
\ee
with $\inp{0}{0} = 1$.
It is then clear that 
the path-integral correlation function~\eqref{aa0}
is related to the time-ordered product of free field operators
$(\hat{a}_0, \hat{a}_0^{\dagger})$
in the Hamiltonian formalism via 
\be
\label{aa_match}
\bigl\langle 
a(t) \, a^{\dagger}(t')
\bigr\rangle_0
= 
\ev{\mathcal{T} \, 
\bigl\{ \hat{a}_0 (t) \, \hat{a}_0^{\dagger}(t') \bigr\}}
{0}
\ee
as desired.

If the nonlocal interaction term in eq.~\eqref{S_a_1}
is treated perturbatively,
the path-integral correlation function~\eqref{aa0} 
can be evaluated as 
\begin{align}
\bigl\langle 
a(t) \, a^{\dagger}(t')
\bigr\rangle
&= 
\bigl\langle 
a(t) \, a^{\dagger}(t')
\bigr\rangle_0 
+ 
\bigl\langle 
a(t) \, a^{\dagger}(t')
\bigr\rangle_1 
+ 
\mathcal{O}(\lambda^2)
\nn \\
&= 
\Theta(t - t') 
+ 
i \lambda 
\int_{t'}^{\infty} dt'' \,
b(t'' + \sigma/2) \, \Theta(t - t'' - \sigma) 
+
\mathcal{O}(\lambda^2)
\nn \\
&= 
\Theta(t - t') 
+ 
i \lambda \, 
\Theta(t - t' - \sigma) 
\int_{t' \, + \, \sigma/2}^{t \, - \, \sigma/2} dt'' \, 
b(t'') 
+
\mathcal{O}(\lambda^2)
\, .
\label{aa1}
\end{align}
Meanwhile,
since the equations of motion become
\begin{align}
i \del_t \, \hat{a}(t) 
+ 
\lambda \, b(t - \sigma/2) \, 
\hat{a}(t - \sigma) 
&= 
0
\, ,
\\
i \del_t \, \hat{a}^{\dagger}(t) 
-
\lambda \, b(t + \sigma/2) \, 
\hat{a}^{\dagger}(t + \sigma) 
&= 
0
\, ,
\end{align}
the operators in the Hamiltonian formalism 
can be obtained perturbatively in $\lambda$ as 
\begin{align}
\hat{a}(t) 
&= 
\hat{a}_0 
+ 
i \lambda \int_{-\infty}^t dt'' \, 
b(t'' - \sigma/2) \, \hat{a}_0 
+
\mathcal{O}(\lambda^2)
\, ,
\\
\hat{a}^{\dagger}(t) 
&= 
\hat{a}^{\dagger}_0 
- 
i \lambda \int_{-\infty}^t dt'' \, 
b(t'' + \sigma/2) \, \hat{a}_0^{\dagger} 
+
\mathcal{O}(\lambda^2)
\, .
\end{align}
Plugging these expressions into the time-ordered product gives
\be
\ev{\mathcal{T}
\bigl\{ \hat{a}(t) \, \hat{a}^{\dagger}(t') \bigr\}}{0} 
=
\Theta(t - t') 
+ 
i \lambda \, \Theta(t - t') 
\int_{t' \, + \, \sigma/2}^{t \, - \, \sigma/2} dt'' \, 
b(t'') 
+
\mathcal{O}(\lambda^2)
\, .
\ee 
Notice that due to a nonzero scale of nonlocality $\sigma$,
the time-ordered product in the Heisenberg picture does not agree with 
the path-integral correlation function~\eqref{aa1},
as the $\mathcal{O}(\lambda)$ term does not vanish when $0 < t - t' < \sigma$.
As a consequence, 
we find that the standard Hamiltonian formalism for $a(t)$ and $a^{\dagger}(t)$
fails to reproduce the path-integral result
in the presence of nonlocal interactions:
\be 
\ev{\mathcal{T}
\bigl\{ \hat{a}(t) \, \hat{a}^{\dagger}(t') \bigr\}}{0} 
\neq 
\bigl\langle 
a(t) \, a^{\dagger}(t')
\bigr\rangle
\, .
\ee 
The origin of the discrepancy comes from the time independence of the operators $\hat{a}_0$ and $\hat{a}^{\dagger}_0$ in the free theory. 
As we saw above, 
their time dependence is crucial in adjusting the argument of the step function $\Theta(t - t')$ in the correction terms to possibly achieve a successful Hamiltonian formalism. 

\section{Derivation of $g_1(t, t')$ and $g_1^{c}(t, t')$}
\label{app:derive_g1}

For the nonlocal 1D model~\eqref{S_a_tilde}, the Schwinger-Dyson equation~\eqref{SD2} at $\mathcal{O}(\lambda)$ is given by 
\be 
\del_{t'} \, 
\bigl\langle 
\tilde{a}(t) \, \tilde{a}^{\dagger}(t')
\bigr\rangle_1
=
- i \lambda \, b(t' + \sigma) \, 
\bigl\langle 
\tilde{a}(t) \, \tilde{a}^{\dagger}(t' + \sigma)
\bigr\rangle_0
\, ,
\ee 
where the free propagator $\ev{\tilde{a}(t) \, \tilde{a}^{\dagger}(t')}_0$ 
takes the form~\eqref{D0}.
In order for the operator formalism to be consistent with 
the path-integral correlation function,
$\hat{\tilde{a}}_1 (t)$ has to obey 
\be 
\del_{t'}
\left[ 
\Theta(t - t')
\left(
\ev{\hat{\tilde{a}}_0 (t) \, \hat{\tilde{a}}_1^{\dagger} (t')}{0}
+
\ev{\hat{\tilde{a}}_1 (t) \, \hat{\tilde{a}}_0^{\dagger} (t')}{0}
\right)
\right]
=
- i \lambda \, 
b(t' + \sigma) \, 
\Theta(t - t' - 2 \sigma)
\, .
\ee 
Matching both sides of the equation yields 
the following two conditions:
\begin{align}
\ev{\hat{\tilde{a}}_0 (t) \, \hat{\tilde{a}}_1^{\dagger} (t)}{0}
+
\ev{\hat{\tilde{a}}_1 (t) \, \hat{\tilde{a}}_0^{\dagger} (t)}{0}
&=
0
\, ,
\label{cond1}
\\
\Theta(t - t')
\left(
\ev{\hat{\tilde{a}}_0 (t) \, \del_{t'} \, \hat{\tilde{a}}_1^{\dagger} (t')}{0}
+
\ev{\hat{\tilde{a}}_1 (t) \, \del_{t'} \, \hat{\tilde{a}}_0^{\dagger} (t')}{0}
\right)
&=
- i \lambda \, 
b(t' + \sigma) \, 
\Theta(t - t' - 2 \sigma)
\, .
\label{cond2}
\end{align}
Plugging in the ansatz~\eqref{a1}--\eqref{a1_dagger} 
for $\hat{\tilde{a}}_1 (t)$ and $\hat{\tilde{a}}_1^{\dagger} (t)$
and making use of the zeroth-order commutator~\eqref{comm0},
the first condition~\eqref{cond1} demands that 
\be 
\int_{-\infty}^{\infty} dt'' 
\left[ g_1 (t \, , t'') - g_1^{c} (t \, , t'') \right]
b(t'') \, 
\Theta(\abs{t' - t''} - \sigma)
=
0
\, ,
\ee 
which leads to 
\be 
\label{g1 = g1_star}
g_1^{c} (t \, , t')
=
g_1 (t \, , t')
\, . 
\ee 
On the other hand,
the second condition~\eqref{cond2} can be expressed as 
\be 
\label{g1_eq}
\begin{aligned}
- \Theta(t - t' - 2 \sigma) \, 
b(t' + \sigma)
=
\Theta(t - t') \, 
\biggl[
&
g_1 (t \, , t' - \sigma) \, b(t' - \sigma)
-
g_1 (t \, , t' + \sigma) \, b(t' + \sigma)
\\
&-
\int_{-\infty}^{\infty} dt'' \, 
\del_{t'} \, g_1 (t' , t'') \, 
b(t'') \, 
\Theta(\abs{t - t''} - \sigma)
\biggr]
\, ,
\end{aligned}
\ee 
where we have utilized eq.~\eqref{g1 = g1_star}. 

To solve for $g_1 (t \, , t')$ from eq.~\eqref{g1_eq}, 
we adopt the ansatz 
\be 
\label{g1_ansatz}
g_1 (t, t')
=
\xi_+ \, 
\Theta( t - t' - \Delta_+ )
+
\xi_- \, 
\Theta( t' - t - \Delta_- )
\, ,
\ee 
where $\Delta_+ (\sigma) \geq 0$ and $\Delta_- (\sigma) \geq 0$ are constants.\footnote{
In principle, 
one could start with a more general ansatz 
\be 
g_1 (t \, , t')
=
F_+ (t \, , t') \, 
\Theta( t - t' - \Delta_+ )
+
F_- (t \, , t') \, 
\Theta( t' - t - \Delta_- )
\, .
\ee 
However, 
a closer inspection of eq.~\eqref{g1_eq} reveals that 
$F_+ (t \, , t')$ and $F_- (t \, , t')$ must be step functions, 
which can then either be absorbed 
into the existing terms in eq.~\eqref{g1_ansatz},
or else produce terms in eq.~\eqref{g1_eq} 
that do not match the left-hand side of the equation.
}
Subsequently,
eq.~\eqref{g1_eq} is satisfied if 
\be 
b(t' - \sigma) \, 
\Theta(t - t' - \Delta_+ + \sigma)
=
b(t' - \Delta_+) \, 
\Theta(t - t' + \Delta_+ - \sigma)
\ee 
and 
\be 
\begin{aligned}
&
\Theta(t - t' - 2 \sigma) \, 
b(t' + \sigma)
\\
& 
= 
b(t' + \sigma) 
\left[ 
\xi_+ \, 
\Theta( t - t' - \Delta_+ - \sigma)
+
\xi_- \, \Theta(t - t') \, 
\Theta( t' - t - \Delta_- + \sigma)
\right]
\\
& \quad \,
- \xi_- \, 
b(t' + \Delta_-)
\left[ 
\Theta( t - t' - \Delta_- - \sigma )
+
\Theta(t - t') \, 
\Theta( t' - t + \Delta_- - \sigma )
\right]
.
\end{aligned}
\ee 
From these equations we obtain 
\be 
\Delta_+ = \Delta_- = \sigma 
\qquad 
\text{and}
\qquad 
\xi_+ - \xi_- = 1
\, .
\ee
Hence, we arrive at 
\be 
g_1 (t, t')
=
g_1^{c} (t, t')
=
\xi \, 
\Theta(t - t' - \sigma)
+
\left( \xi - 1 \right)
\Theta(t' - t - \sigma)
\, , 
\ee 
where $\xi \in \mathbb{R}$ is a constant parameter. 
It can be readily verified that the other Schwinger-Dyson equation~\eqref{SD1} 
is obeyed up to $\mathcal{O}(\lambda)$ as well.

\section{Operators to All Orders}
\label{app:a_allorder}

In this appendix,
we show that the correspondence~\eqref{path_op_corres_2pt}
with the path-integral formalism can be achieved 
to all orders in $\lambda$ 
through the recurrence relations
\begin{align}
\hat{\tilde{a}}_j (t)
&=
i \lambda 
\int_{-\infty}^{\infty} dt'' \, 
\Theta(t - t'' - \sigma) \, 
b(t'') \, 
\hat{\tilde{a}}_{j - 1} (t'')
\, ,
\label{aj_recurr}
\\
\hat{\tilde{a}}_j^{\dagger} (t)
&=
- i \lambda 
\int_{-\infty}^{\infty} dt'' \, 
\Theta(t - t'' - \sigma) \, 
b(t'') \, 
\hat{\tilde{a}}_{j - 1}^{\dagger} (t'')
\label{aj_dag_recurr}
\end{align}
for all $j \geq 1$.

Recall that the $\mathcal{O}(\lambda^n)$ correction
to the correlation function is given in eq.~\eqref{Dn},
and thus based on eq.~\eqref{comm_n},
we demand that the operators 
$\hat{\tilde{a}}_j (t)$ and $\hat{\tilde{a}}_j^{\dagger} (t)$
($1 \leq j \leq n$) satisfy
\begin{align}
&
\sum_{j = 0}^n
\ev{\hat{\tilde{a}}_j (t) \, \hat{\tilde{a}}_{n - j}^{\dagger} (t')}{0}
\nn \\
&=
\left( i \lambda \right)^n
\bigintssss_{-\infty}^{\infty} 
\Biggl[ \, 
\prod_{k = 1}^n 
d t_k \, b(t_k) 
\Biggr] 
\Theta(t - t_n - \sigma) \, 
\Theta(t_n - t_{n - 1} - \sigma)
\cdots 
\Theta(t_2 - t_1 - \sigma) \, 
\Theta(t_1 - t' - \sigma) 
\nn \\
& \quad 
+
\left( - i \lambda \right)^n
\bigintssss_{-\infty}^{\infty} 
\Biggl[ \, 
\prod_{k = 1}^n 
d t_k \, b(t_k) 
\Biggr] 
\Theta(t' - t_n - \sigma) \, 
\Theta(t_n - t_{n - 1} - \sigma)
\cdots 
\Theta(t_2 - t_1 - \sigma) \, 
\Theta(t_1 - t - \sigma) 
\, .
\label{sum_aj_an-j}
\end{align}
We will now verify that the ansatzes~\eqref{aj_recurr} and~\eqref{aj_dag_recurr}
are valid by substituting them into the left-hand side of the above equation.

Starting with the first two terms of the summation ($j = 0$ and $j = 1$)
on the left-hand side of eq.~\eqref{sum_aj_an-j}, 
we obtain
\begin{align}
\ev{\hat{\tilde{a}}_0 (t) \, \hat{\tilde{a}}_n^{\dagger} (t')}{0}
=
\left( - i \lambda \right)^n 
\bigintssss_{-\infty}^{\infty} 
\Biggl[ 
&
\prod_{k = 1}^n 
d t_k \, 
b(t_k) 
\Biggr] 
\Theta(t' - t_n - \sigma) \, 
\Theta(t_n - t_{n - 1} - \sigma) \cdots 
\Theta(t_2 - t_1 - \sigma) 
\nn \\
&\times 
\comm*{\hat{\tilde{a}}_0 (t)}{\hat{\tilde{a}}_0^{\dagger} (t_1)}
\nn \\
=
\left( - i \lambda \right)^n 
\bigintssss_{-\infty}^{\infty} 
\Biggl[ 
&
\prod_{k = 1}^n 
d t_k \, 
b(t_k) 
\Biggr] 
\Theta(t' - t_n - \sigma) \, 
\Theta(t_n - t_{n - 1} - \sigma) \cdots 
\Theta(t_2 - t_1 - \sigma) 
\nn \\
&\times 
\left[ 
\underline{\Theta(t - t_1 - \sigma)}
+
\Theta(t_1 - t - \sigma)
\right] 
,
\label{a0_an}
\end{align}
and 
\begin{align}
\ev{\hat{\tilde{a}}_1 (t) \, \hat{\tilde{a}}_{n - 1}^{\dagger} (t')}{0}
=
\left( i \lambda \right)^n 
(-1)^{n - 1}
\bigintssss_{-\infty}^{\infty} 
\Biggl[ 
&
\prod_{k = 1}^n 
d t_k \, 
b(t_k) 
\Biggr] 
\Theta(t - t_n - \sigma) \, 
\Theta(t' - t_{n - 1} - \sigma) 
\nn \\
&\times 
\Theta(t_{n - 1} - t_{n - 2} - \sigma) \cdots 
\Theta(t_2 - t_1 - \sigma) \, 
\comm*{\hat{\tilde{a}}_0 (t_n)}{\hat{\tilde{a}}_0^{\dagger} (t_1)}
\nn \\
=
\left( i \lambda \right)^n 
(-1)^{n - 1}
\bigintssss_{-\infty}^{\infty} 
\Biggl[ 
&
\prod_{k = 1}^n 
d t_k \, 
b(t_k) 
\Biggr] 
\Theta(t - t_n - \sigma) \, 
\Theta(t' - t_{n - 1} - \sigma) 
\nn \\
&\times 
\Theta(t_{n - 1} - t_{n - 2} - \sigma) \cdots 
\Theta(t_2 - t_1 - \sigma) 
\nn \\
& \times 
\left[ 
\Theta(t_n - t_1 - \sigma)
+
\underline{\Theta(t_1 - t_n - \sigma)}
\right] 
.
\label{a1_an-1}
\end{align}
Notice that the term involving the underlined part in eq.~\eqref{a0_an} 
cancels the corresponding term involving the underlined part in eq.~\eqref{a1_an-1}
after relabeling the integration variables as
$\left\{ t_1, t_2, t_3, \cdots, t_n \right\} \rightarrow 
\left\{ t_n, t_1, t_2, \cdots, t_{n-1} \right\}$.

For a general $j \geq 2$ term in the summation,
we can express it as 
\begin{alignat}{2}
&
\ev{\hat{\tilde{a}}_j (t) \, \hat{\tilde{a}}_{n - j}^{\dagger} (t')}{0}
&&
\nn \\
&
=
\left( i \lambda \right)^n 
(-1)^{n - j}
\bigintssss_{-\infty}^{\infty} 
\Biggl[ 
&&
\prod_{k = 1}^n 
d t_k \, 
b(t_k) 
\Biggr] 
\Theta(t - t_n - \sigma) \, 
\Theta(t_n - t_{n - 1} - \sigma) 
\cdots 
\Theta(t_{n - j + 2} - t_{n - j + 1} - \sigma) 
\nn \\
&
&&
\times 
\Theta(t' - t_{n - j} - \sigma) \, 
\Theta(t_{n - j} - t_{n - j - 1} - \sigma) 
\cdots 
\Theta(t_2 - t_1 - \sigma) 
\nn \\
&
&&
\times 
\comm*{\hat{\tilde{a}}_0 (t_{n - j + 1})}{\hat{\tilde{a}}_0^{\dagger} (t_1)}
\nn \\
&=
\left( i \lambda \right)^n 
(-1)^{n - j}
\bigintssss_{-\infty}^{\infty} 
\Biggl[ 
&&
\prod_{k = 1}^n 
d t_k \, 
b(t_k) 
\Biggr] 
\underbrace{\Theta(t - t_n - \sigma) \, 
\Theta(t_n - t_{n - 1} - \sigma) 
\cdots}_{j - 1}
\Theta(t_{n - j + 2} - t_{n - j + 1} - \sigma) 
\nn \\
& 
&&
\times 
\underbrace{\Theta(t' - t_{n - j} - \sigma) \, 
\Theta(t_{n - j} - t_{n - j - 1} - \sigma) 
\cdots}_{n - j - 1}
\Theta(t_2 - t_1 - \sigma) 
\nn \\
&
&& 
\times 
\left[ 
\underline{\Theta(t_{n - j + 1} - t_1 - \sigma)}
+
\Theta(t_1 - t_{n - j + 1} - \sigma) 
\right] 
,
\label{aj_an-j}
\end{alignat}
with the subsequent ($j + 1$)-th term in the summation given by
\begin{alignat}{2}
&
\ev{\hat{\tilde{a}}_{j + 1} (t) \, \hat{\tilde{a}}_{n - j - 1}^{\dagger} (t')}{0}
&&
\nn \\
&
=
\left( i \lambda \right)^n 
(-1)^{n - j - 1}
\bigintssss_{-\infty}^{\infty} 
\Biggl[ 
&&
\prod_{k = 1}^n 
d t_k \, 
b(t_k) 
\Biggr] 
\Theta(t - t_n - \sigma) \, 
\Theta(t_n - t_{n - 1} - \sigma) 
\cdots 
\Theta(t_{n - j + 1} - t_{n - j} - \sigma) 
\nn \\
&
&&
\times 
\Theta(t' - t_{n - j - 1} - \sigma) \, 
\Theta(t_{n - j - 1} - t_{n - j - 2} - \sigma) 
\cdots 
\Theta(t_2 - t_1 - \sigma) 
\nn \\
&
&&
\times 
\comm*{\hat{\tilde{a}}_0 (t_{n - j})}{\hat{\tilde{a}}_0^{\dagger} (t_1)}
\nn \\
&=
\left( i \lambda \right)^n 
(-1)^{n - j - 1}
\bigintssss_{-\infty}^{\infty} 
\Biggl[ 
&&
\prod_{k = 1}^n 
d t_k \, 
b(t_k) 
\Biggr] 
\underbrace{\Theta(t - t_n - \sigma) \, 
\Theta(t_n - t_{n - 1} - \sigma) 
\cdots}_{j}
\Theta(t_{n - j + 1} - t_{n - j} - \sigma) 
\nn \\
& 
&&
\times 
\underbrace{\Theta(t' - t_{n - j - 1} - \sigma) \, 
\Theta(t_{n - j - 1} - t_{n - j - 2} - \sigma) 
\cdots}_{n - j - 2}
\Theta(t_2 - t_1 - \sigma) 
\nn \\
&
&& 
\times 
\left[ 
\underline{\Theta(t_{n - j} - t_1 - \sigma)}
+
\Theta(t_1 - t_{n - j} - \sigma) 
\right] 
.
\label{aj+1_an-j-1}
\end{alignat}
Once again, 
the term containing the underlined part in eq.~\eqref{aj_an-j} 
cancels the corresponding underlined term in eq.~\eqref{aj+1_an-j-1}. 
This cancellation can be seen by relabeling the first 
$(n - j)$ integration variables as 
$\{ t_1, t_2, t_3, \cdots, t_{n-j} \} 
\rightarrow 
\{ t_{n-j}, t_1, t_2, \cdots, t_{n-j-1} \}$,
while leaving the remaining variables 
$\{ t_{n-j+1}, \cdots, t_n \}$ unchanged.

After summing over all terms on the left-hand side of eq.~\eqref{sum_aj_an-j}, 
the only remaining contributions are the non-underlined terms 
from the $j = 0$ contribution in eq.~\eqref{a0_an} 
and from the $j = n$ contribution below:
\begin{align}
\ev{\hat{\tilde{a}}_n (t) \, \hat{\tilde{a}}_0^{\dagger} (t')}{0}
=
\left( i \lambda \right)^n 
\bigintssss_{-\infty}^{\infty} 
\Biggl[ 
&
\prod_{k = 1}^n 
d t_k \, 
b(t_k) 
\Biggr] 
\Theta(t - t_n - \sigma) \, 
\Theta(t_n - t_{n - 1} - \sigma) \cdots 
\Theta(t_2 - t_1 - \sigma) 
\nn \\
&\times 
\comm*{\hat{\tilde{a}}_0 (t_1)}{\hat{\tilde{a}}_0^{\dagger} (t')}
\nn \\
=
\left( i \lambda \right)^n 
\bigintssss_{-\infty}^{\infty} 
\Biggl[ 
&
\prod_{k = 1}^n 
d t_k \, 
b(t_k) 
\Biggr] 
\Theta(t - t_n - \sigma) \, 
\Theta(t_n - t_{n - 1} - \sigma) \cdots 
\Theta(t_2 - t_1 - \sigma) 
\nn \\
&\times 
\left[ 
\Theta(t_1 - t' - \sigma)
+
\underline{\Theta(t' - t_1 - \sigma)}
\right] 
.
\end{align}
As a result, 
we arrive at 
\begin{align}
&
\sum_{j = 0}^n
\ev{\hat{\tilde{a}}_j (t) \, \hat{\tilde{a}}_{n - j}^{\dagger} (t')}{0}
\nn \\
&=
\left( i \lambda \right)^n
\bigintssss_{-\infty}^{\infty} 
\Biggl[ \, 
\prod_{k = 1}^n 
d t_k \, b(t_k) 
\Biggr] 
\Theta(t - t_n - \sigma) \, 
\Theta(t_n - t_{n - 1} - \sigma)
\cdots 
\Theta(t_2 - t_1 - \sigma) \, 
\Theta(t_1 - t' - \sigma) 
\nn \\
& \quad 
+
\left( - i \lambda \right)^n
\bigintssss_{-\infty}^{\infty} 
\Biggl[ \, 
\prod_{k = 1}^n 
d t_k \, b(t_k) 
\Biggr] 
\Theta(t' - t_n - \sigma) \, 
\Theta(t_n - t_{n - 1} - \sigma)
\cdots 
\Theta(t_2 - t_1 - \sigma) \, 
\Theta(t_1 - t - \sigma) 
\, ,
\end{align}
reproducing the desired outcome~\eqref{sum_aj_an-j} 
consistent with the path-integral correlation function.

\section{Dual Wave Function to All Orders}
\label{app:psi_n}

Up to order $\mathcal{O}(\lambda^n)$, 
the constraint~\eqref{eom_a_dag_xi=1} has the explicit form
\be 
\begin{aligned}
i 
\left[ \psi_n^c (t) - \psi_n^c (t - 2 \sigma) \right]
=
& - \lambda
\sum_{j = 0}^{n - 2}
\int_{-\infty}^{\infty} d \tau \, 
\psi_{n - j - 1}^c (\tau) \, 
\bigl[ 
\hat{\tilde{a}}_0 (\tau)
\, , \, 
b(t) \, \hat{\tilde{a}}_j^{\dagger} (t)
-
b(t - 2 \sigma) \, \hat{\tilde{a}}_j^{\dagger} (t - 2 \sigma)
\bigr]
\\
& - \lambda
\int_{-\infty}^{\infty} d \tau \, 
\frac{\psi_0^{\ast} (\tau)}{2 N \sigma} \, 
\bigl[ 
\hat{\tilde{a}}_0 (\tau)
\, , \, 
b(t) \, \hat{\tilde{a}}_{n - 1}^{\dagger} (t)
-
b(t - 2 \sigma) \, \hat{\tilde{a}}_{n - 1}^{\dagger} (t - 2 \sigma)
\bigr]
\, .
\end{aligned}
\ee 
For $n = 1$,
it leads to 
\be 
\psi_1^c (t)
=
i \lambda \, b(t) 
\int_{-\infty}^{\infty} d \tau \, 
\frac{\psi_0^{\ast} (\tau)}{2 N \sigma} \, 
\bigl[ 
\hat{\tilde{a}}_0 (\tau) \, , 
\hat{\tilde{a}}_0^{\dagger} (t) 
\bigr]
+
\frac{\lambda \, c_1^{\prime}}{2 N \sigma}
=
i \lambda \, \overbar{\psi}_0^{\ast} \, b(t)
+
\frac{\lambda \, c_1^{\prime}}{2 N \sigma}
\, ,
\ee 
which matches the previously obtained result~\eqref{psi_1_star} (with $\xi = 1$).
For $n \geq 2$,
we arrive at the expression
\be 
\label{psi^c_iter}
\psi_n^c (t)
=
i \lambda \, b(t) 
\left[ 
\sum_{j = 0}^{n - 2} 
F_n^{(j)} (t)
+
\frac{1}{2 N \sigma} \, 
F_n^{(n - 1)} (t) 
\right]
+
\frac{\lambda^n \, c_n^{\prime}}{2 N \sigma}
\, ,
\ee 
where $c_n^{\prime}$ is an arbitrary constant 
allowed by the boundary conditions $\psi_n^c (\pm \infty) = \mathrm{constant}$,
and we defined
\be 
\label{F}
F_n^{(j)} (t)
\equiv 
\int_{-\infty}^{\infty} d \tau \, 
\psi_{n - j - 1}^c (\tau) \, 
\bigl[ 
\hat{\tilde{a}}_0 (\tau) \, , 
\hat{\tilde{a}}_j^{\dagger} (t) 
\bigr]
\, .
\ee 
The dual wavefunction $\Psi^c (t)$~\eqref{psi_expand_star} 
can be iteratively constructed from eq.~\eqref{psi^c_iter}
since the $\mathcal{O}(\lambda^n)$ term ${\displaystyle \psi_n^c} (t)$
depends on the previous $n$ lower-order corrections 
$\left\{ {\displaystyle \psi_j^c} \right\}_{j = 0}^{n - 1}$
(with ${\displaystyle \psi_0^c}$ identified as $\psi_0^{\ast}$).
In particular, 
the solution of $\psi_n^c (t)$ for $n \geq 2$ is given by~\eqref{psi_n^c}
if and only if the following equation holds:
\begin{align}
& 
\sum_{j = 0}^{n - 2} 
F_n^{(j)} (t)
+
\frac{1}{2 N \sigma} \, 
F_n^{(n - 1)} (t) 
\nn \\
&=
(i \lambda)^{n - 1} \, \overbar{\psi}_0^{\ast}
\bigintssss_{-\infty}^{\infty}
\Biggl[ 
\prod_{k = 0}^{n - 1}
d t_k \, b(t_k)
\Biggr]
\Theta(t_{n - 1} - t_{n - 2} - \sigma) 
\cdots 
\Theta(t_2 - t_1 - \sigma) \, 
\Theta(t_1 - t - \sigma) 
\nn \\
& \quad \ 
+
\sum_{j = 0}^{n - 2} 
\lambda^{n - j - 1} \, c_{n - j - 1}^{\prime} 
\int_{-\infty}^{\infty} \frac{d \tau}{2 N \sigma} \, 
\bigl[ 
\hat{\tilde{a}}_0 (\tau) \, , 
\hat{\tilde{a}}_j^{\dagger} (t) 
\bigr]
\label{psi^c_ansatz}
\, .
\end{align}
We shall prove this equality using mathematical induction.

For $n = 2$, the equality is satisfied because
\begin{align}
F_2^{(0)} (t)
+
\frac{1}{2 N \sigma} \, 
F_2^{(1)} (t) 
&=
\int_{-\infty}^{\infty} d t_1 \, 
\psi_1^c (t_1) \, 
\bigl[ 
\hat{\tilde{a}}_0 (t_1) \, ,
\hat{\tilde{a}}_0^{\dagger} (t)
\bigr]
+
\int_{-\infty}^{\infty} d t_1 \, 
\frac{\psi_0^{\ast} (t_1)}{2 N \sigma} \, 
\bigl[ 
\hat{\tilde{a}}_0 (t_1) \, ,
\hat{\tilde{a}}_1^{\dagger} (t)
\bigr]
\nn \\
&=
i \lambda \, 
\overbar{\psi}_0^{\ast} 
\int_{-\infty}^{\infty} d t_1 \,
b(t_1) \, 
\Theta(t_1 - t - \sigma) 
+
\lambda \, c_1^{\prime}
\, .
\end{align}
Having shown as the base case that the equality~\eqref{psi^c_ansatz} 
is true up to $n = 2$,
we assume that it is valid up to $\mathcal{O}(\lambda^{n - 1})$,
and show below that it holds true at $\mathcal{O}(\lambda^n)$ as well.

Let us examine the summation $\sum_{j = 0}^{n - 2} {\displaystyle F_n^{(j)}} (t)$
on the left-hand side of eq.~\eqref{psi^c_ansatz}.
We substitute the wave functions $\bigl\{ \psi_j^c \bigr\}_{j = 0}^{n - 1}$
into the expression~\eqref{F} for ${\displaystyle F_n^{(j)}} (t)$
using the ansatz~\eqref{psi_n^c}.
For the $j = 0$ term, 
this yields
\begin{align}
F_n^{(0)} (t)
= 
\ &
\lambda^{n - 1} \, c_{n - 1}^{\prime} 
\int_{-\infty}^{\infty} 
\frac{d \tau}{2 N \sigma} \, 
\bigl[ 
\hat{\tilde{a}}_0 (\tau) \, , 
\hat{\tilde{a}}_0^{\dagger} (t) 
\bigr]
\nn \\
& 
+
(i \lambda)^{n - 1} \, 
\overbar{\psi}_0^{\ast} 
\bigintssss_{-\infty}^{\infty}
\Biggl[ \, 
\prod_{k = 1}^{n - 1}
d t_k \, b(t_k)
\Biggr]
\underbrace{\Theta(t_{n - 1} - t_{n - 2} - \sigma) 
\cdots 
\Theta(t_3 - t_2 - \sigma) \, 
\Theta(t_2 - t_1 - \sigma)}_{n - 2}
\nn \\
& \qquad \qquad \qquad \qquad \ 
\times 
\left[ 
\Theta(t_1 - t - \sigma)
+
\underline{\Theta(t - t_1 - \sigma)}
\right]
,
\label{F0}
\end{align}
and for the $j = 1$ term, we obtain
\begin{align}
F_n^{(1)} (t)
= 
\ &
\lambda^{n - 2} \, c_{n - 2}^{\prime} 
\int_{-\infty}^{\infty} 
\frac{d \tau}{2 N \sigma} \, 
\bigl[ 
\hat{\tilde{a}}_0 (\tau) \, , 
\hat{\tilde{a}}_1^{\dagger} (t) 
\bigr]
\nn \\
&
- (i \lambda)^{n - 1} \, 
\overbar{\psi}_0^{\ast} 
\bigintssss_{-\infty}^{\infty}
\Biggl[ \, 
\prod_{k = 1}^{n - 1}
d t_k \, b(t_k)
\Biggr]
\Theta(t - t_{n - 1} - \sigma)
\nn \\
& \qquad \qquad \qquad \qquad \ 
\times 
\underbrace{\Theta(t_{n - 2} - t_{n - 3} - \sigma) 
\cdots 
\Theta(t_3 - t_2 - \sigma) \, 
\Theta(t_2 - t_1 - \sigma)}_{n - 3}
\nn \\
& \qquad \qquad \qquad \qquad \ 
\times 
\left[ 
\underline{\Theta(t_1 - t_{n - 1} - \sigma)}
+
\Theta(t_{n - 1} - t_1 - \sigma)
\right]
.
\label{F1}
\end{align}
From the expressions above, 
we observe that the terms involving the underlined step functions in 
${\displaystyle F_n^{(0)}} (t)$ and ${\displaystyle F_n^{(1)}} (t)$ 
cancel each other out.
Similarly, for the $j$-th term in the summation,
we have 
\begin{align}
F_n^{(j)} (t)
= 
\ & 
\lambda^{n - j - 1} \, c_{n - j - 1}^{\prime} 
\int_{-\infty}^{\infty} \frac{d \tau}{2 N \sigma} \, 
\bigl[ 
\hat{\tilde{a}}_0 (\tau) \, , 
\hat{\tilde{a}}_j^{\dagger} (t) 
\bigr]
\nn \\
& +
(i \lambda)^{n - 1} 
\left( -1 \right)^j
\overbar{\psi}_0^{\ast} 
\bigintssss_{-\infty}^{\infty}
\Biggl[ \, 
\prod_{k = 1}^{n - 1}
d t_k \, b(t_k)
\Biggr]
\Theta(t - t_{n - 1} - \sigma)
\nn \\
& \qquad \qquad \qquad \qquad \qquad \quad 
\times 
\underbrace{\Theta(t_{n - 1} - t_{n - 2} - \sigma) 
\cdots
\Theta(t_{n - j + 1} - t_{n - j} - \sigma)}_{j - 1}
\nn \\
& \qquad \qquad \qquad \qquad \qquad \quad 
\times 
\underbrace{\Theta(t_{n - j - 1} - t_{n - j - 2} - \sigma) \cdots 
\Theta(t_2 - t_1 - \sigma)}_{n - j - 2} 
\nn \\
& \qquad \qquad \qquad \qquad \qquad \quad 
\times 
\left[ 
\Theta(t_1 - t_{n - j} - \sigma)
+
\underline{\Theta(t_{n - j} - t_1 - \sigma)}
\right]
.
\label{Fj}
\end{align}
Here, 
the term with the underlined part 
cancels out with the corresponding underlined term in the 
$(j + 1)$-th contribution to the summation:
\begin{align}
F_n^{(j + 1)} (t)
= 
\ & 
\lambda^{n - j - 2} \, c_{n - j - 2}^{\prime} 
\int_{-\infty}^{\infty} \frac{d \tau}{2 N \sigma} \, 
\bigl[ 
\hat{\tilde{a}}_0 (\tau) \, , 
\hat{\tilde{a}}_{j + 1}^{\dagger} (t) 
\bigr]
\nn \\
&
- (i \lambda)^{n - 1} 
\left( -1 \right)^j
\overbar{\psi}_0^{\ast} 
\bigintssss_{-\infty}^{\infty}
\Biggl[ \, 
\prod_{k = 1}^{n - 1}
d t_k \, b(t_k)
\Biggr]
\Theta(t - t_{n - 1} - \sigma)
\nn \\
& \qquad \qquad \qquad \qquad \qquad \quad 
\times 
\underbrace{\Theta(t_{n - 1} - t_{n - 2} - \sigma) 
\cdots
\Theta(t_{n - j} - t_{n - j - 1} - \sigma)}_{j}
\nn \\
& \qquad \qquad \qquad \qquad \qquad \quad 
\times 
\underbrace{\Theta(t_{n - j - 2} - t_{n - j - 3} - \sigma) \cdots 
\Theta(t_2 - t_1 - \sigma)}_{n - j - 3} 
\nn \\
& \qquad \qquad \qquad \qquad \qquad \quad 
\times 
\left[ 
\underline{\Theta(t_1 - t_{n - j - 1} - \sigma)}
+
\Theta(t_{n - j - 1} - t_1 - \sigma)
\right]
.
\label{Fj+1}
\end{align}
Finally, the last term ($j = n - 2$) in the summation is given by 
\begin{align}
F_n^{(n - 2)} (t)
=
\ & 
\lambda \, c_1^{\prime} 
\int_{-\infty}^{\infty} \frac{d \tau}{2 N \sigma} \, 
\bigl[ 
\hat{\tilde{a}}_0 (\tau) \, , 
\hat{\tilde{a}}_{n - 2}^{\dagger} (t) 
\bigr]
\nn \\
& +
(i \lambda)^{n - 1} \, (-1)^{n - 2} \, 
\overbar{\psi}_0^{\ast} 
\bigintssss_{-\infty}^{\infty} 
\Biggl[ \, 
\prod_{k = 1}^{n - 1}
d t_k \, b(t_k)
\Biggr] 
\Theta(t - t_{n - 1} - \sigma) 
\nn \\
& \qquad \qquad \qquad \qquad \qquad \qquad 
\times 
\Theta(t_{n - 1} - t_{n - 2} - \sigma) 
\cdots 
\Theta(t_3 - t_2 - \sigma) 
\nn \\
& \qquad \qquad \qquad \qquad \qquad \qquad 
\times 
\left[ 
\underline{\Theta(t_1 - t_2 - \sigma)}
+
\Theta(t_2 - t_1 - \sigma)
\right]
.
\end{align}
Based on the preceding analysis, 
the term containing the underlined part 
is canceled out by a contribution from ${\displaystyle F_n^{(n - 3)}} (t)$. 
Furthermore, the remaining part of the second term in ${\displaystyle F_n^{(n - 2)}} (t)$,
which does not contain an underlined step function, 
cancels with the second term 
on the left-hand side of eq.~\eqref{psi^c_ansatz} since
\be 
\begin{aligned}
\frac{F_n^{(n - 1)} (t)}{2 N \sigma}
=
(- i \lambda)^{n - 1} \, 
\overbar{\psi}_0^{\ast} 
\bigintssss_{-\infty}^{\infty} 
\Biggl[ 
&
\prod_{k = 1}^{n - 1}
d t_k \, b(t_k)
\Biggr] 
\Theta(t - t_{n - 1} - \sigma) 
\\
& \times 
\Theta(t_{n - 1} - t_{n - 2} - \sigma) 
\cdots 
\Theta(t_2 - t_1 - \sigma) 
\, .
\end{aligned}
\ee 

To summarize, after performing the summation, 
the only terms that remain on the left-hand side of eq.~\eqref{psi^c_ansatz}
are contributions from the constants $\bigl\{ c_j^{\prime} \bigr\}_{j = 1}^{n - 1}$:
\be 
\sum_{j = 0}^{n - 2} 
\lambda^{n - j - 1} \, c_{n - j - 1}^{\prime} 
\int_{-\infty}^{\infty} \frac{d \tau}{2 N \sigma} \, 
\bigl[ 
\hat{\tilde{a}}_0 (\tau) \, , 
\hat{\tilde{a}}_j^{\dagger} (t) 
\bigr]
\, ,
\ee 
as well as the part of the second term in eq.~\eqref{F0} that is not underlined:
\be 
(i \lambda)^{n - 1} \, \overbar{\psi}_0^{\ast}
\bigintssss_{-\infty}^{\infty}
\Biggl[ \, 
\prod_{k = 0}^{n - 1}
d t_k \, b(t_k)
\Biggr]
\Theta(t_{n - 1} - t_{n - 2} - \sigma) 
\cdots 
\Theta(t_2 - t_1 - \sigma) \, 
\Theta(t_1 - t - \sigma) 
\, .
\ee 
All other terms are eliminated through cancellations.
Thus, we have established the validity of 
the equality~\eqref{psi^c_ansatz} at order $\lambda^n$, 
completing the induction step.

\section{Hamiltonian to All Orders}
\label{app:Ham}

With the zeroth-order Hamiltonian 
deduced in eq.~\eqref{H0_2},
we show in this appendix that the interaction Hamiltonian 
that generates the operator evolution 
consistent with the path-integral formalism 
is indeed the form given by eq.~\eqref{H_int}. 

At order $\lambda^n$,
the interaction Hamiltonian 
$\hat{H}_{\mathrm{int}} (t) 
= \sum_{j = 1}^n \hat{H}_j (t) + \mathcal{O}(\lambda^{n + 1})$
is required to satisfy the Heisenberg equation
\be 
\label{heisenberg_n}
\del_t \, \hat{\tilde{a}}_n (t)
=
- i 
\sum_{j = 0}^{n - 1} \, 
\bigl[ 
\hat{\tilde{a}}_j (t) \, ,
\hat{H}_{n - j} (t)
\bigr]
\, ,
\ee 
where the operators $\hat{\tilde{a}}_n (t)$~\eqref{an}
follow the time evolution governed by 
\be 
\label{an_del_t}
\del_t \, \hat{\tilde{a}}_n (t)
=
i \lambda \, b(t - \sigma) \, \hat{\tilde{a}}_{n - 1} (t - \sigma)
\qquad 
\forall \ 
n \geq 1
\, ,
\ee 
while $\hat{\tilde{a}}_n^{\dagger} (t)$
is simply the Hermitian adjoint of $\hat{\tilde{a}}_n (t)$
with $\sigma^* = \sigma$. 

In what follows, 
we explicitly verify that eq.~\eqref{H_int}, 
in which each $\mathcal{O}(\lambda^n)$ term 
has the form
\be 
\hat{H}_n (t)
=
- \lambda \, b(t - \sigma) 
\sum_{j = 0}^{n - 1} 
\hat{\tilde{a}}_j^{\dagger} (t - \sigma) \, 
\hat{\tilde{a}}_{n - 1 - j} (t - \sigma)
\qquad 
\forall \ 
n \geq 1
\, ,
\ee 
satisfies eq.~\eqref{heisenberg_n}.
To establish this, 
it suffices to prove that the operators defined by
eqs.~\eqref{an} and~\eqref{an_dag} obey the identity
\be 
\label{a_n-1_heisenberg}
\sum_{j = 0}^{n - 1} 
\left[ 
\hat{\tilde{a}}_j (t) \, ,
\sum_{k = 0}^{n - j - 1}
\hat{\tilde{a}}_k^{\dagger} (t - \sigma) \, 
\hat{\tilde{a}}_{n - j - 1 - k} (t - \sigma)
\right]
=
\hat{\tilde{a}}_{n - 1} (t - \sigma)
\, 
,
\ee 
which is what we will confirm below.

We shall focus on how the commutators 
with different indices $j$ 
on the left-hand side of eq.~\eqref{a_n-1_heisenberg}
combine and cancel out each other, 
leaving just $\hat{\tilde{a}}_{n - 1} (t - \sigma)$.
Starting with $j = 0$, we have 
\begin{align}
&
\left[ 
\hat{\tilde{a}}_0 (t) \, ,
\sum_{k = 0}^{n - 1}
\hat{\tilde{a}}_k^{\dagger} (t - \sigma) \, 
\hat{\tilde{a}}_{n - 1 - k} (t - \sigma)
\right]
\nn \\
&
=
\hat{\tilde{a}}_{n - 1} (t - \sigma) 
-
i \lambda \, 
\hat{\tilde{a}}_{n - 2} (t - \sigma) 
\int_{-\infty}^{\infty} d t_1 \, b(t_1) \, 
\Theta(t - t_1 - 2 \sigma) 
\nn \\
&
\quad 
+
\sum_{k = 2}^{n - 1}
\hat{\tilde{a}}_{n - 1 - k} (t - \sigma)
\left( - i \lambda \right)^k
\bigintssss_{-\infty}^{\infty} 
\Biggl[ \, 
\prod_{m = 1}^k 
d t_m \, b(t_m) 
\Biggr]
\Theta(t - t_k - 2 \sigma) \,
\Theta(t - t_1 - \sigma) 
\nn \\
& \qquad \qquad \qquad \qquad \qquad \qquad \qquad \quad \ \ 
\times 
\Theta(t_k - t_{k - 1} - \sigma)
\cdots 
\Theta(t_2 - t_1 - \sigma)
\, ,
\label{j = 0}
\end{align}
and for $j = 1$ we have 
\begin{align}
&
\left[ 
\hat{\tilde{a}}_1 (t) \, ,
\sum_{k = 0}^{n - 2}
\hat{\tilde{a}}_k^{\dagger} (t - \sigma) \, 
\hat{\tilde{a}}_{n - 2 - k} (t - \sigma)
\right]
\nn \\
&=
i \lambda \, 
\hat{\tilde{a}}_{n - 2} (t - \sigma)
\int_{-\infty}^{\infty} d t_1 \, b(t_1) \, 
\Theta(t - t_1 - 2 \sigma)
\nn \\
& \quad 
+
\sum_{k = 1}^{n - 2} 
\hat{\tilde{a}}_{n - 2 - k} (t - \sigma) 
\cdot 
\left(i \lambda \right)^{k + 1} 
\left( - 1 \right)^k
\bigintssss_{-\infty}^{\infty}
\Biggl[ \, 
\prod_{m = 1}^{k + 1}
d t_m \, b(t_m) 
\Biggr] 
\Theta(t - t_{k + 1} - 2 \sigma) \, 
\Theta(t - t_1 - \sigma) 
\nn \\
& \qquad \qquad \qquad \qquad \qquad \qquad \qquad \qquad 
\qquad \quad
\times 
\Theta(t_{k + 1} - t_k - \sigma) \cdots 
\Theta(t_3 - t_2 - \sigma) 
\nn \\
& \qquad \qquad \qquad \qquad \qquad \qquad \qquad \qquad 
\qquad \quad
\times 
\Bigl[
\Theta(t_1 - t_2 - \sigma)
+
\underline{\Theta(t_2 - t_1 - \sigma)}
\Bigr] 
\, .
\label{j = 1}
\end{align}
The second term on the right-hand side of eq.~\eqref{j = 0} 
cancels with the first term on the right-hand side of eq.~\eqref{j = 1}. 
Additionally, 
by shifting the index $k$ of the summation in eq.~\eqref{j = 0} by 1, 
it becomes evident that the summation over $k$
cancels the terms involving the underlined part in eq.~\eqref{j = 1}.

For general $j$, we write 
\begin{align}
&
\left[ 
\hat{\tilde{a}}_j (t) \, ,
\sum_{k = 0}^{n - j - 1}
\hat{\tilde{a}}_k^{\dagger} (t - \sigma) \, 
\hat{\tilde{a}}_{n - j - 1 - k} (t - \sigma)
\right]
\nn \\
&=
\left( i \lambda \right)^j 
\hat{\tilde{a}}_{n - j - 1} (t - \sigma)
\bigintssss_{-\infty}^{\infty} 
\Biggl[ \, 
\prod_{k = 1}^j
d t_k \, b(t_k) 
\Biggr] 
\Theta(t - t_j - \sigma) 
\cdots 
\Theta(t_2 - t_1 - \sigma) \, 
\Theta(t - t_1 - 2 \sigma)
\nn \\
& \quad 
+
\sum_{k = 1}^{n - j - 1} 
\hat{\tilde{a}}_{n - j - 1 - k} (t - \sigma) 
\cdot 
\left(i \lambda \right)^{j + k}
\left( - 1 \right)^k
\bigintssss_{-\infty}^{\infty} 
\Biggl[ \, 
\prod_{m = 1}^{j + k}
d t_m \, b(t_m) 
\Biggr] 
\Theta(t - t_{j + k} - 2 \sigma) \, 
\Theta(t - t_j - \sigma) 
\nn \\
& \qquad \qquad \qquad \qquad \qquad \qquad \qquad \qquad 
\qquad \qquad \ \ 
\times 
\Theta(t_{j + k} - t_{j + k - 1} - \sigma) 
\cdots 
\Theta(t_{j + 2} - t_{j + 1} - \sigma)
\nn \\
& \qquad \qquad \qquad \qquad \qquad \qquad \qquad \qquad 
\qquad \qquad \ \ 
\times 
\Theta(t_j - t_{j - 1} - \sigma) 
\cdots 
\Theta(t_2 - t_1 - \sigma)
\nn \\
& \qquad \qquad \qquad \qquad \qquad \qquad \qquad \qquad 
\qquad \qquad \ \  
\times 
\left[ 
\underline{\Theta(t_1 - t_{j + 1} - \sigma)}
+
\Theta(t_{j + 1} - t_1 - \sigma)
\right]
.
\label{j}
\end{align}
Meanwhile, 
for $j + 1$, we have 
\begin{align}
&
\left[ 
\hat{\tilde{a}}_{j + 1} (t) \, ,
\sum_{k = 0}^{n - j - 2}
\hat{\tilde{a}}_k^{\dagger} (t - \sigma) \, 
\hat{\tilde{a}}_{n - j - 2 - k} (t - \sigma)
\right]
\nn \\
&=
\left( i \lambda \right)^{j + 1} 
\hat{\tilde{a}}_{n - j - 2} (t - \sigma)
\bigintssss_{-\infty}^{\infty} 
\Biggl[ \, 
\prod_{k = 1}^{j + 1} 
d t_k \, b(t_k) 
\Biggr] 
\Theta(t - t_{j + 1} - \sigma) 
\cdots 
\Theta(t_2 - t_1 - \sigma) \, 
\Theta(t - t_1 - 2 \sigma)
\nn \\
& \quad 
+
\sum_{k = 1}^{n - j - 2} 
\hat{\tilde{a}}_{n - j - 2 - k} (t - \sigma) 
\cdot 
\left(i \lambda \right)^{j + k + 1}
\left( - 1 \right)^k
\bigintssss_{-\infty}^{\infty} 
\Biggl[
\prod_{m = 1}^{j + k + 1}
d t_m \, b(t_m) 
\Biggr] 
\Theta(t - t_{j + k + 1} - 2 \sigma) \, 
\Theta(t - t_{j + 1} - \sigma)
\nn \\
& \qquad \qquad \qquad \qquad \qquad \qquad \qquad \qquad 
\qquad \qquad \qquad 
\times 
\Theta(t_{j + k + 1} - t_{j + k} - \sigma) 
\cdots 
\Theta(t_{j + 3} - t_{j + 2} - \sigma)
\nn \\
& \qquad \qquad \qquad \qquad \qquad \qquad \qquad \qquad 
\qquad \qquad \qquad 
\times 
\Theta(t_{j + 1} - t_j - \sigma)
\cdots 
\Theta(t_2 - t_1 - \sigma)
\nn \\
& \qquad \qquad \qquad \qquad \qquad \qquad \qquad \qquad 
\qquad \qquad \qquad  
\times 
\left[ 
\Theta(t_1 - t_{j + 2} - \sigma)
+
\underline{\Theta(t_{j + 2} - t_1 - \sigma)}
\right]
.
\label{j + 1}
\end{align}
Again, we observe that the terms in eq.~\eqref{j} 
involving the underlined part cancel with 
both the first term on the right-hand side of eq.~\eqref{j + 1} 
and the underlined summation terms in the same equation.

Finally, the last two contributions 
to the summation in eq.~\eqref{a_n-1_heisenberg} 
are the $j = n - 2$ term:
\begin{align}
&\left[ 
\hat{\tilde{a}}_{n - 2} (t)
\, ,
\sum_{k = 0}^1 
\hat{\tilde{a}}_k^{\dagger} (t - \sigma) \, 
\hat{\tilde{a}}_{1 - k} (t - \sigma)
\right]
\nn \\
&= 
\uwave{( k = 0 \ \text{term})}
-
(i \lambda)^{n - 1} \, 
\hat{\tilde{a}}_0 (t - \sigma)
\bigintssss_{-\infty}^{\infty} 
\Biggl[ \, 
\prod_{k = 1}^{n - 1}
d t_k \, b(t_k) 
\Biggr] 
\Theta(t - t_{n - 1} - 2 \sigma) \, 
\Theta(t - t_{n - 2} - \sigma) 
\nn \\
& \qquad \qquad \qquad \qquad \qquad \qquad
\qquad \qquad \qquad  
\times 
\Theta(t_{n - 2} - t_{n - 3} - \sigma) 
\cdots \Theta(t_2 - t_1 - \sigma) 
\nn \\
& \qquad \qquad \qquad \qquad \qquad \qquad
\qquad \qquad \qquad 
\times 
\Bigl[
\underline{\Theta(t_1 - t_{n - 1} - \sigma)}
+
\uwave{\Theta(t_{n - 1} - t_1 - \sigma)}
\Bigr] 
\, ,
\label{n - 2}
\end{align}
and the $j = n - 1$ term: 
\begin{align}
&
\left[ 
\hat{\tilde{a}}_{n - 1} (t)
\, , 
\hat{\tilde{a}}_0^{\dagger} (t - \sigma) \, 
\hat{\tilde{a}}_0 (t - \sigma)
\right]
\nn \\
&=
\left( i \lambda \right)^{n - 1}  
\hat{\tilde{a}}_0 (t - \sigma) 
\bigintssss_{-\infty}^{\infty} 
\Biggl[ \, 
\prod_{k = 1}^{n - 1}
d t_k \, b(t_k) 
\Biggr] 
\Theta(t - t_{n - 1} - \sigma) \, 
\Theta(t_{n - 1} - t_{n - 2} - \sigma)
\cdots 
\Theta(t_2 - t_1 - \sigma) 
\nn \\
& \qquad \qquad \qquad \qquad \qquad \quad \ 
\times 
\Theta(t - t_1 - 2 \sigma)
\, .
\label{n - 1}
\end{align}
Similar to what was demonstrated earlier in the case of general $j$, 
the parts of eq.~\eqref{n - 2} underlined with wiggly lines 
are canceled by contributions from the $j = n - 3$ term.
Furthermore, 
the part of eq.~\eqref{n - 2} underlined with a straight line 
exactly cancels the contribution from eq.~\eqref{n - 1}. 

In summary, after the entire summation, 
only the first term $\hat{\tilde{a}}_{n - 1} (t - \sigma)$
on the right-hand side of eq.~\eqref{j = 0} remains, i.e.,
\be 
\sum_{j = 0}^{n - 1} 
\left[ 
\hat{\tilde{a}}_j (t) \, ,
\sum_{k = 0}^{n - j - 1}
\hat{\tilde{a}}_k^{\dagger} (t - \sigma) \, 
\hat{\tilde{a}}_{n - j - 1 - k} (t - \sigma)
\right]
=
\hat{\tilde{a}}_{n - 1} (t - \sigma)
\, .
\ee 
This verifies that the interaction Hamiltonian is indeed given by
\be 
\hat{H}_{\mathrm{int}} (t)
=
- \lambda \, b(t - \sigma) \, 
\hat{\tilde{a}}^{\dagger} (t - \sigma) \, 
\hat{\tilde{a}} (t - \sigma)
\ee 
to all orders in $\lambda$. 

\section{Comparison With Conventional Approach}
\label{app:Llosa}

In this appendix, we apply the nonlocal Hamiltonian formalism put forward in ref.~\cite{Llosa94} to the free-field part of the nonlocal 1D model~\eqref{S_a_tilde}.

Following ref.~\cite{Llosa94}, the primary constraint $C_A (t \, , \tau)$ on the conjugate momentum $P_A(t \, , \tau)$ is given by
\begin{align}
0
\approx 
C_A (t \, , \tau)
&
\equiv 
P_A (t \, , \tau) 
-
\int_{-\infty}^{\infty} d \tau' 
\left[ \Theta(\tau) - \Theta(\tau') \right]
\frac{\delta \mathcal{L}[\tilde{A} \, , \tilde{A}^{\dagger}](t \, , \tau')}
{\delta \tilde{A} (t \, , \tau)}
\nn \\
&= 
P_A (t \, , \tau) 
-
i \, \delta(\tau) \, 
\tilde{A}^{\dagger} (t \, , \tau - \sigma)
\, .
\label{C1}
\end{align}
Likewise, the primary constraint $C_{A^{\dagger}} (t \, , \tau)$
on $P_{A^{\dagger}}(t \, , \tau)$ is 
\begin{align}
0
\approx 
C_{A^{\dagger}} (t \, , \tau)
&
\equiv 
P_{A^{\dagger}} (t \, , \tau) 
-
\int_{-\infty}^{\infty} d \tau' 
\left[ \Theta(\tau) - \Theta(\tau') \right]
\frac{\delta \mathcal{L}[\tilde{A} \, , \tilde{A}^{\dagger}](t \, , \tau')}
{\delta \tilde{A}^{\dagger} (t \, , \tau)}
\nn \\
&= 
P_{A^{\dagger}} (t \, , \tau) 
+
i 
\left[ \Theta( \tau + \sigma) - \Theta(\tau) \right]
\del_{\tau} 
\tilde{A} (t \, , \tau + \sigma)
\, .
\label{C2}
\end{align}
Note that the extra coordinate $\tau \in (-\infty, \infty)$
introduced in this framework serves as a continuous parameter
indexing the constraints.

In the conventional approach,
the natural next step is to 
require the preservation of the primary constraints 
$C_A (t \, , \tau) \approx 0$ and $C_{A^{\dagger}} (t \, , \tau) \approx 0$
under time evolution governed by the Hamiltonian~\eqref{H_A}.
This procedure would yield secondary constraints~\cite{Llosa94}
\begin{align}
C_A^{(2)}(t \, , \tau)
&=
-i \, \del_{\tau} 
\tilde{A}^{\dagger} (t \, , \tau - \sigma)
\approx 
0
\, ,
\label{secondary_A}
\\
C_{A^{\dagger}}^{(2)}(t \, , \tau)
&=
i \, \del_{\tau} 
\tilde{A} (t \, , \tau + \sigma)
\approx 
0
\label{secondary_A_dag}
\end{align}
that correspond to the original equations of motion
for $\tilde{a}(t)$ and $\tilde{a}^{\dagger}(t)$
when the chirality condition~\eqref{A=a} is applied.
Together with the primary constraints~\eqref{C1} and~\eqref{C2},
they form a set of second-class constraints for all values of $\tau$.
This establishes the equivalence between 
the dynamics of this constrained Hamiltonian system 
and the dynamics of the original nonlocal Lagrangian system
$L[\tilde{a} \, , \tilde{a}^{\dagger}]$
in the subspace of physical trajectories 
defined by the Euler-Lagrange equations~\cite{Llosa94}. 

On the other hand, recall that in the Hamiltonian formalism developed in section~\ref{sec:toy_Ham}, the operator algebra $[ \hat{\tilde{a}}(t) , \hat{\tilde{a}}^{\dagger}(t')]$~\eqref{comm0} was constructed solely based on the path-integral formalism, without imposing the equation-of-motion constraints on the fields themselves. 
Instead, these constraints were imposed on the Fock space~\eqref{fock} after quantization (see section~\ref{sec:phys}). 
Therefore, to facilitate a comparison with the operator algebra derived from our approach, we shall evaluate below the Dirac bracket 
$\left\{ 
\tilde{A} (t \, , \tau) \, , 
\tilde{A}^{\dagger} (t \, , \tau') 
\right\}_{\mathrm{D}}$
in the $(1 + 1)$-dimensional formalism 
but without employing the equation-of-motion constraints~\eqref{secondary_A} and~\eqref{secondary_A_dag}.

According to the Poisson algebras
\be 
\left\{ 
\tilde{A} (t \, , \tau) \, , 
P_A (t \, , \tau')
\right\}
=
\delta(\tau - \tau')
\qquad 
\text{and}
\qquad 
\bigl\{ 
\tilde{A}^{\dagger} (t \, , \tau) \, , 
P_{A^{\dagger}} (t \, , \tau')
\bigr\}
=
\delta(\tau - \tau')
\ee 
of the phase-space variables in the Hamiltonian system~\eqref{H_A},
the Poisson bracket between 
the primary constraints~\eqref{C1} and~\eqref{C2}
can be evaluated as 
\be 
\left\{ 
C_A (t \, , \tau) \, , 
C_{A^{\dagger}} (t \, , \tau')
\right\}
=
i 
\left[ \Theta(\tau') - \Theta(\tau' + \sigma) \right]
\del_{\tau'} \, 
\delta( \tau' - \tau + \sigma )
- i \, 
\delta(\tau) \, \delta(\tau - \tau' - \sigma)
\, .
\ee 
By setting $\Theta(0) \equiv 1$, 
it becomes clear that the Poisson bracket vanishes 
outside the domains $\tau \in [0, \sigma)$ or $\tau' \in [- \sigma, 0)$.
Keeping in mind that only the momentum constraints are implemented, 
this suggests that the subsets of constraints 
\be 
\label{1st_class}
\left\{ 
C_A (t \, , \tau) 
\ \vert \ 
\tau \notin \Sigma_A
\right\}
\qquad \text{and} \qquad 
\left\{ 
C_{A^{\dagger}} (t \, , \tau) 
\ \vert \ 
\tau \notin \Sigma_{A^{\dagger}}
\right\}
\ee 
are \textit{first class},
where we have defined the intervals $\Sigma_A \equiv [0, \sigma)$ and 
$\Sigma_{A^{\dagger}} \equiv [- \sigma, 0)$.

\begin{figure}[t]
\centering
\includegraphics[scale=0.85]{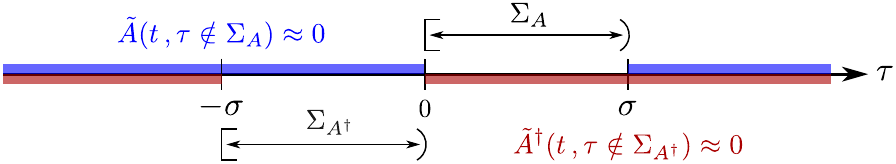}
\caption{\label{fig:gauge_constraint}
A figure illustrating the intervals $\Sigma_A = [0, \sigma)$
and $\Sigma_{A^{\dagger}} = [- \sigma , 0)$
along the $\tau$-axis, 
as well as the domains covered by the 
gauge-fixing constraints~\eqref{C5_gauge} (blue region)
and~\eqref{C6_gauge} (red region). 
}
\end{figure}

The infinitesimal gauge transformations generated by 
the first-class constraints~\eqref{1st_class} are given by
\be
\delta \tilde{A}(t \, , \tau)
=
\left\{ 
\tilde{A}(t \, , \tau) 
\, , 
\int_{-\infty}^{\infty} 
d \tau' \, \epsilon(t \, , \tau') \, 
C_A (t \, , \tau' \notin \Sigma_A)
\right\}
=
\epsilon(t \, , \tau) 
\qquad 
\text{for}
\quad 
\tau \notin \Sigma_A
\, ,
\ee 
and 
\be 
\delta \tilde{A}^{\dagger}(t \, , \tau)
=
\left\{ 
\tilde{A}^{\dagger}(t \, , \tau) 
\, , 
\int_{-\infty}^{\infty} 
d \tau' \, \epsilon^{\dagger} (t \, , \tau') \, 
C_{A^{\dagger}} (t \, , \tau' \notin \Sigma_{A^{\dagger}})
\right\}
=
\epsilon^{\dagger} (t \, , \tau) 
\qquad 
\text{for}
\quad 
\tau \notin \Sigma_{A^{\dagger}}
\, ,
\ee 
respectively.
These results imply that the value of the field
$\tilde{A}(t, \tau)$ outside the domain $\tau \in \Sigma_A = [0, \sigma)$,
and similarly the value of $\tilde{A}^{\dagger}(t, \tau)$
outside $\tau \in \Sigma_{A^{\dagger}} = [- \sigma, 0)$,
has no physical content,
as depicted in figure~\ref{fig:gauge_constraint}.
They represent gauge degrees of freedom of the system.
To proceed, we separate the first-class constraints~\eqref{1st_class}
from $C_A(t, \tau)$ and $C_{A^{\dagger}}(t, \tau)$,
and relabel the primary constraints as 
\be 
\begin{dcases}
C_1 (t \, , \tau \in \Sigma_A)
\equiv 
C_A(t \, , \tau \in \Sigma_A) 
\\ 
C_2 (t \, , \tau \notin \Sigma_A)
\equiv 
C_A(t \, , \tau \notin \Sigma_A) 
\end{dcases}
\, , 
\qquad 
\begin{dcases}
C_3 (t \, , \tau \in \Sigma_{A^{\dagger}})
\equiv 
C_{A^{\dagger}}(t \, , \tau \in \Sigma_{A^{\dagger}}) 
\\ 
C_4 (t \, , \tau \notin \Sigma_{A^{\dagger}})
\equiv 
C_{A^{\dagger}}(t \, , \tau \notin \Sigma_{A^{\dagger}}) 
\end{dcases}
\, .
\ee 
Moreover, 
we introduce extra gauge-fixing constraints of the form 
\begin{align}
C_5 (t \, , \tau \notin \Sigma_A) 
&\equiv 
\sigma^{-1} 
\tilde{A}(t \, , \tau \notin \Sigma_A)
\approx 
0
\, , 
\label{C5_gauge}
\\ 
C_6 (t \, , \tau \notin \Sigma_{A^{\dagger}}) 
&\equiv 
\sigma^{-1} 
\tilde{A}^{\dagger}(t \, , \tau \notin \Sigma_{A^{\dagger}})
\approx 
0
\, .
\label{C6_gauge}
\end{align}
These additional constraints are sufficient to convert 
$\left\{ C_{\alpha} \right\}_{\alpha = 1}^6$
into second-class constraints.

Let $M_{\alpha \beta}(\tau \, , \tau') 
\equiv 
\left\{ C_{\alpha} (t \, , \tau) \, , C_{\beta} (t \, , \tau') \right\}$ 
be the elements of a matrix $\mathbf{M}$ composed of the 
Poisson brackets between the constraints listed above.
Schematically,
the matrix $\mathbf{M}$ takes the antisymmetric form 
\be 
\mathbf{M} (\tau \, , \tau')
=
\begin{bmatrix}
0 & 0 & M_{13} & 0 & 0 & 0
\\
0 & 0 & 0 & 0 & M_{25} & 0
\\
M_{31} & 0 & 0 & 0 & 0 & 0
\\
0 & 0 & 0 & 0 & 0 & M_{46}
\\
0 & M_{52} & 0 & 0 & 0 & 0
\\
0 & 0 & 0 & M_{64} & 0 & 0
\end{bmatrix}
,
\ee 
where the nontrivial elements are given by 
\begin{align}
M_{13}(\tau \in \Sigma_A \, , \tau' \in \Sigma_{A^{\dagger}})
&=
\left\{ 
C_1 (t \, , \tau \in \Sigma_A) \, , 
C_3 (t \, , \tau' \in \Sigma_{A^{\dagger}})
\right\}
\nn \\
&=
- i \, 
\del_{\tau'} \, 
\delta( \tau' - \tau + \sigma )
- i \, 
\delta(\tau) \, \delta(\tau - \tau' - \sigma)
\, , 
\\ 
M_{25}(\tau \notin \Sigma_A \, , \tau' \notin \Sigma_A)
&=
\left\{ C_2 (t, \tau \notin \Sigma_A) \, , 
C_5 (t, \tau' \notin \Sigma_A) \right\}
=
- \sigma^{-1} \, \delta(\tau - \tau')
\, , 
\\
M_{46}(\tau \notin \Sigma_{A^{\dagger}} 
\, , 
\tau' \notin \Sigma_{A^{\dagger}})
&=
\left\{ C_4 (t, \tau \notin \Sigma_{A^{\dagger}}) \, , 
C_6 (t, \tau' \notin \Sigma_{A^{\dagger}}) \right\}
=
- \sigma^{-1} \, \delta(\tau - \tau')
\, .
\end{align}
The elements of the inverse matrix $\mathbf{M}^{-1}$ 
can be subsequently obtained from the relation
\be 
\int_{-\infty}^{\infty} d \tau'' \, 
M_{\alpha \gamma} (\tau \, , \tau'') \, 
M_{\gamma \beta}^{-1}(\tau'', \tau')
=
\int_{-\infty}^{\infty} d \tau'' \, 
M_{\alpha \gamma}^{-1}(\tau \, , \tau'') \, 
M_{\gamma \beta} (\tau'', \tau')
=
\delta_{\alpha \beta} \, 
\delta(\tau - \tau')
\, .
\ee 
In particular,
the element $M_{31}^{-1}(\tau \, , \tau')$
satisfies the equation
\begin{align}
\delta(\tau - \tau')
&=
\int_{\Sigma_A} d \tau'' \, 
M_{31}^{-1} (\tau \, , \tau'') \, 
M_{13} (\tau'' \in \Sigma_A \, , \tau' \in \Sigma_{A^{\dagger}})
\nn \\
&=
- i \, \del_{\tau'} 
M_{31}^{-1} (\tau \, , \tau' + \sigma) 
-
i \, \delta(\tau' + \sigma) \, 
M_{31}^{-1} (\tau \, , 0)
\, ,
\end{align}
which yields 
\be 
M_{31}^{-1}(\tau \, , \tau' + \sigma)
=
i \Theta(\tau' - \tau)
\qquad 
\text{for}
\quad 
\tau \, , \tau' \in [ - \sigma , 0 )
\, .
\ee 
This can be further expressed as 
$M_{31}^{-1}(\tau \in \Sigma_{A^{\dagger}} \, , \tau' \in \Sigma_A)
= i \Theta(\tau' - \tau - \sigma)$,
and thus $M_{13}^{-1}$ is given by  
\be 
\label{M_13_inv}
M_{13}^{-1} (\tau \in \Sigma_A \, , \tau' \in \Sigma_{A^{\dagger}})
=
- i \Theta(\tau - \tau' - \sigma) 
\, .
\ee 
The remaining nontrivial inverse matrix elements are 
\begin{align}
M_{25}^{-1} (\tau \notin \Sigma_A \, , \tau' \notin \Sigma_A)
&=
\sigma \, \delta(\tau - \tau') 
\, , 
\\ 
M_{46}^{-1} ( \tau \notin \Sigma_{A^{\dagger}} 
\, , 
\tau' \notin \Sigma_{A^{\dagger}} )
&=
\sigma \, \delta(\tau - \tau') 
\, .
\end{align}

As stated previously,
we are interested in the Dirac bracket
$\left\{ 
\tilde{A} (t \, , \tau) \, , 
\tilde{A}^{\dagger} (t \, , \tau') 
\right\}_{\mathrm{D}}$
in the reduced phase space under the set of constraints 
$\left\{ C_{\alpha} \right\}_{\alpha = 1}^6$.
It is defined as~\cite{Dirac:1950pj, Henneaux:1992ig} 
\be 
\begin{aligned}
\bigl\{ 
\tilde{A} (t \, , \tau) \, , 
\tilde{A}^{\dagger} (t \, , \tau') 
\bigr\}_{\mathrm{D}}
=
\, &
\bigl\{ 
\tilde{A} (t \, , \tau) \, , 
\tilde{A}^{\dagger} (t \, , \tau') 
\bigr\}
\\
&-
\int d \tau_1 
\int d \tau_2 
\left\{ 
\tilde{A} (t \, , \tau) \, , 
C_{\alpha}(t \, , \tau_1)
\right\}
M_{\alpha \beta}^{-1} (\tau_1 \, , \tau_2) \, 
\bigl\{  
C_{\beta}(t \, , \tau_2) \, ,
\tilde{A}^{\dagger} (t \, , \tau')
\bigr\}
. 
\end{aligned}
\ee 
Eq.~\eqref{M_13_inv}, combined with the fact that 
\begin{align}
\left\{ 
\tilde{A} (t \, , \tau) \, , 
C_{\alpha}(t \, , \tau_1)
\right\}
&=
\delta_{\alpha 1} \, 
\delta(\tau - \tau_1 \ \vert \ \tau_1 \in \Sigma_A)
+
\delta_{\alpha 2} \, 
\delta(\tau - \tau_1 \ \vert \ \tau_1 \notin \Sigma_A)
\, , 
\\
\bigl\{ 
C_{\alpha}(t \, , \tau_2) \, , 
\tilde{A}^{\dagger} (t \, , \tau')
\bigr\}
&=
- \delta_{\alpha 3} \, 
\delta(\tau' - \tau_2 \ \vert \ \tau_2 \in \Sigma_{A^{\dagger}})
- \delta_{\alpha 4} \, 
\delta(\tau' - \tau_2 \ \vert \ \tau_2 \notin \Sigma_{A^{\dagger}})
\, ,
\end{align}
results in 
\be 
\bigl\{ 
\tilde{A} (t \, , \tau) \, , 
\tilde{A}^{\dagger} (t \, , \tau') 
\bigr\}_{\mathrm{D}}
=
\begin{dcases}
- i \Theta(\tau - \tau' - \sigma) 
&\quad 
\text{for} \ 
\tau \in \Sigma_A 
\wedge 
\tau' \in \Sigma_{A^{\dagger}}
\\ 
0
& \quad 
\text{otherwise}
\end{dcases}
\, , 
\ee 
which is consistent with the domains 
where the physical degrees of freedom of 
$\tilde{A}(t, \tau)$ and $\tilde{A}^{\dagger}(t, \tau)$ 
respectively reside. 
It can also be verified that the Dirac bracket obtained above
is preserved under time $t$ evolution 
as dictated by the Hamiltonian~\eqref{H_A}.
When expressed in terms of the original fields 
$\tilde{a}(t)$ and $\tilde{a}^{\dagger}(t)$
using Hamilton's equation~\eqref{A=a},
we obtain
\be 
\bigl\{ 
\tilde{a} (t) \, , 
\tilde{a}^{\dagger} (t') 
\bigr\}_{\mathrm{D}}
=
\begin{dcases}
- i \Theta(t - t' - \sigma) 
&\quad 
\text{for} \ 
0 < t - t' < 2 \sigma 
\\ 
0
& \quad 
\text{otherwise}
\end{dcases}
\, .
\ee 
As a consequence, 
applying canonical quantization 
on this constrained Hamiltonian system
leads to the commutator 
\be 
\bigl[
\hat{\tilde{a}} (t) \, , 
\hat{\tilde{a}}^{\dagger} (t') 
\bigr] 
=
\begin{dcases}
\Theta(t - t' - \sigma) 
&\quad 
\text{for} \ 
0 < t - t' < 2 \sigma 
\\ 
0
& \quad 
\text{otherwise}
\end{dcases}
\, ,
\ee 
which is similar but not entirely equivalent to 
either eq.~\eqref{aa_comm_t>t'} or eq.~\eqref{comm0} in our construction, 
primarily due to the lack of translation symmetry along the $\tau$-direction
in the $(1 + 1)$-dimensional Hamiltonian system~\eqref{H_A}. 

\section{Space-Time Uncertainty Relation}
\label{app:STUR}

In this appendix, we show that the light-cone uncertainty relation~\eqref{stur_UV} implies an uncertainty relation between space $x$ and time $t$.

Consider a particle state in the 2D toy model composed of purely positive-frequency outgoing modes.
Suppose that the state is characterized by a wave packet with size $T$ in the temporal $t$-direction and size $L$ in the spatial $x$-direction.
Any two physical events involving this state should occur within this rectangular region of size $T \times L$, and thus the magnitude of their separations in time and space are bounded from above as 
\be 
\Delta t \leq T
\, , 
\qquad
\Delta x \leq L
\, .
\ee 
In terms of the light-cone coordinates~\eqref{UandV},
we have 
\be
\Delta U 
\, , \,  
\Delta V \leq T + L
\, .
\ee
From the UV/IR relation 
$\Omega_U \leq \Delta V / 4 \ell_E^2$
and its ingoing counterpart
$\Omega_V \leq \Delta U / 4 \ell_E^2$,
it follows that
\be
\Omega_U \, , \, 
\Omega_V \leq \frac{T + L}{4 \ell_E^2}
\, .
\ee
Since $k^0 = \Omega_U + \Omega_V$~\eqref{lightcone_freq}, 
we obtain 
\be 
k^0 \leq \frac{T + L}{2 \ell_E^2}
\, .
\ee 
Combined with the time-energy uncertainty principle 
$\Delta t \, \Delta k^0 \geq 1$,
this implies
\be 
\label{ineq}
T 
\, \geq \, 
\Delta t 
\, \geq \,  
\frac{1}{\Delta k^0} 
\, \gtrsim \,  
\frac{1}{k^0}
\, \geq \, 
\frac{2 \ell_E^2}{T + L}
\, ,
\ee 
leading to 
\be 
\label{T(T+L)}
T \left( T + L \right) 
\geq 
2 \ell_E^2
\, .
\ee 
Similarly,
by interchanging the roles of $t$ and $x$ in 
the sequence of inequalities in eq.~\eqref{ineq},
we find
\be
\label{L(T+L)}
L \left( T + L \right) \geq 2 \ell_E^2
\, .
\ee
When $T < L$, we have $2L > T + L$, and thus 
eq.~\eqref{T(T+L)} can be written as 
\be
\label{TL_max-1}
TL \geq \ell_E^2
\, .
\ee
When $T > L$, eq.~\eqref{L(T+L)} also implies eq.~\eqref{TL_max-1}. 

In conclusion, the uncertainty relation~\eqref{stur_UV} in the light-cone frame suggests that the width and duration of a wave packet in the 2D toy model obeys a similar condition~\eqref{TL_max-1} in the $(t, x)$ coordinates.
In this sense, the 2D toy model offers an explicit realization of the space-time uncertainty principle
\be
\Delta t \, \Delta x \gtrsim \ell_E^2
\ee
proposed by Yoneya~\cite{Yoneya:1987gb, Yoneya:1989ai, Yoneya:1997gs, Yoneya:2000bt} 
as a fundamental principle underlying string theory.

\small

\bibliographystyle{myJHEP}
\bibliography{bibliography}

\end{document}